\documentclass[%
 reprint,
 twocolumn,
 nofootinbib,
 superscriptaddress,
 showpacs,
 preprintnumbers,
 amsmath,
 amssymb,
 aps,
 prd,
 floatfix,
]{revtex4-1}

\usepackage{latexsym} 	
\usepackage{amssymb} 
\usepackage{amsbsy}
\usepackage{amsmath}
\usepackage{epsfig}     
\usepackage{graphics,color}
\usepackage{multirow}
\usepackage[caption=false]{subfig}
\usepackage{hyperref}

\newcommand{\be}{\begin{equation}}
\newcommand{\ee}{\end{equation}}
\newcommand{\bea}{\begin{eqnarray}}
\newcommand{\eea}{\end{eqnarray}}
\newcommand{\bean}{\begin{eqnarray*}}
\newcommand{\eean}{\end{eqnarray*}}
\newcommand{\bit}{\begin{itemize}}
\newcommand{\eit}{\end{itemize}}

\newcommand{\tr}{\mbox{tr}\,}
\newcommand{\nn}{\nonumber}
\newcommand{\half}{\frac{1}{2}}
\newcommand{\id}{1\!\!1}

\newcommand{\hm}{\hspace*{-0.35cm}}

\newcommand{\qqquad}{\qquad\qquad}

\newcommand{\ket}{\rangle}
\newcommand{\bra}{\langle}
\newcommand{\Nconf}{{N_{\rm conf}}}
\newcommand{\veck}{\mathbf{k}}

\providecommand{\expv}[1]{\left\langle#1 \right\rangle}

\graphicspath{{./figs/}}

\begin{document}

\title{
Scalar field restricted Boltzmann machine \\as an ultraviolet regulator
}

\author{Gert Aarts}
\email{g.aarts@swansea.ac.uk}
\affiliation{Department of Physics, Swansea University, Swansea SA2 8PP, United Kingdom}
\affiliation{European Centre for Theoretical Studies in Nuclear Physics and Related Areas (ECT*) \& Fondazione Bruno Kessler, 
38123 Villazzano (TN), Italy}

\author{Biagio Lucini}
\email{b.lucini@swansea.ac.uk}
\affiliation{Department of Mathematics, Swansea University, Swansea, SA2 8PP,  United Kingdom}

\author{Chanju Park}
\email{c.j.park@swansea.ac.uk}
\affiliation{Department of Physics, Swansea University, Swansea SA2 8PP, United Kingdom}

\date{\today}

\begin{abstract}
  Restricted Boltzmann machines (RBMs) are well-known tools used in machine
  learning to learn probability distribution functions from data. We analyse
  RBMs with scalar fields on the nodes from the perspective of lattice field
  theory. Starting with the simplest case of Gaussian fields, we show that the
  RBM acts as an ultraviolet regulator, with the cutoff determined by either
  the number of hidden nodes or a model mass parameter. We verify these ideas
  in the scalar field case, where the target distribution is known, and explore
  implications for cases where it is not known using the MNIST dataset. We
  also demonstrate that infrared modes are learnt quickest.
\end{abstract}


\maketitle


\section{Introduction}
 \label{sec:intro}

  In recent years machine learning (ML) has gained tremendous popularity in the
  physical sciences \cite{Carleo_2019}. In theoretical nuclear and high-energy
  physics, ML is applied to a wide range of problems, see e.g.\ the reviews
  \cite{Mehta:2018dln,Zhou:2023pti}. In lattice field theory (LFT), there are
  applications to all aspects of LFT computations \cite{Boyda:2022nmh}, with
  the development of normalising flows to generate field configurations a
  particularly active area of research \cite{Cramner_2023,gerdes2022learning}.
  From a theoretical perspective, it is of interest to explore synergies
  between ML on the one hand and statistical physics and LFT on the other hand,
  as many ML problems can be studied using the tools of the latter, see e.g.\
  Ref.\ \cite{ariosto2023statistical}. The connection between neural networks,
  Markov random fields and (Euclidean) lattice field theory have indeed not gone
  unnoticed, leading to the notions of quantum field-theoretic machine learning
  (QFT/ML) \cite{Bachtis:2021xoh} and neural network/QFT
  correspondence \cite{Halverson:2020trp,Erbin:2021kqf}. Further
  exploration of this connection may be fruitful in both directions, providing
  potential insights relevant to both the ML and the LFT/QFT communities. 

  In this paper, we take a step in this direction by considering one of the
  simplest generative ML models, the restricted Boltzmann machine (RBM)
  \cite{smolensky,10.1162/089976602760128018}. We analyse the RBM with
  continuous fields as degrees of freedom from the perspective of a Euclidean
  LFT and give a complete understanding in the case of Gaussian fields. We
  verify our analytical insights using simple scalar field theories in one and
  two dimensions, for which the target distribution is known, and also the
  MNIST dataset, to demonstrate that our findings are indeed relevant for
  typical ML datasets without known target distributions. We are in particular
  interested in the choice of ``architecture,'' which admittedly is quite
  straightforward for an RBM, namely the number of hidden nodes as well as the
  choice of certain hyperparameters. Our main conclusion is that the scalar
  field RBM acts as an ultraviolet regulator, with the cutoff determined by
  either the number of hidden nodes or a model mass parameter. We will make
  clear what this implies for the MNIST dataset, but note here already that in
  QFT language the MNIST dataset is ultraviolet divergent and infrared safe.  

  The paper is organised as follows. In Sec.\ \ref{sec:scalar} we introduce
  scalar field RBMs from the perspective of LFT and give some exact solutions
  for the Gaussian case. The standard equations to train an RBM are summarised
  in Sec.\ \ref{sec:training}. In Sec.\ \ref{sec:SVD} we analyse these
  equations analytically and work out some simple examples in detail. The
  findings of this section will be further explored in the two following
  sections. First, we consider as target theories free scalar fields in one and
  two dimensions in Sec.\ \ref{sec:gaussian}, for which the target distribution
  is known. In Sec.\ \ref{sec:MNIST} we validate our findings for a dataset
  with an unknown distribution, namely the MNIST dataset. Options to add
  interactions are discussed in Sec.\ \ref{sec:interact}. A summary is given in
  the final section. Appendix \ref{app:pcd} contains some more details on the
  algorithm employed, while in Appendix \ref{sec:KL} the Kullback-Leibler
  divergence is evaluated in the Gaussian case.

\section{Scalar fields on a bipartite graph}
\label{sec:scalar}

\begin{figure}[t]
    \begin{center}
      \includegraphics[width=0.9\columnwidth]{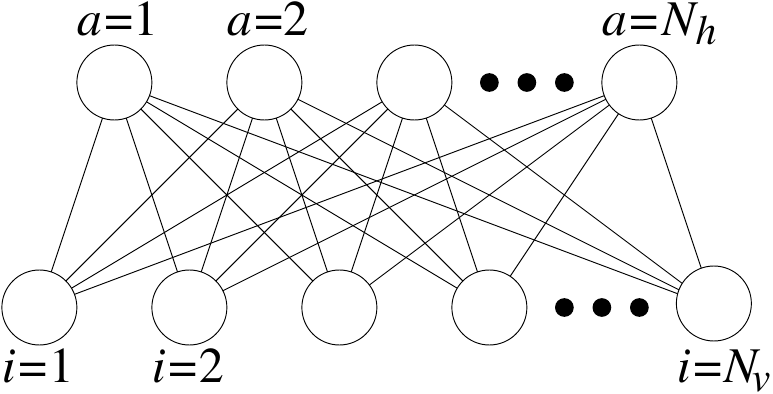}
    \end{center}
    \caption{Bipartite graph, with $N_v$ ($N_h$) nodes in the visible (hidden) layer.}
    \label{fig:bipartite}
  \end{figure}

  RBMs are defined on a bipartite graph, consisting of one visible layer 
  (with $N_v$ nodes) and one hidden layer (with $N_h$ nodes), see Fig.\ \ref{fig:bipartite}. 
  Importantly, there are no connections within each layer, only between the two layers. 
  The degrees of freedom living on the nodes can be discrete, as in an Ising model, 
  continuous or mixed; Ref.\ \cite{Decelle_2021} is a useful review.

  In this section, we consider an RBM from the viewpoint of lattice field
  theory. We consider continuous fields and denote these as $\phi_i$ ($i=1,
  \ldots, N_v$) for the visible layer and $h_a$ ($a=1, \ldots, N_h$) for the
  hidden layer. The layers are coupled via bilinear terms and involve the
  $N_v\times N_h$ weight matrix $W$, as
  \be
    \phi^T W h = \sum_{i=1}^{N_v} \sum_{a=1}^{N_h} \phi_i W_{ia} h_a.
  \ee
  The aim is to describe a probability distribution $p(\phi)$ on the visible
  layer, constructed by integrating over the hidden nodes in the joint
  probability distribution $p(\phi,h)$, as follows, 
  \be
    p(\phi) = \int Dh\, p(\phi,h),
    \quad
    p(\phi,h) = \frac{\exp(-S(\phi,h))}{Z},
  \ee
  where we have denoted the ``energy'' in the exponential as an action (following
  LFT notation) and the partition function reads
  \be
    Z = \int D\phi Dh \exp(-S(\phi,h)).
  \ee
  The integrals are over all nodes,
  \be
    \int D\phi = \prod_{i=1}^{N_v} \int_{-\infty}^\infty d\phi_i,
    \qquad
    \int Dh = \prod_{a=1}^{N_h} \int_{-\infty}^\infty dh_a.
  \ee
  Due to the absence of intralayer connections, the action takes a simple form in general,
  \be \label{eq:action}
    S(\phi,h) = V_\phi(\phi) + V_h(h) - \phi^T W h,
  \ee
  where the two potentials can be any function (as long as the integrals are
  well defined) and be node dependent, i.e.,
  \be \label{eq:potential}
    V_\phi(\phi) = \sum_i V_i^{(\phi)}(\phi_i),
    \qquad
    V_h(h) = \sum_a V_a^{(h)}(h_a).
  \ee
  Since there is no coupling between nodes within a layer, there is no
  ``kinetic'' or nearest-neighbour term; these are only generated via the
  coupling to the other layer. 

  To proceed, a natural starting point is to consider quadratic potentials, i.e.\
  free fields (we discuss interactions in Sec. \ref{sec:interact}). We hence
  consider as action,
  \bea
    S(\phi,h) = &&\hm\sum_i \half \mu^2 \phi_i^2 + \sum_a \frac{1}{2\sigma_h^2} 
    \left(h_a - \eta_a\right)^2 -\sum_{i,a} \phi_i W_{ia} h_a \nn\\
    = &&\hm\half \mu^2\phi^T\phi + \frac{1}{2\sigma_h^2} 
    \left(h - \eta\right)^T\left(h - \eta\right) - \phi^T W h.
  \eea
  A few comments are in order. We have denoted the prefactor as a mass term
  ($\mu^2$) in the case of $\phi$ and as a variance ($1/\sigma_h^2$) in the
  case of $h$; this is inessential, but emphasises that the model on the
  visible layer is ultimately the one we are interested in. Both $\mu^2$ and
  $\sigma_h^2$ are independent of the node; this is sufficient, as node
  dependence can be introduced via the weight matrix $W$, as we will see
  shortly. Finally, a source (or bias) $\eta_a$ is introduced in the hidden
  layer but not in the visible layer; again this is sufficient, as a nonzero
  bias breaks both symmetries, $h\to -h$, $\phi\to -\phi$. 

  Integrating out the hidden nodes then leads to the following distribution on
  the visible layer, 
  \bea 
     \hm p(\phi) &&\hm= \int Dh \, p(\phi,h) \nn\\ 
     &&\hm=\frac{1}{Z} \exp
        \left(
          -\frac{1}{2} \phi^T K \phi +  \phi^T J
        \right),\label{eq:phi_pdf}
  \eea
  with
  \be \label{eq:kernel}
    K \equiv \mu^2 \id - \sigma_h^2 WW^T,
    \qquad
    J \equiv W\eta,
  \ee
  and where $Z$ now reads
  \be
      Z = \int D\phi \, \exp
      \left(
        -\frac{1}{2} \phi^T K \phi + \phi^T J
      \right).
  \ee
  We note therefore that the distribution on the visible layer resembles a
  generating function for a scalar field theory, with the possibility of
  all-to-all bilinear interactions between the fields via the nonlocal kernel
  $K$, and the bias resulting in a source term $J$ coupled to $\phi$. The
  connected two-point function or propagator is given by 
  \be
    \bra\phi_i\phi_j\ket - \bra\phi_i\ket\bra\phi_j\ket = K_{ij}^{-1}.
  \ee
  The hidden layer has provided auxiliary degrees of freedom to establish
  correlations between the visible nodes.

  To continue the discussion we now assume the target probability distribution
  $p_{\rm target}(\phi)$ is known and Gaussian, such that we can solve the RBM
  explicitly, i.e.\ we give explicit expressions for the weight matrix $W$ and
  the bias $\eta$. We denote the target kernel as $K^\phi$ and consider the
  symmetric case ($\phi\to -\phi$, $\eta=J=0$) for simplicity. Since $K^\phi$
  is a real and symmetric matrix, it can be diagonalised; for the theory to
  exist, all its eigenvalues are assumed to be semipositive. The RBM is then
  solved by equating the two kernels, $K^\phi=K$, i.e.,
  \be 
    K^\phi = \mu^2 \id - \sigma_h^2 WW^T
  \ee
  which implies
  \be \label{eq:WWT} 
    WW^T =  \frac{1}{\sigma_h^2}\left( \mu^2 \id - K^\phi \right) \equiv {\cal K}.
  \ee
  Since $WW^T$ is semipositive, we find conditions on the parameter $\mu^2$,
  namely 
  \be \label{eq:mu2}
  \begin{aligned}
    \mu^2/\sigma_h^2 \ge& \max\left[
    \mbox{\rm eigenvalues}\left(WW^T\right)\right],
    \\
    \mu^2 \ge& \max\left[{\rm eigenvalues}\left(K^\phi\right)\right].
  \end{aligned}
  \ee
  Consider now the case that $N_h=N_v$. It is then easy to find some solutions
  for $W$, given that the rhs of Eq.~(\ref{eq:WWT}) is symmetric and positive:
  \bit
    \item 
    The rhs of Eq.\ (\ref{eq:WWT}) can be decomposed in a Cholesky
    decomposition, ${\cal K} = LL^T$, where $L$ is a lower triangular matrix
    with real and positive diagonal entries. The solution is then simply $W=L$.
    The triangular structure means that hidden node $a$ connects to visible
    nodes with $a\leq i$ only.
    \item 
    The rhs of Eq.\ (\ref{eq:WWT}) can be diagonalised via an orthogonal transformation, 
    \be
      {\cal K} = ODO^T = O\sqrt{D} O^T O\sqrt{D} O^T,
    \ee
    yielding the symmetric solution 
    \be
      W=W^T = O\sqrt{D} O^T.
    \ee
  \end{itemize}
  Hence we have found two explicit solutions. Additional solutions are found
  from either of the above by a right multiplication of $W$ by an orthogonal
  transformation, rotating the hidden nodes, 
  \be
    W \to WO_R^T, \qquad h\to O_Rh, \qquad  O_R^T O_R=\id,
  \ee
  since $O_R$ drops out of the combination $WW^T$. 

  We conclude therefore that an infinite number of solutions is present. These
  can be constrained by imposing further conditions on $W$, as in the first two
  cases above.  We will discuss this degeneracy further below.

  Next, we may consider the case where $N_h<N_v$. From Eq.\ (\ref{eq:WWT}) it
  is clear that the accuracy of reproducing the target distribution depends on
  the ranks of the matrices involved. We find
  \be
    \mbox{rank}\left(WW^T\right) \le \min\left(N_v, N_h\right),
    \quad
    \mbox{rank}\left({\cal K}\right) \le N_v.
    \label{eq:dof_reg}
  \ee
  Only when the ranks are equal will the target distribution be reproducible;
  this is particularly relevant when choosing $N_h\ll N_v$. Below we will
  consider in detail what happens of either of the two conditions found so far,
  i.e.\ Eq.\ (\ref{eq:mu2}) and 
  $
  \mbox{rank}\left(WW^T\right) =
  \mbox{rank}\left({\cal K}\right) 
  $
  is not valid.

\section{Training RBM parameters}
\label{sec:training}

  The exact solutions above are only useful when the target model is a known
  Gaussian model and $N_h=N_v$. In general, the target distribution is not
  known and one has to learn from a finite dataset. The training of the model
  is then done by maximising the log-likelihood function $\mathcal{L(\theta |
  \phi)}$. The learnable parameters are collectively indicated as $\theta=\{W,
  \eta, \mu^2\}$. Note that we will consider the case of unbroken symmetry and
  hence the bias is taken to be zero throughout, $\eta_a=0$. We are hence
  concerned with determining the weight matrix $W$ and the mass parameter
  $\mu^2$.

  The model distribution is given by Eq.\ (\ref{eq:phi_pdf}), with $J=0$. Given
  data consisting of $\Nconf$ configurations, labelled as $\phi^{(d)},
  d=1,\ldots, \Nconf$, the log-likelihood function of the model is written as
  \bea \label{eq:loglike}
    \mathcal{L}(\phi | \theta) 
    &&\hm= \frac{1}{\Nconf} \sum_{d=1}^\Nconf 
    \log  p_{\rm model}\left(\phi^{(d)};\theta\right) \nn \\
    &&\hm= -\frac{1}{\Nconf} \sum_{d=1}^\Nconf
    \left(
        \frac{1}{2} \phi^{(d)T} K \phi^{(d)} + \ln Z
    \right).
  \eea
  This log-likelihood function can be optimised with gradient ascent
  algorithms, where the gradient is taken with respect to the coupling matrix
  $W$ and the mass parameter $\mu^2$. Explicitly, 
  \begin{widetext}
  \begin{align} \label{eq:loglike_grad}
    \frac{\partial \mathcal{L}}{\partial W_{ia}} 
    &= \frac{1}{\Nconf} \sum_d  \sum_j 
       \sigma_h^2 \phi_i^{(d)} W_{aj} \phi_j^{(d)}
    - \sum_j\sigma_h^2 \left\langle \phi_i W_{aj} \phi_j \right\rangle_{\rm model}
    \nn \\
    &= \sigma_h^2 \sum_j
    \left(
        \frac{1}{\Nconf} \sum_d
            \phi_i^{(d)} W_{aj} \phi_j^{(d)} 
        -\left\langle \phi_i  W_{aj} \phi_j  \right\rangle_{\rm model}
    \right) 
    =\sigma_h^2 \sum_j
    \left(
        C_{ij}^{\rm target} - C_{ij}^{\rm model}
    \right) W_{ja},
  \end{align}
  %
  where the two-point correlation matrices for the data (i.e.\ the target) and
  the model are given, respectively, by 
  \be
    C_{ij}^{\rm target} = \frac{1}{\Nconf} \sum_{d=1}^\Nconf\phi_i^{(d)} \phi_j^{(d)} 
    = \expv{\phi_i \phi_j}_{\rm target} \equiv K^{-1}_{\phi ij}, 
    \qqquad
    C_{ij}^{\rm model} = \left\langle \phi_i \phi_j \right\rangle_{\rm model} = K^{-1}_{ij}.
  \ee
  \end{widetext}
  Similarly, for $\mu^2$ one finds
  \be \label{eq:mu2update}
    \frac{\partial \mathcal{L}}{\partial \mu^2} = 
    -\half\sum_i \left( \bra\phi_i^2\ket_{\rm target} -
    \bra\phi_i^2\ket_{\rm model} \right). 
  \ee
  When all the data is available, one is able to evaluate the two-point function by summing over configurations before training the RBM. This would yield the target two-point function, computed via the data. In the numerical implementations below, we will analyse the properties of this two-point function further, since the matrix sizes are such that this is feasible. 
  Alternatively, we may consider the case where the target distribution $p_{\rm
  target}(\phi)$ is known and the correlation matrix $C_{ij}^{\rm target}$ of
  the target theory is obtainable. In that case, there is no need to use data
  but one can use the correlation function directly. It should be noted that in
  general the correlation matrix $C_{ij}^{\rm target}$ is not directly
  accessible due to computational complexity, even if the analytical form of
  the target distribution is known.

  If the target distribution is known, then the same equations can also be derived
  by extremising the Kullback-Leibler (KL) divergence,
  \be
    KL(p_{\rm target} || p_{\rm model}) 
    = \int D\phi\, p_{\rm target}(\phi) \log
    \frac{p_{\rm target} (\phi)}{p_{\rm model}(\phi,\theta)},
  \ee
  keeping in mind that only the model distribution depends on the learnable
  parameters $\theta$. With the distribution given by Eq.\ (\ref{eq:phi_pdf})
  and the $\theta$ dependence contained in the kernel $K$ only (recall that
  $J=0$), extremising with respect to $\theta$ then yields
  \begin{align}
    & \frac{\partial}{\partial \theta}KL(p_{\rm target} || p_{\rm model}) 
    = 
    \nn \\
    & \half\left\bra \phi^T \frac{\partial K}{\partial\theta} \phi\right\ket_{\rm target}
     -\half\left\bra \phi^T \frac{\partial K}{\partial\theta} \phi\right\ket_{\rm model}, 
  \end{align}
  which yields the same equations for $W$ and $\mu^2$ as above, but with the
  opposite sign, as the KL divergence is minimised.

  In actual applications, the gradients are used in a discretised update of the
  form
  \be
    \theta_{n+1} = \theta_n +\eta_n \frac{\partial{\cal L}}{\partial \theta},
  \ee
  where $\eta_n$ is the, possibly time-dependent, learning rate. Details of the
  commonly used persistent contrastive divergence algorithm and time-dependent
  learning rate can be found in Appendix \ref{app:pcd}.

\section{Semianalytical solution}
\label{sec:SVD}

\subsection{Singular value decomposition}

  Before solving the RBM numerically, we aim to gain analytical insight in the
  update equations using a singular decomposition for the weight matrix in the
  continuous time limit \cite{Decelle_2021}. 

  The update equations for the weight matrix $W$ and the mass term $\mu^2$, in
  the continuous time limit, read [see Eqs.\
  (\ref{eq:loglike_grad})-(\ref{eq:mu2update})],  
  \bea
    \label{eq:Wevol}
    \dot W =&&\hm \sigma_h^2\left[ K_\phi^{-1} - K^{-1} \right] W \\
    \label{eq:muevol}
    \dot \mu^2 = &&\hm - \half\tr K_\phi^{-1}  + \half\tr  K^{-1}, 
  \eea
  with the two-point functions (or propagators) 
  \bea
    \hspace*{-0.5cm} {K_\phi}_{ij}^{-1} = \bra\phi_i\phi_j\ket_{\rm target}, 
    &\ \tr K_\phi^{-1} = \sum_{i=1}^{N_v} \bra\phi_i\phi_i\ket_{\rm target},\\
    \hspace*{-0.5cm} K_{ij}^{-1} =\bra\phi_i\phi_j\ket_{\rm model}, 
    &\ \tr K^{-1} = \sum_{i=1}^{N_v} \bra\phi_i\phi_i\ket_{\rm model}.
  \eea
  Recall that $\bra\phi_i\ket=0$. The dot denotes the time derivative. We
  remind the reader that both $K$ and $K_\phi$ are symmetric $N_v\times N_v$
  matrices and that the weight matrix $W$ is of size $N_v\times N_h$. We assume
  $N_h\leq N_v$. The RBM (model) kernel is
  \be
    K = \mu^2\id - \sigma_h^2 WW^T,
  \ee
  where $\sigma_h^2$ is the variance of the hidden nodes. 

  We use the singular value decomposition to write $W$ as
  \be
    W = U \Xi V^T, 
    \ UU^T=\id_{N_v\times N_v}, \ VV^T = \id_{N_h\times N_h},
  \ee
  where $U$ is an orthogonal $N_v\times N_v$ matrix, $V$ is an orthogonal
  $N_h\times N_h$ matrix, and $\Xi$ is the rectangular   $N_v\times N_h$ matrix
  with the (ordered) singular values  $\xi_a$ ($a=1,\ldots, N_h$) on the
  diagonal. The RBM kernel then takes the form
  \bea
    K &&\hm=  \mu^2\id - \sigma_h^2 U \Xi\Xi^T U^T \nn \\
    &&\hm= U\left[  \mu^2\id - \sigma_h^2  \Xi\Xi^T\right]  U^T \equiv U D_K U^T,
  \eea
  with the diagonal matrix
  \bea  \label{eq:DK}
    &&\hm D_K = \\ \nn
    &&\hm
    \mbox{diag}\big(\underbrace{\mu^2-\sigma_h^2\xi_1^2, 
    \mu^2-\sigma_h^2\xi_2^2, \ldots, \mu^2-\sigma_h^2\xi_{N_h}^2}_{N_h}, 
    \underbrace{\mu^2_{ {}_{} }, \ldots, \mu^2}_{N_v-N_h}\big).
  \eea
  Note that the existence of the model requires that $\mu^2>\sigma_h^2\xi_1^2$,
  with $\xi_1$ the largest singular value of $W$. Equation (\ref{eq:DK})
  demonstrates that only the first $N_h$ eigenvalues can potentially be learnt,
  with the remaining $N_v-N_h$ eigenvalues frozen at the higher scale $\mu^2$.
   
  The symmetric target kernel $K_\phi$ and the corresponding two-point
  function $K_\phi^{-1}$ can be diagonalised via an orthogonal transformation
  as 
  \begin{align}
  \nn
    & K_\phi = O_\phi D_\phi O_\phi^T, 
    \qqquad 
    K_\phi^{-1} = O_\phi D_\phi^{-1} O_\phi^T, 
    \\
    &
    O_\phi O_\phi^T=\id_{N_v\times N_v},
   \end{align}
  where the eigenvalues of $K_\phi$ are assumed to be positive again.
  
  The rhs of Eq.\ (\ref{eq:Wevol}) can now be written as
  \begin{align}
    & \sigma_h^2 
    \left[ K_\phi^{-1} - K^{-1} \right] 
     W \nn\\
     & =
    U \sigma_h^2 \left[ U^T O_\phi D_\phi^{-1} O_\phi^T U
     -  D_K^{-1} \right]
     \Xi V^T.
  \end{align}
  The term within the brackets will be encountered frequently below and hence
  we honour it with a new symbol,
  \be \label{eq:Lambda}
    \Lambda \equiv U^T O_\phi D_\phi^{-1} O_\phi^T U - D_K^{-1} = \Lambda^T.
  \ee
  The evolution equation for $W$ can then be compactly written as
  \be \label{eq:Wevol2}
    \dot W = \sigma_h^2 U \Lambda \Xi V^T , \qquad \dot W^T = \sigma_h^2 V\Xi^T\Lambda U^T.
  \ee
  We note that $\Lambda$ drives the evolution in the learning process: it
  vanishes when the basis on the visible layer is aligned with the basis for
  the data $(U\to O_\phi$) and the eigenvalues, or widths of the Gaussians, are
  correctly determined ($D_K\to D_\phi$). One may note that $\Lambda$ does not
  depend on $V$, which acts on the hidden nodes, resulting in the degeneracy
  discussed in Sec.\ \ref{sec:scalar}: any rotation of the hidden nodes leaves
  the solution on the visible layer invariant and the learning stops when
  $\Lambda\to 0$, irrespective of what $V$ is.
  \vspace{-0.4cm}

\subsection{Learning dynamics}
 
  Having defined the needed quantities, we are now in a position to determine
  the learning dynamics of $W$ in detail, i.e.\ the evolution of $U, V$, and
  the singular values $\Xi$.  We consider separately
  \be
    WW^T = U\Xi\Xi^T U^T, \qquad
    W^TW = V\Xi^T\Xi V^T.
  \ee 
  Taking the derivative of the first product gives
  \be
    \frac{d}{dt}\left(WW^T\right) 
    = \dot U\, \Xi\Xi^T U^T + U \Xi\Xi^T \dot U^T + U \frac{d}{dt}
    \left(\Xi\Xi^T\right)U^T.
  \ee
  On the other hand, Eq.\ (\ref{eq:Wevol2}) gives
  \bea
    \frac{d}{dt}WW^T &&\hm= \dot W W^T + W \dot W^T \nn \\
    &&\hm= \sigma_h^2 U \Lambda \Xi \Xi^T U^T + \sigma_h^2 U\Xi\Xi^T\Lambda U^T.
  \eea
  Conjugating both equations with $U^T$ and $U$ then yields 
  \be \label{eq:UUXX}
    U^T\dot U\, \Xi\Xi^T +  \Xi\Xi^T \dot U^T U +  \frac{d}{dt}\left(\Xi\Xi^T\right)
    = \sigma_h^2  \Lambda \Xi \Xi^T +  \sigma_h^2 \Xi\Xi^T\Lambda.
  \ee
  Since $U^T\dot U=-\dot U^T U$ is skew symmetric (due to $U$ being orthogonal)
  and  $\Xi \Xi^T$ is diagonal, it is easy to see that 
  \be
    U^T\dot U\, \Xi\Xi^T +  \Xi\Xi^T \dot U^T U 
    = U^T\dot U\, \Xi\Xi^T -  \Xi\Xi^T U^T\dot U
  \ee
  is a symmetric matrix with zeroes on the diagonal. Equation ~(\ref{eq:UUXX}) then
  decomposes into one equation for the diagonal elements, determining the
  singular values,  and one for the off-diagonal ones, determining $U$, namely
  \bea \label{eq:XiXi}
    \hspace*{-0.6cm} \frac{d}{dt}\left(\Xi\Xi^T\right) 
    = \sigma_h^2 \Lambda_d \Xi \Xi^T + &&\hm \sigma_h^2 \Xi\Xi^T\Lambda_d 
    = 2 \sigma_h^2 \Lambda_d \Xi \Xi^T, \\
    \hspace*{-0.6cm} U^T\dot U\, \Xi\Xi^T - \Xi\Xi^T U^T\dot U 
    =&&\hm \sigma_h^2 \left(\Lambda-\Lambda_d\right) \Xi \Xi^T \nn \\
    &&\hspace*{-0.4cm} + \sigma_h^2 \Xi\Xi^T\left(\Lambda-\Lambda_d\right),  
  \eea
  where
  \be 
    \Lambda_d = \mbox{diag}\left(\Lambda\right).
  \ee
  Repeating the same analysis for $W^TW$ gives nearly identical equations in
  the $N_h\times N_h$ subspace, namely
  \bea
    \label{eq:V}
    \frac{d}{dt}\left(\Xi^T\Xi\right) = &&\hm 2 \sigma_h^2\Xi^T\Lambda_d \Xi, \\
    V^T\dot V\, \Xi^T\Xi -  \Xi^T\Xi V^T\dot V 
    = &&\hm  2 \sigma_h^2\Xi^T\left(\Lambda-\Lambda_d\right) \Xi.
  \eea
  Note that 
  \bea
    \Xi\Xi^T = &&\hm \mbox{diag}(\xi_1^2, \xi_2^2,\ldots, \xi_{N_h}^2, 0,\ldots, 0),\\
    \Xi^T\Xi= &&\hm \mbox{diag}(\xi_1^2, \xi_2^2,\ldots, \xi_{N_h}^2).
  \eea
  The equations for $\Xi\Xi^T$ and $\Xi^T\Xi$ yield identical equations for the
  $N_h$ singular values. 

  The equation for $\mu^2$ finally reads, in this notation, 
  \be
    \dot\mu^2 = -\half\tr \Lambda =  -\half\tr \Lambda_d.
  \ee
  Keeping $\mu^2$ fixed, it is easy to see that $\sigma_h^2$ can be absorbed in
  the time parameter ($\tilde t= t\sigma_h^2$) and the singular values, see
  Eq.\ (\ref{eq:DK}); hence it does not add any freedom to the model. When
  $\mu^2$ is learnt as well, its time evolution will depend on $\sigma_h^2$,
  after rescaling time as $t\to\tilde t$.

  As noted, $V$ does not appear in the driving term $\Lambda$. Hence $V$ simply
  follows the evolution, until $\Lambda-\Lambda_d\to 0$, see Eq.\ (\ref{eq:V}).
  For square matrices, $N_h=N_v$, this redundancy can be removed by choosing
  $W$ to be symmetric ($V=U$) or by enforcing $W$ to be of the lower (or upper)
  triangular form (Cholesky decomposition of $WW^T$), see Sec.\
  \ref{sec:scalar}.
  \vspace{-0.4cm}

\subsection{Simple examples}

\subsubsection*{1. $\boldsymbol{N_v=N_h=2}$}

  The simple example of two visible and two hidden nodes can be worked out in
  detail. We will note a number of characteristics which remain relevant also
  for larger systems.

  First we note that $U, V,$ and $O_\phi$ are all $2\times 2$ rotation matrices;
  we denote the angles as $\theta_U$, $\theta_V$, and $\theta_0$, respectively.
  Then one notes that 
  \be
    U^T\dot U = \dot \theta_U 
    \left(
      \begin{array}{cc}
      0 & -1\\
      1 & 0
      \end{array}
    \right), 
  \ee
  and the same for $V^T\dot V$, with $\dot \theta_V$. Finally, the combination
  $O_\phi^TU$ is a rotation over an angle $\Delta\theta=\theta_U-\theta_0$.

  We denote the two eigenvalues of the target kernel $K_\phi$ with
  $\kappa_{1,2}$ and of the RBM kernel $K$ with $\lambda_{1,2} =
  \mu^2-\sigma_h^2\xi_{1,2}^2$. This yields the driving term, 

  \begin{widetext}
  \be
    \Lambda = \left(
      \begin{array}{cc}
        \frac{1}{\kappa_1} \cos^2\Delta\theta 
        +\frac{1}{\kappa_2}\sin^2\Delta\theta  
        -\frac{1}{\lambda_1} & 
        \left(\frac{1}{\kappa_2}-\frac{1}{\kappa_1}\right)
        \cos\Delta\theta\sin\Delta\theta\\
        \left(\frac{1}{\kappa_2}-\frac{1}{\kappa_1}\right)
        \cos\Delta\theta\sin\Delta\theta & 
        \frac{1}{\kappa_2} \cos^2\Delta\theta 
        +\frac{1}{\kappa_1}\sin^2\Delta\theta  -\frac{1}{\lambda_2}
      \end{array}
    \right), 
  \ee
  \end{widetext}

\begin{figure*}[t]
    \begin{center}
      \includegraphics[width=0.32\textwidth]{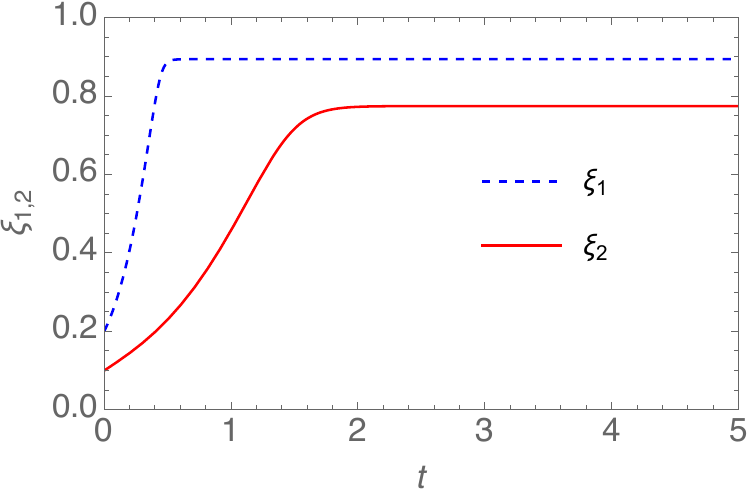}
      \includegraphics[width=0.32\textwidth]{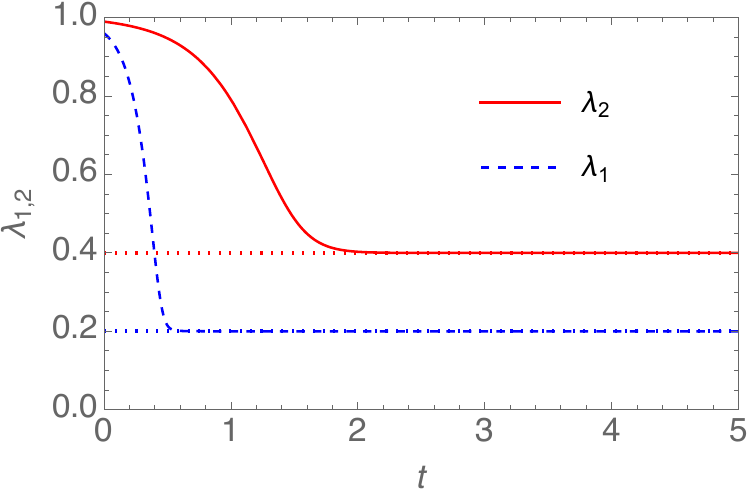}
      \includegraphics[width=0.32\textwidth]{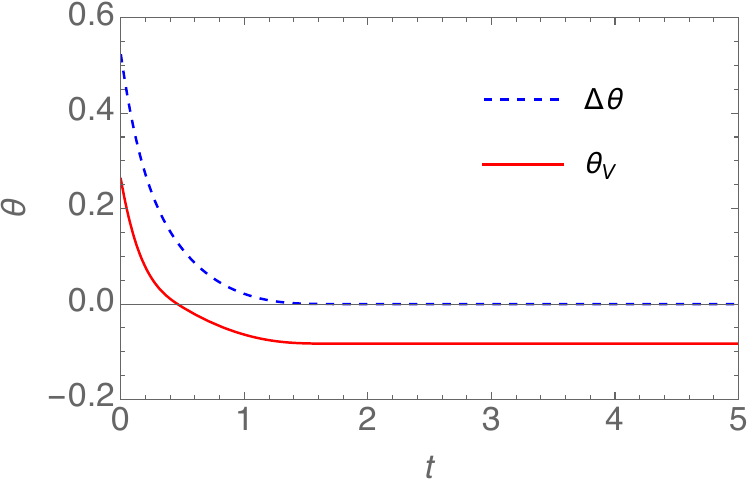}
      \\
      \includegraphics[width=0.32\textwidth]{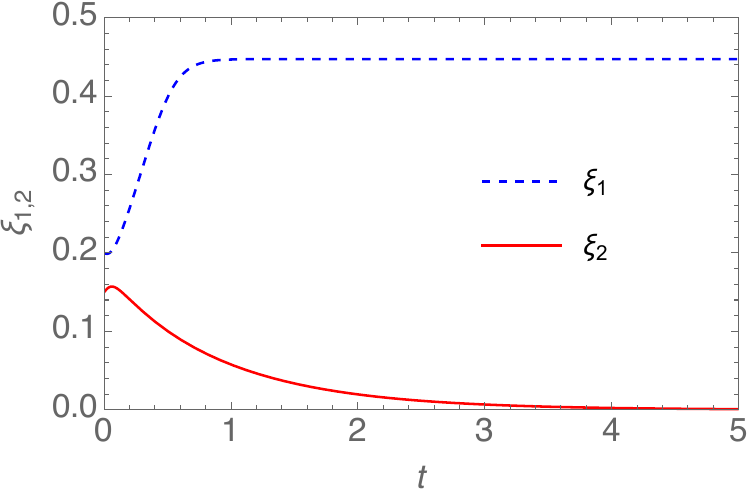}
      \includegraphics[width=0.32\textwidth]{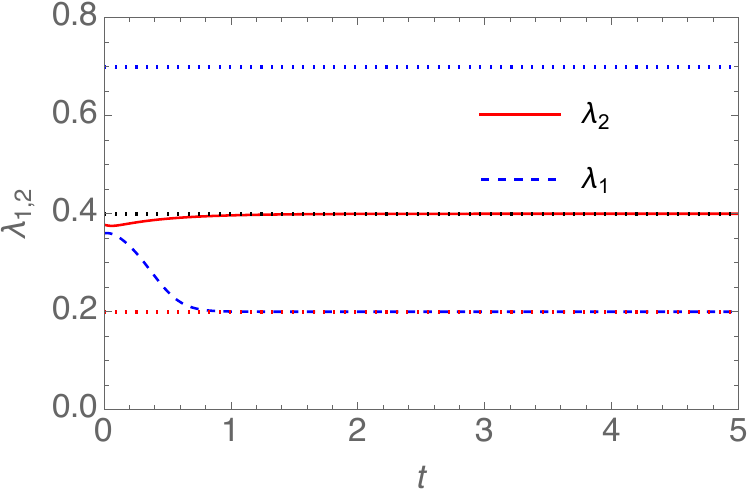}
      \includegraphics[width=0.32\textwidth]{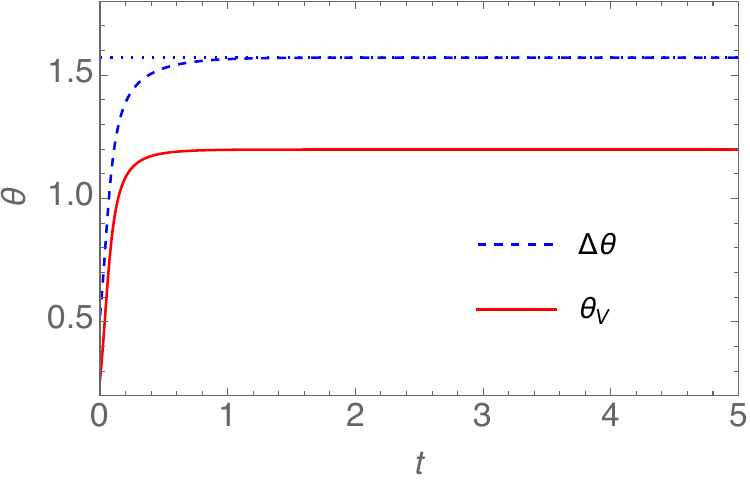}
      \\
      \includegraphics[width=0.32\textwidth]{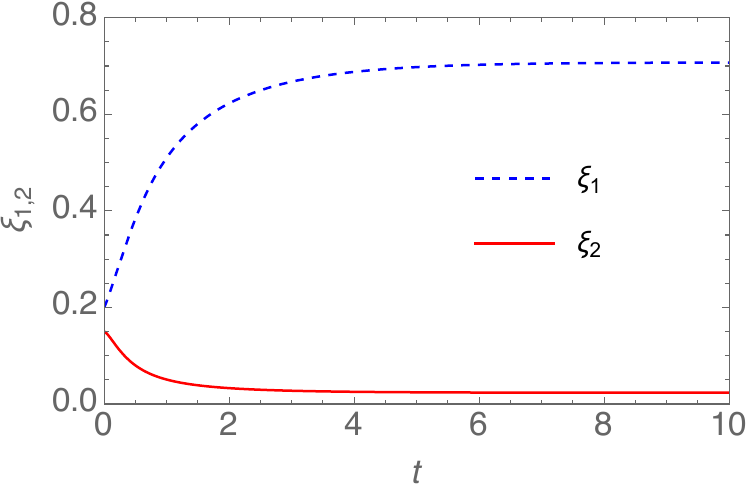}
      \includegraphics[width=0.32\textwidth]{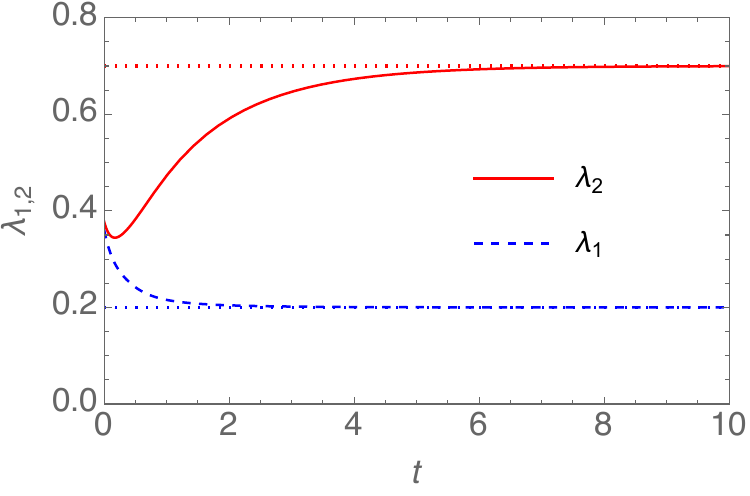}
      \includegraphics[width=0.32\textwidth]{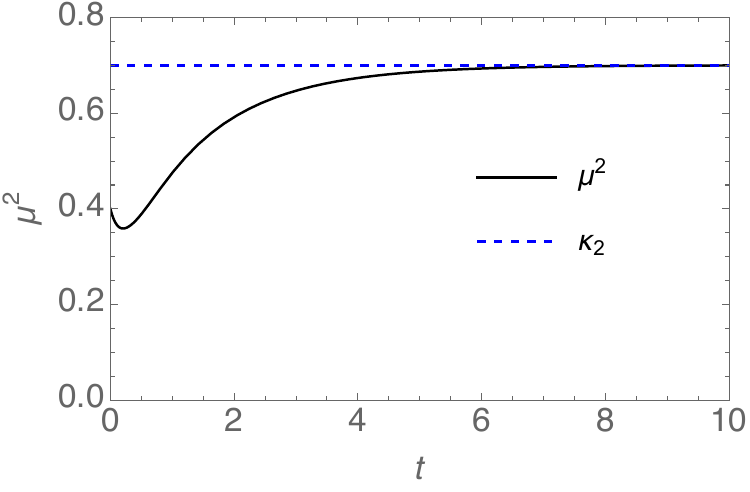}
    \end{center}
    \caption{Top row: learning evolution for the case $\mu^2>\kappa_2>\kappa_1$,
      specifically $\kappa_1=0.2, \kappa_2=0.4, \mu^2=1$, of the singular values
      (left), eigenvalues (middle) and angles (right). 
      Middle row: as above,  for the case $\kappa_1>\mu^2>\kappa_2$, specifically
      $\kappa_1=0.7, \kappa_2=0.2, \mu^2=0.4$.
      Bottom row: as above, including a time dependent $\mu^2$ (right), for the
      case $\kappa_2>\mu^2>\kappa_1$, specifically $\kappa_1=0.2, \kappa_2=0.7,
      \mu^2(0)=0.4$. In all cases, $\sigma_h^2=1$.}
    \label{fig:evol}
  \end{figure*}

  Putting everything together then gives the following equations
  \bea
    \label{eq:xi1}
    \hspace*{-0.7cm} \dot\xi_1 = &&\hm \sigma_h^2 \left( \frac{1}{\kappa_1} \cos^2\Delta\theta 
      +\frac{1}{\kappa_2}\sin^2\Delta\theta  
      -\frac{1}{\mu^2-\sigma_h^2\xi_1^2} 
    \right) \xi_1,\\ 
    %
    \label{eq:xi2}
    \hspace*{-0.7cm} \dot\xi_2 = &&\hm \sigma_h^2 \left( \frac{1}{\kappa_2} \cos^2\Delta\theta 
      +\frac{1}{\kappa_1}\sin^2\Delta\theta  
      -\frac{1}{\mu^2-\sigma_h^2\xi_2^2} 
    \right) \xi_2, 
    %
  \eea
  and
  \bea 
    \label{eq:Deltatheta}
    \dot{\Delta\theta} =&&\hm \sigma_h^2 \frac{\xi_1^2+\xi_2^2}{\xi_1^2-\xi_2^2}\rho, \\
    \dot{\theta}_V =&&\hm 2\sigma_h^2 \frac{\xi_1\xi_2}{\xi_1^2-\xi_2^2}\rho, \\
    \label{eq:mu21}
    \dot\mu^2 = &&\hm \half\left(\frac{1}{\mu^2-\sigma_h^2\xi_1^2} 
      +\frac{1}{\mu^2-\sigma_h^2\xi_2^2} 
      - \frac{1}{\kappa_1}-\frac{1}{\kappa_2} 
    \right), 
  \eea
  where
  \be
    \rho = \left(\frac{1}{\kappa_2}-\frac{1}{\kappa_1}\right)
    \cos\Delta\theta\sin\Delta\theta.
  \ee
  These equations have several fixed points. The difference of angles is given
  by $\Delta\theta=0, \pi/2$. Which of these is selected depends on which
  eigenvalue $\kappa_{1,2}$ is smaller. Note that the SVD decomposition orders
  the singular values, $\xi_1>\xi_2$. 
  The equations have fixed points at $\sigma_h^2\xi^2_{1,2} = \mu^2 - \kappa_{1,2}$ and at $\xi^2_{1,2}=0$.
  We consider first the case of fixed
  $\mu^2$. The actual realisation depends on the ordering of $\kappa_{1,2}$ and
  $\mu^2$. We find
  \bea
    \label{eq:case1}
    \mu^2> \kappa_2>\kappa_1:  &&\Delta\theta= 0, \\
    && \mu^2-\sigma_h^2\xi_1^2=\kappa_1, \quad \mu^2-\sigma_h^2\xi_2^2=\kappa_2, \nn\\
    \mu^2 > \kappa_1>\kappa_2: && \Delta\theta= \pi/2, \\
    && \mu^2-\sigma_h^2\xi_1^2=\kappa_2, \quad \mu^2-\sigma_h^2\xi_2^2=\kappa_1.
    \nn
  \eea
  (The fixed points at $\xi^2_{1,2}=0$ are unstable.)
  This is illustrated in Fig.\ \ref{fig:evol} (top row). In this case, both
  eigenvalues are learnt correctly. If $\mu^2$ is smaller than an eigenvalue,
  then it cannot be reproduced and is replaced by $\mu^2$, 
  \bea
    \hspace*{-0.7cm} \kappa_2>\mu^2> \kappa_1: \Delta\theta &\hm= 0, &
     \mu^2-\sigma_h^2\xi_1^2=\kappa_1, \  \xi_2=0, 
     \\
    \hspace*{-0.7cm} \kappa_1>\mu^2>\kappa_2: \Delta\theta &= \pi/2, &
     \mu^2-\sigma_h^2\xi_1^2=\kappa_2, \ \xi_2=0,
  \eea
  %
  see Fig.\ \ref{fig:evol} (middle row). In this case, only the smallest
  eigenvalue is learnt, while the other one evolves to $\mu^2$ (see also Eq.\
  (\ref{eq:DK})).

  In case $\mu^2$ is smaller than all eigenvalues,  $\mu^2<\kappa_{1,2}$, the
  eigenmodes cannot be reproduced and are replaced by $\mu^2$, with
  $\xi_1=\xi_2=0$. Finally, we remark again that $\theta_V$ simply evolves
  until $\rho\to 0$, but it does not influence the learning of the other
  parameters. 

  The actual eigenvalues may not be known, and one may choose $\mu^2$ to be too
  low, as in the second example above. This can be evaded by learning $\mu^2$
  itself, using Eq.~(\ref{eq:mu21}). The system is now overparameterised,
  with $\xi_{1,2}$ and $\mu^2$ being learnt to reproduce $\kappa_{1,2}$. In
  this case one finds that the eigenvalues are reproduced, irrespective of the
  initial value of $\mu^2$, see Fig.\ \ref{fig:evol} (bottom row). Note that
  one of the singular values decreases since $\mu^2$ itself increases towards
  the largest eigenvalue.

\subsubsection*{2. Approach to the fixed point}

  To understand the evolution towards the fixed point, a simple linearisation
  suffices. We consider the case of fixed $\mu^2$. Taking concretely case
  (\ref{eq:case1}) above, we expand about the fixed point and write
  \be
    \sigma_h^2\xi_{i}^2 = \mu^2-\kappa_{i} + x_{i}, \qquad (\Delta\theta)^2 = 0 + y.
  \ee
  Expanding Eqs.\ (\ref{eq:xi1}), (\ref{eq:xi2}), and (\ref{eq:Deltatheta}) in $x_i$
  and $y$ and absorbing $\sigma_h^2$ in the time parameter (denoting the
  derivative with respect to $\tilde t=\sigma_h^2 t$ with a prime) then yields
  the linearised equations 
  \begin{align}
    x_1' = & -2 \left(\mu^2-\kappa_1\right) \left[ \frac{x_1}{\kappa_1^2} 
    +\left(\frac{1}{\kappa_1}-\frac{1}{\kappa_2}\right) y\right], \\
    x_2' = & -2 \left(\mu^2-\kappa_2\right) \left[ \frac{x_2}{\kappa_2^2} 
    +\left(\frac{1}{\kappa_2}-\frac{1}{\kappa_1}\right) y\right], \\
    y' = & -2\frac{2\mu^2-\kappa_1-\kappa_2}{\kappa_1\kappa_2}y.
  \end{align}
  We hence find exponential convergence, controlled by the relaxation rates
  \be
    \gamma_{i} = \frac{\mu^2-\kappa_{i}}{\kappa_{i}^2},
    \qquad
    \gamma_{\Delta\theta} = \frac{\kappa_1}{\kappa_2}\gamma_1 
    + \frac{\kappa_2}{\kappa_1}\gamma_2.
  \ee
  The angle $\Delta\theta(\tilde t)$ relaxes with $\gamma_{\Delta\theta}$,
  whereas the singular values $\xi_i(\tilde t)$ decay with $\min(\gamma_i,
  \gamma_{\Delta\theta})$. Interestingly, the relaxation rates are set by the
  difference between the RBM mass parameter $\mu^2$ and the eigenvalues
  $\kappa_i$ in the spectrum. Irrespective of the actual values of $\mu^2$ and
  the eigenvalues $\kappa_i$, the mode corresponding to the higher eigenvalue
  relaxes the slowest. We hence conclude the following:
  \begin{itemize}
    \item infrared modes, i.e.\ those corresponding to the smallest eigenvalues
      will converge faster, this can indeed be observed in Fig.\ \ref{fig:evol}
      (top row);
    \item increasing the value of $\mu^2$ will lead to more rapid convergence
      for all modes. This will be explored below in more realistic cases. 
  \end{itemize}

\subsubsection*{3. $\boldsymbol{N_v=2, N_h=1}$}

  The case of one hidden node serves to demonstrate what happens when
  $N_h<N_v$. It is particularly simple as $V$ is replaced by $v=1$ and we only
  need to consider one angle and one singular value, determined by the
  following equations
  \bea 
    \label{eq:xi11}
    \hspace*{-0.7cm} \dot\xi_1 = &&\hm \sigma_h^2 \left( \frac{1}{\kappa_1} \cos^2\Delta\theta 
    +\frac{1}{\kappa_2}\sin^2\Delta\theta  
    -\frac{1}{\mu^2-\sigma_h^2\xi_1^2} \right) \xi_1, 
  \eea
  and
  \bea
    \dot{\Delta\theta} =&&\hm \sigma_h^2 \rho, \\
    \label{eq:mu22}
    \dot\mu^2 = &&\hm \half\left( \frac{1}{\mu^2-\sigma_h^2\xi_1^2} 
    +\frac{1}{\mu^2} - \frac{1}{\kappa_1}-\frac{1}{\kappa_2}\right), 
  \eea
  where
  \be
    \rho = \tilde \rho\cos\Delta\theta\sin\Delta\theta,
    \quad
    \tilde\rho = \left(\frac{1}{\kappa_2}-\frac{1}{\kappa_1}\right)
  \ee
  The equation for the angle is now decoupled and can be solved, as
  \be
    \tan\left[\Delta\theta(t)\right] = 
    \tan\left[\Delta\theta(0)\right] e^{\sigma_h^2\tilde\rho t},
  \ee
  such that
  \bea
    \kappa_2>\kappa_1 &&\hm \ \Leftrightarrow \ \tilde\rho<0 
    \ \Leftrightarrow \ \lim_{t\to\infty} \Delta\theta(t)=0, \\
    \kappa_2<\kappa_1 &&\hm \ \Leftrightarrow \ \tilde\rho>0 
    \ \Leftrightarrow \ \lim_{t\to\infty} \Delta\theta(t)=\frac{\pi}{2}.
  \eea
  Using this in Eq.\ (\ref{eq:xi11}) confirms that the smallest eigenvalue of
  $K_\phi$ is reproduced (for constant $\mu^2$). If $\mu^2$ is learnt as well,
  then Eq.\ (\ref{eq:mu22}) ensures it becomes equal to the largest of the two
  eigenvalues.

\begin{figure*}[t]
  \centering
    \includegraphics[width=0.48\textwidth]{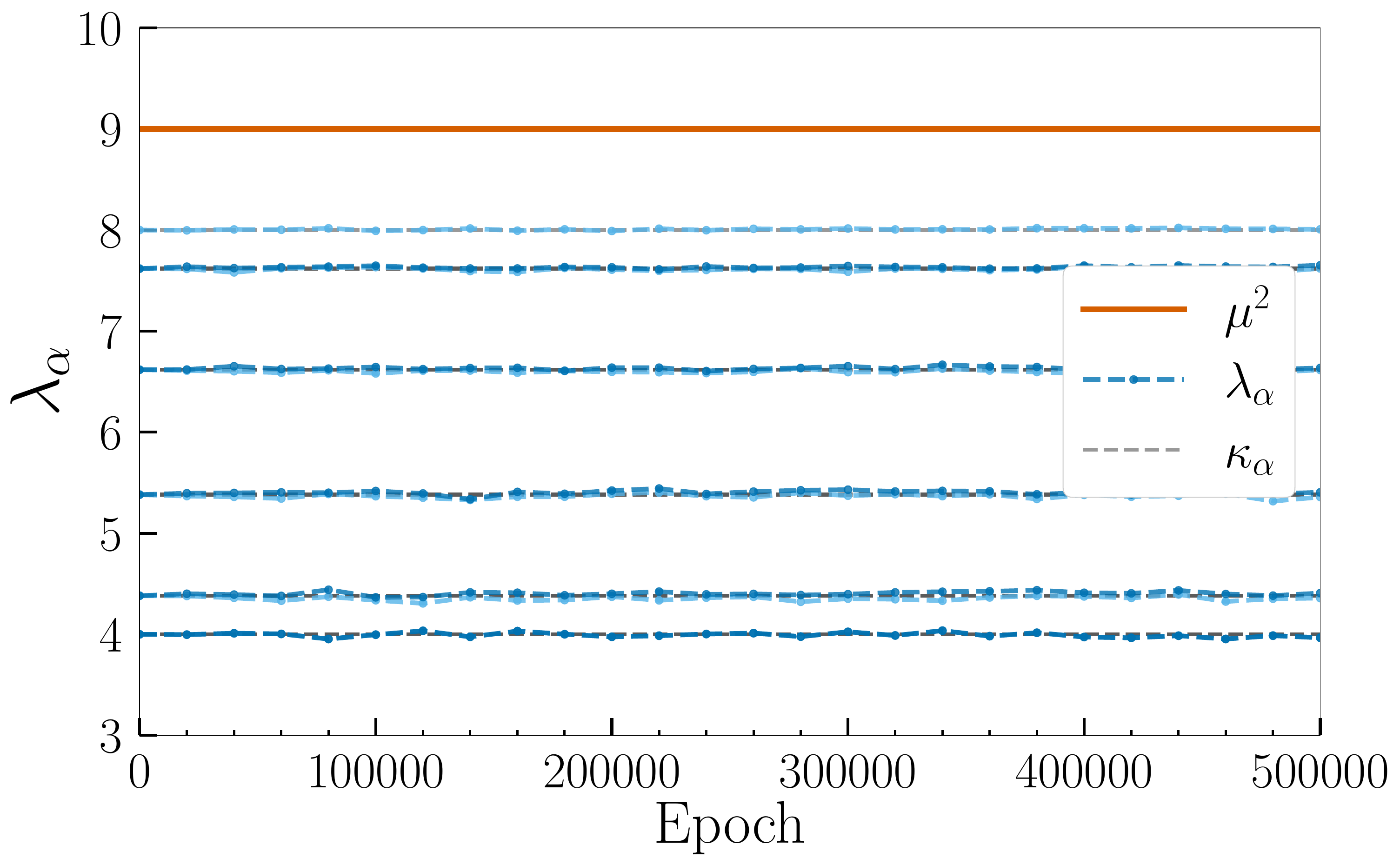}
    \includegraphics[width=0.48\textwidth]{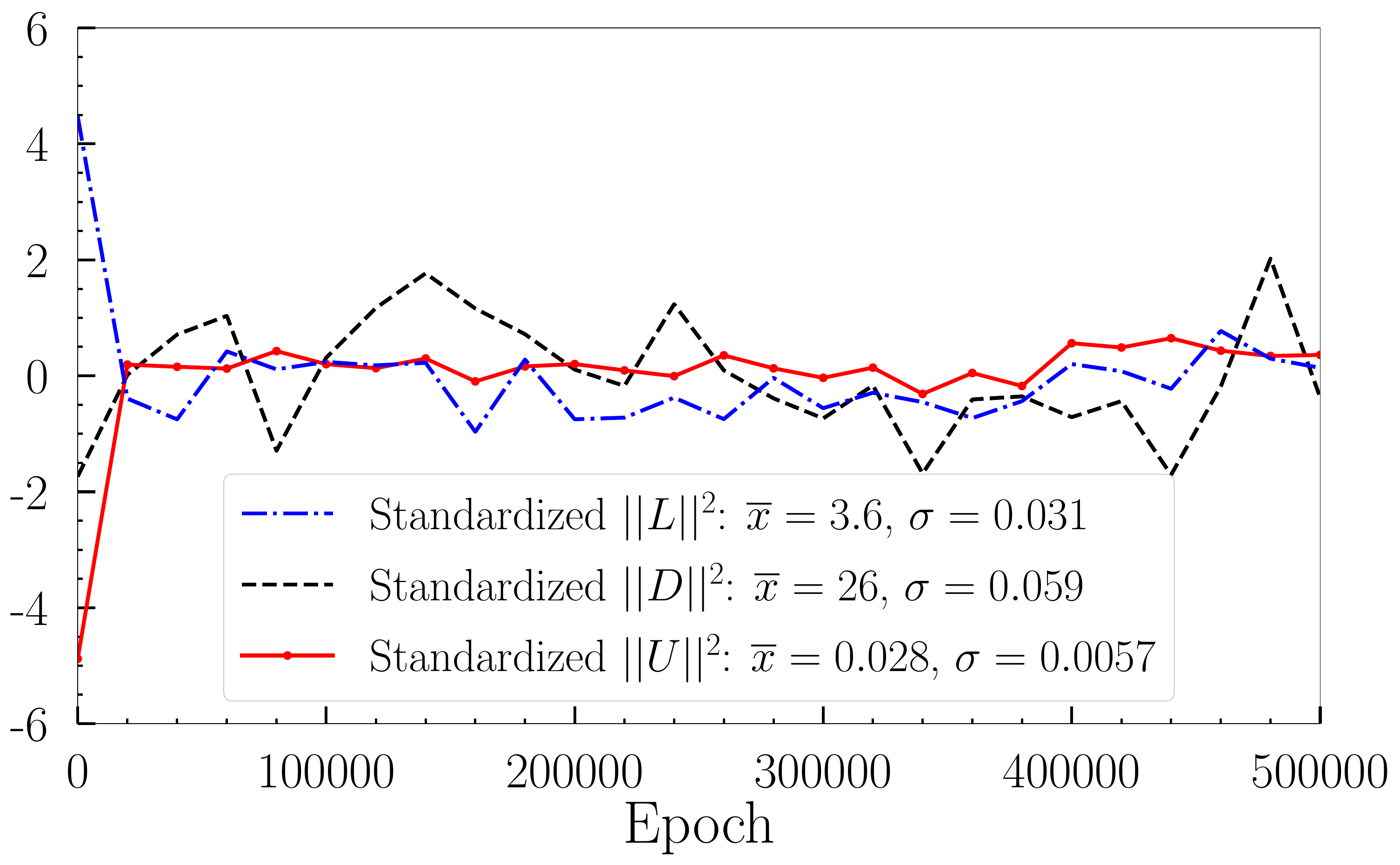}
    \caption{Cholesky initialisation. 
      Left: evolution of RBM eigenvalues $\lambda_\alpha$ during training. Note
      that adjacent eigenvalues are coloured alternatively. Exact eigenvalues
      $\kappa_\alpha$ are shown with horizontal dashed lines and the RBM mass
      parameter $\mu^2$ with the horizontal full line. After the Cholesky
      initialisation, the RBM eigenvalues fluctuate around the correct values.
      Right: the $L_2$ norm of each part of the coupling matrix, diagonal ($D$),
      upper ($U$) triangular, and lower ($L$) triangular. Values are standardised, with
      $\overline{x}$ ($\sigma$) the mean value (standard deviation) along the
      training interval. Each part fluctuates around its average value}
    \label{fig:Cholesky_lc}
  \end{figure*}

  To summarise, we note the following observations: if either the number of
  hidden nodes or the mass parameter $\mu^2$ is chosen too small, the infrared
  part of the spectrum (lowest eigenvalue) is reproduced, while the ultraviolet
  part (highest eigenvalue) evolves to $\mu^2$; making $\mu^2$ a learnable
  parameter yields one more degree of freedom to correctly reproduce the next
  eigenvalue; infrared modes are learnt quicker than ultraviolet modes. These
  observations for the simple case considered here remain relevant for more
  interesting systems, as we will demonstrate now.

\section{Learning Gaussian distributions}
\label{sec:gaussian}

  We continue with the case for which the target distribution is known and
  Gaussian, namely free scalar fields discretised on a lattice in one and two
  dimensions. The continuum action reads, in $n$ Euclidean dimensions,
  \be
    S(\phi) = \int d^nx\, \half\left( \partial_\mu\phi \partial_\mu\phi +m^2\phi^2 \right).
  \ee
  The simplest lattice-discretised equivalent is, on a one-dimensional lattice
  with $N_v$ nodes and with periodic boundary conditions,
  \be
    S(\phi) = \half\sum_{i,j=1}^{N_v} \phi_i K_{ij}^\phi \phi_j, 
   \ee
   where 
   \be  \label{eq:K_phi}
    K_{ij}^\phi = (2+m^2)\delta_{ij} -\delta_{i,j+1} -\delta_{i,j-1}.
  \ee
  We use ``lattice units'', $a=1$, throughout. The spectrum of the target kernel
  $K^\phi$ is easy to compute analytically and reads
  \be
    \kappa_k = m^2 + p_{{\rm lat}, k}^2 
    = m^2 + 2-2\cos\left(\frac{2\pi k}{N_v}\right),
    \ee
    with $-N_v/2 < k \leq N_v/2$.
  Each eigenvalue is doubly degenerate, except the minimum ($k=0, \kappa_{\rm
  min}=m^2$) and the maximal ($k=N_v/2, \kappa_{\rm max}=m^2+4$) ones.
  Referring back to Sec.\ \ref{sec:scalar}, the exact spectrum can only be
  learnt when $N_h=N_v$ and when the RBM mass parameter 
  \be \label{eq:mass_cond}
    \mu^2> \kappa_{\rm max} =m^2+4.
  \ee
  Since the target theory is known, we can train the model directly from the
  correlation matrix of the target theory without the need for pregenerated
  training data. Then each term of the gradient is estimated by persistent
  contrastive divergence (PCD) to train the RBM, see Appendix ~\ref{app:pcd} for
  details. The scalar field mass parameter is chosen as $m^2=4$ and the
  variance on the hidden layer equals $\sigma_h^2=1$ throughout.

  \begin{figure*}[t]
  \centering             
    \includegraphics[width=0.48\textwidth]{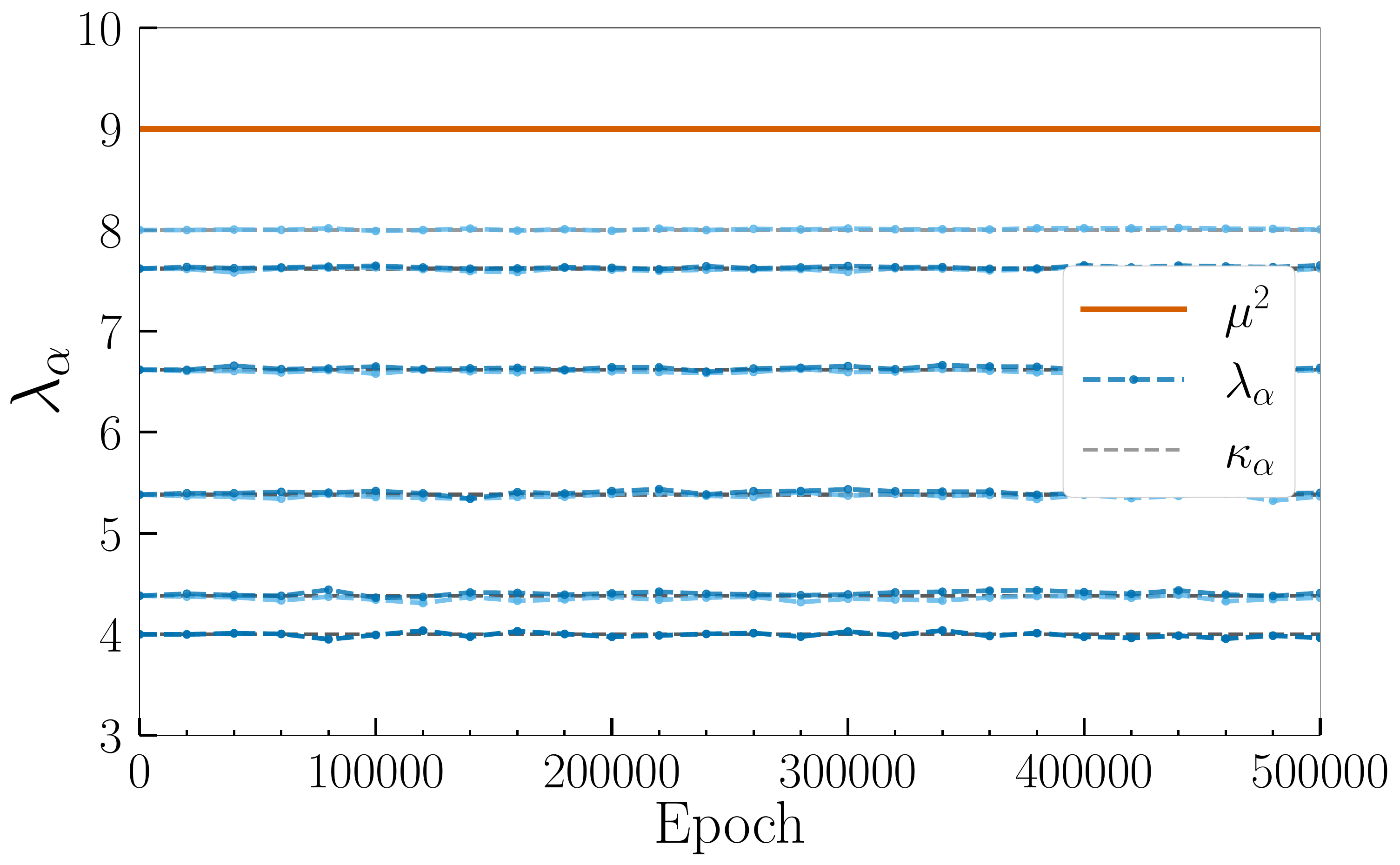}
    \includegraphics[width=0.48\textwidth]{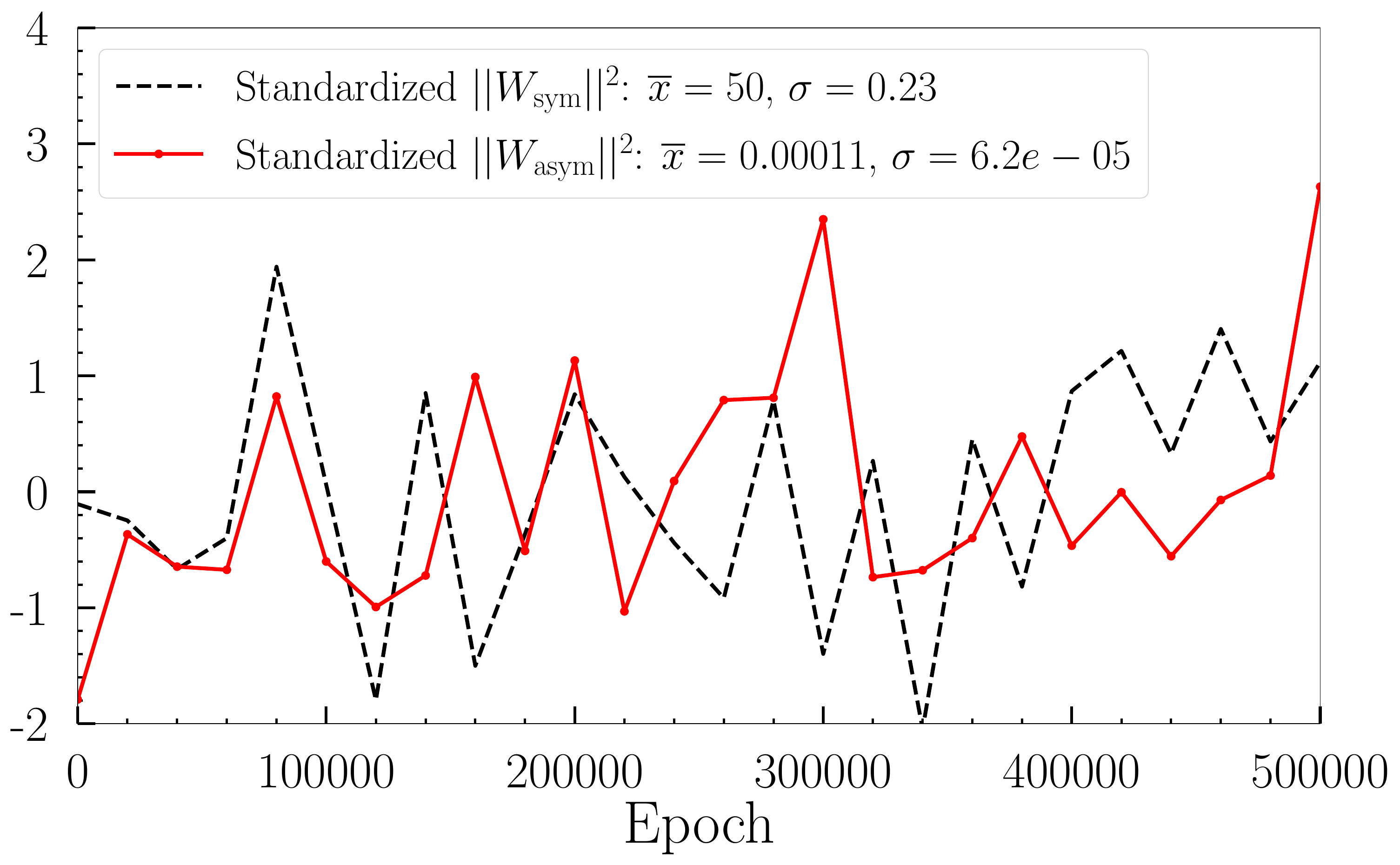}
    \caption{Left: as in Fig.\ \ref{fig:Cholesky_lc}, for the symmetric initialisation. 
      Right: standardised $L_2$ norm of symmetric and asymmetric parts of the
      coupling matrix. The latter remains small during updates.}
  \label{fig:symmetric_lc}
 \vspace{0.2cm}
  \centering
    \includegraphics[width=0.48\textwidth]{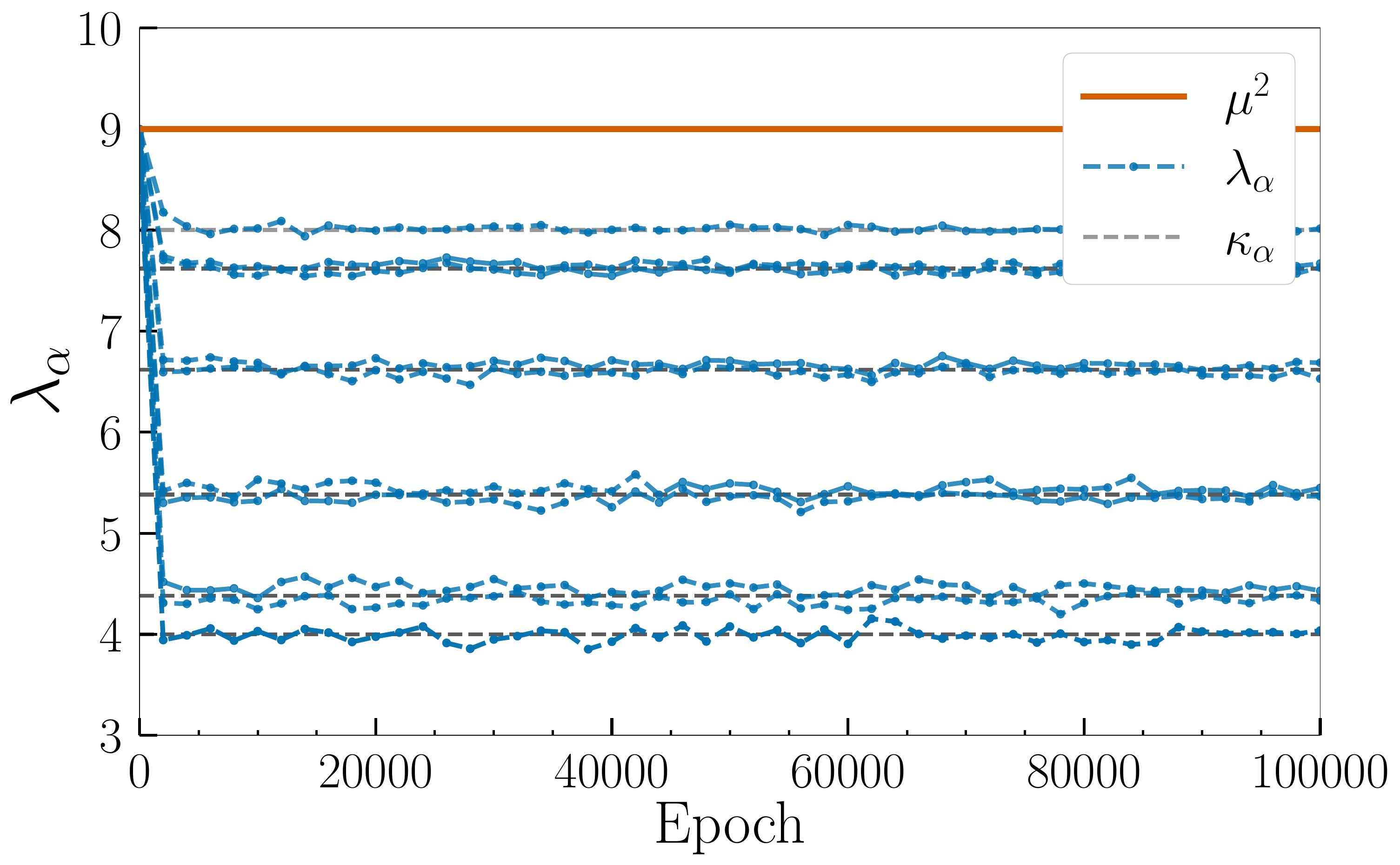}
    \includegraphics[width=0.475\textwidth]{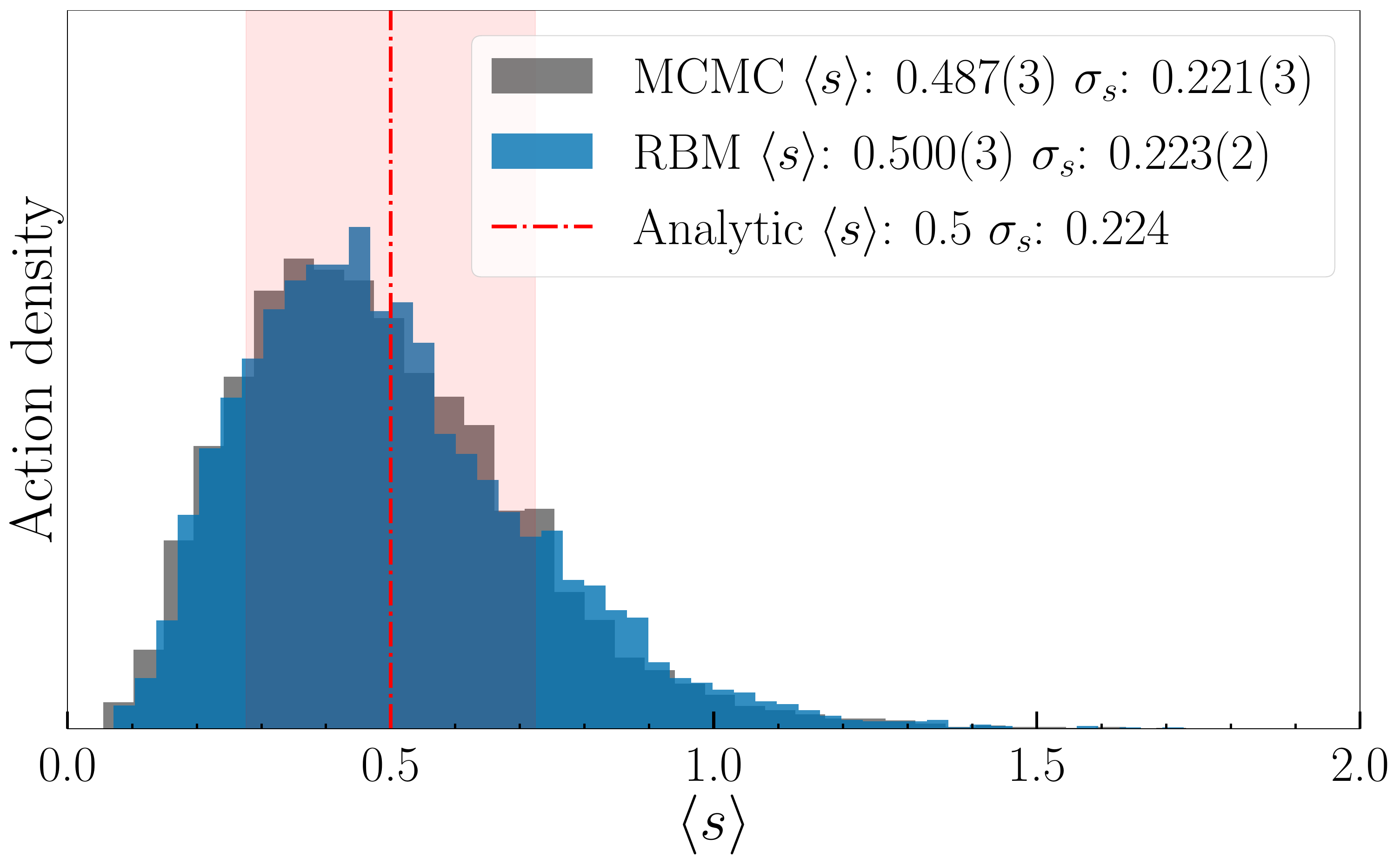}
    \caption{Left: evolution of RBM eigenvalues $\lambda_\alpha$ during training,
      starting from a random coupling matrix $W$. Presentation as in Fig.\
      \ref{fig:Cholesky_lc} (left). Right: histogram density of action density from
      Monte Carlo generated and RBM generated samples.}
    \label{fig:scalar_result}
  \end{figure*}

\subsection{Initialisation with an exact solution}
  
  We start with the case of a constant RBM mass parameter $\mu^2=9 >
  \kappa_{\rm max}=8$, with $N_v=N_h=10$. To test the numerical code, we may
  initialise the weight matrix $W$ according to one of the exact solutions
  found in Sec.~\ref{sec:scalar}: the Cholesky (lower triangular) solution and
  the symmetric solution. The results are shown in Figs.\ \ref{fig:Cholesky_lc}
  and \ref{fig:symmetric_lc}. Here and throughout we denote the exact
  eigenvalues of the target distribution with $\kappa_\alpha$
  ($\alpha=1,\ldots, N_v$) and the eigenvalues of the model kernel $K$ with
  $\lambda_\alpha = \mu^2-\sigma_h^2\xi_\alpha^2$. We will refer to these as
  the RBM eigenvalues. The latter depends on the training stage, indicated by
  epochs, see Appendix \ref{app:pcd}. As can be seen in Figs.\
  \ref{fig:Cholesky_lc} and \ref{fig:symmetric_lc} (left), the RBM eigenvalues
  are correctly initialised for both choices and fluctuate around the correct
  values during training.
  
  To indicate the size of the fluctuations, we do the
  following. In the Cholesky case, we consider separately the $L_2$ norm of the
  lower triangular elements, of the upper triangular elements (which are
  initialised at zero) and of the elements on the diagonal. We then standard
  normalise these to compare the amplitudes of the fluctuations, see Fig.\
  \ref{fig:Cholesky_lc} (right). We observe that the sum of each part
  fluctuates around the average value during training, whose size is set by the
  initial value, demonstrating the stability of the PCD updates.    
  
  For the symmetric initialisation, we show the $L_2$ norms of the symmetric
  and asymmetric parts, $W_{\rm sym} = (W + W^T)/2, W_{\rm asym} = (W -
  W^T)/2$. Since the initial coupling matrix $W$ is symmetric, we expect the
  norm of the asymmetric part to remain significantly smaller during training.
  This can indeed be seen in Fig.\ \ref{fig:symmetric_lc} (right), where we
  show the evolution after standard normalisation. The norm of the symmetric
  part of the coupling matrix is six orders of magnitude larger than that of
  the asymmetric part. As with the Cholesky initialisation, we observe that the
  overall structure of the coupling matrix is approximately preserved. Note
  there is no reason for it to be {\em exactly} preserved, as this is neither
  imposed nor required.

\subsection{Initialisation with a random coupling matrix}

  In practical applications, the coupling matrix $W$ is not initialised at an
  exact solution, but with random entries, drawn e.g.\ from a Gaussian
  distribution. In Fig.\ \ref{fig:scalar_result} we show the results obtained
  with elements of $W$ sampled from a normal distribution $\mathcal{N}(0,0.1)$.
  Other parameters are as above within particular $N_h=N_v$ and
  $\mu^2>\kappa_{\rm max}$; hence there are no obstructions to learning the
  target distribution exactly. This can indeed be seen in Fig.\
  \ref{fig:scalar_result}, where both the eigenvalues (left) and the action
  density (right) are seen to match. For the latter, configurations are
  generated using the trained RBM; the same number of Monte Carlo (Metropolis)
  generated configurations are shown, using binning to remove
  autocorrelations. The analytical result follows from the equipartition.
  It is noted that possible instabilities, due to $\lambda_{\alpha}$ turning 
  negative either initially or during the learning stage, are not encountered with
  this initialization. If they are encountered, then they can be avoided by tuning 
  the width of the initial coupling matrix and learning rate.

  \begin{figure*}[t]
  \centering
    \includegraphics[width=0.48\textwidth]{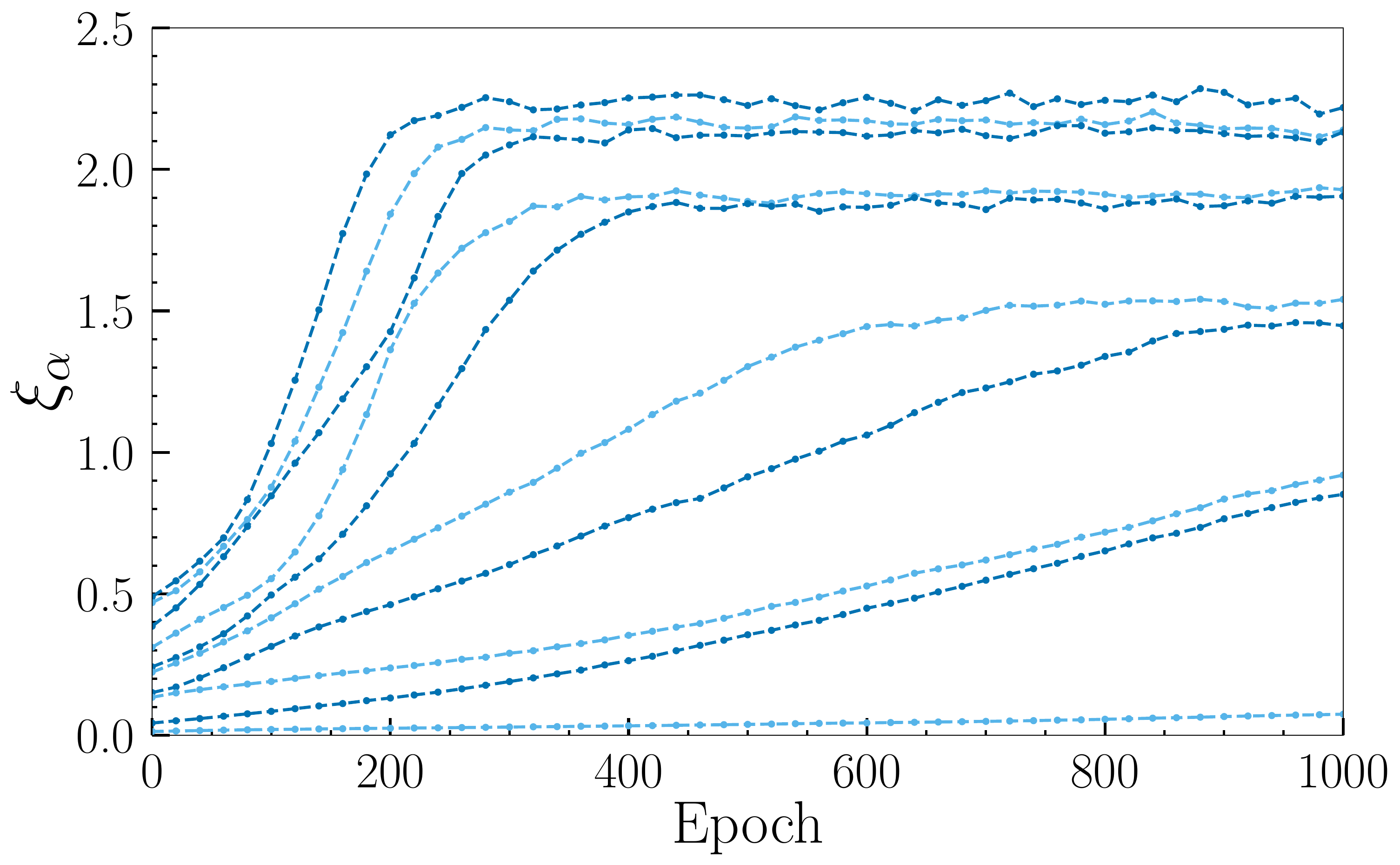}
    \includegraphics[width=0.48\textwidth]{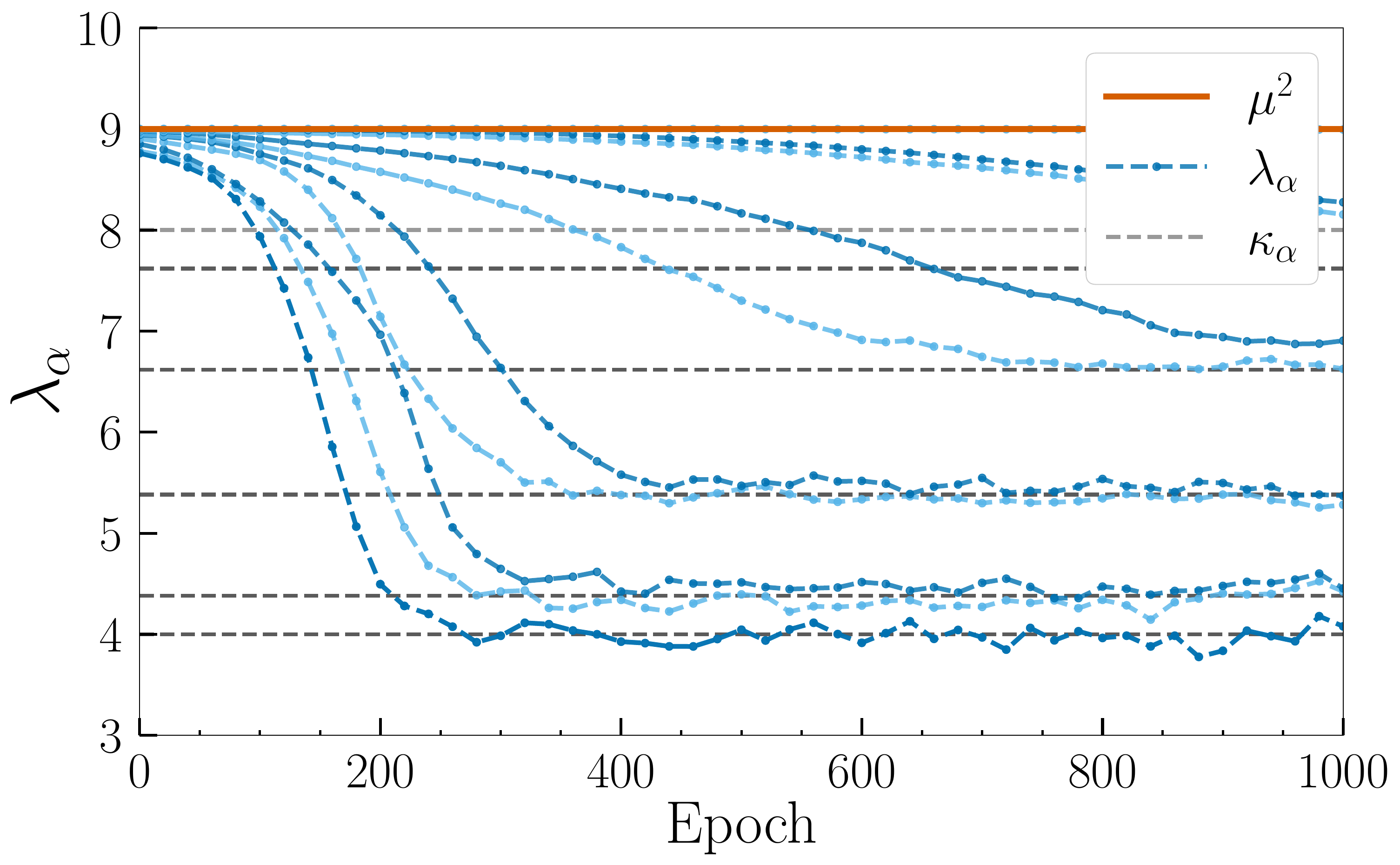}
    \caption{Convergence of the singular values $\xi_\alpha$ (left) and the
      eigenvalues $\lambda_\alpha$ (right) for the system of Fig.\
      \ref{fig:scalar_result}. Infrared modes are learnt the quickest.}
    \label{fig:scalar_convergence}
  \vspace*{0.2cm}
  \centering
    \includegraphics[width=0.48\textwidth]{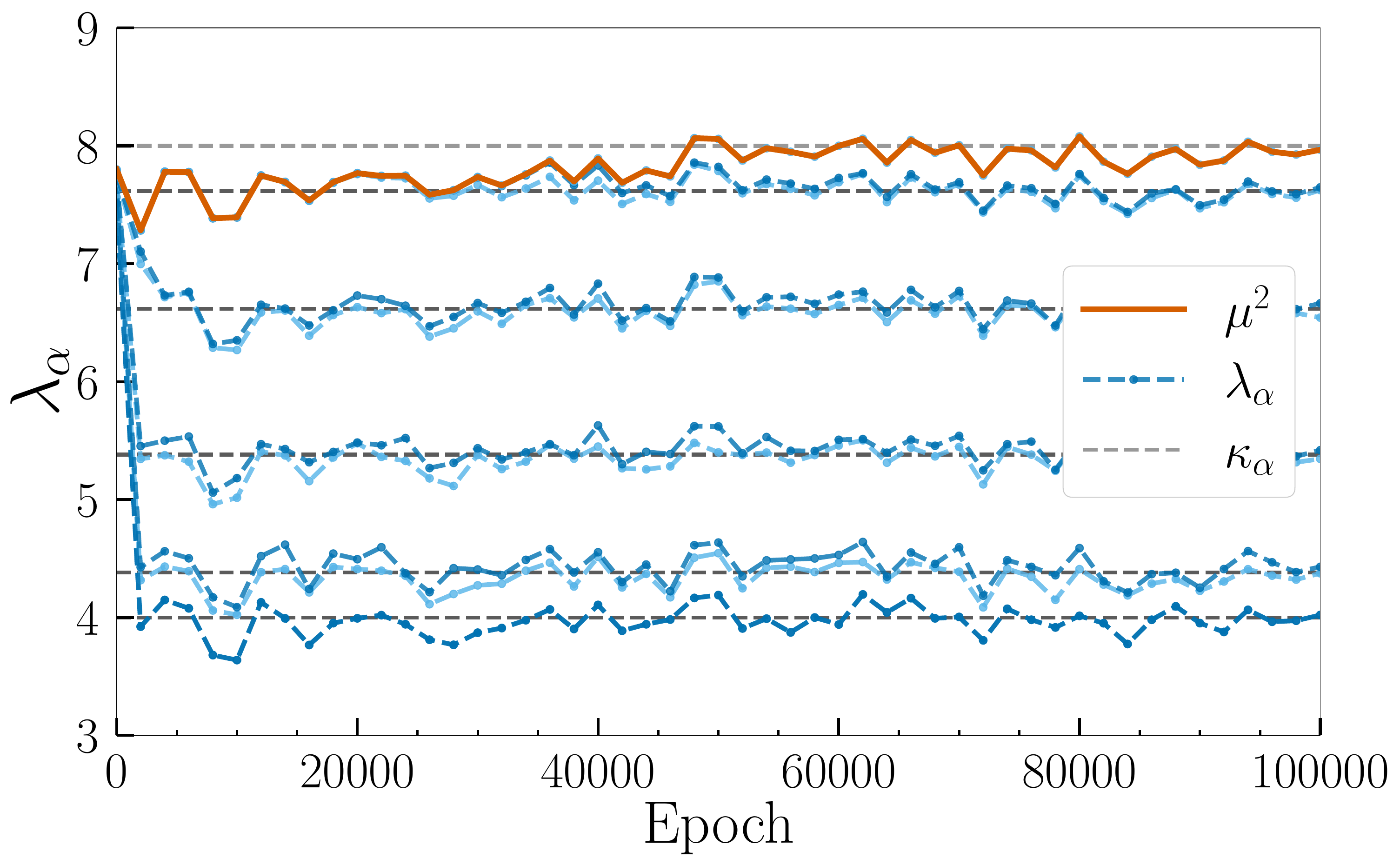}
    \includegraphics[width=0.48\textwidth]{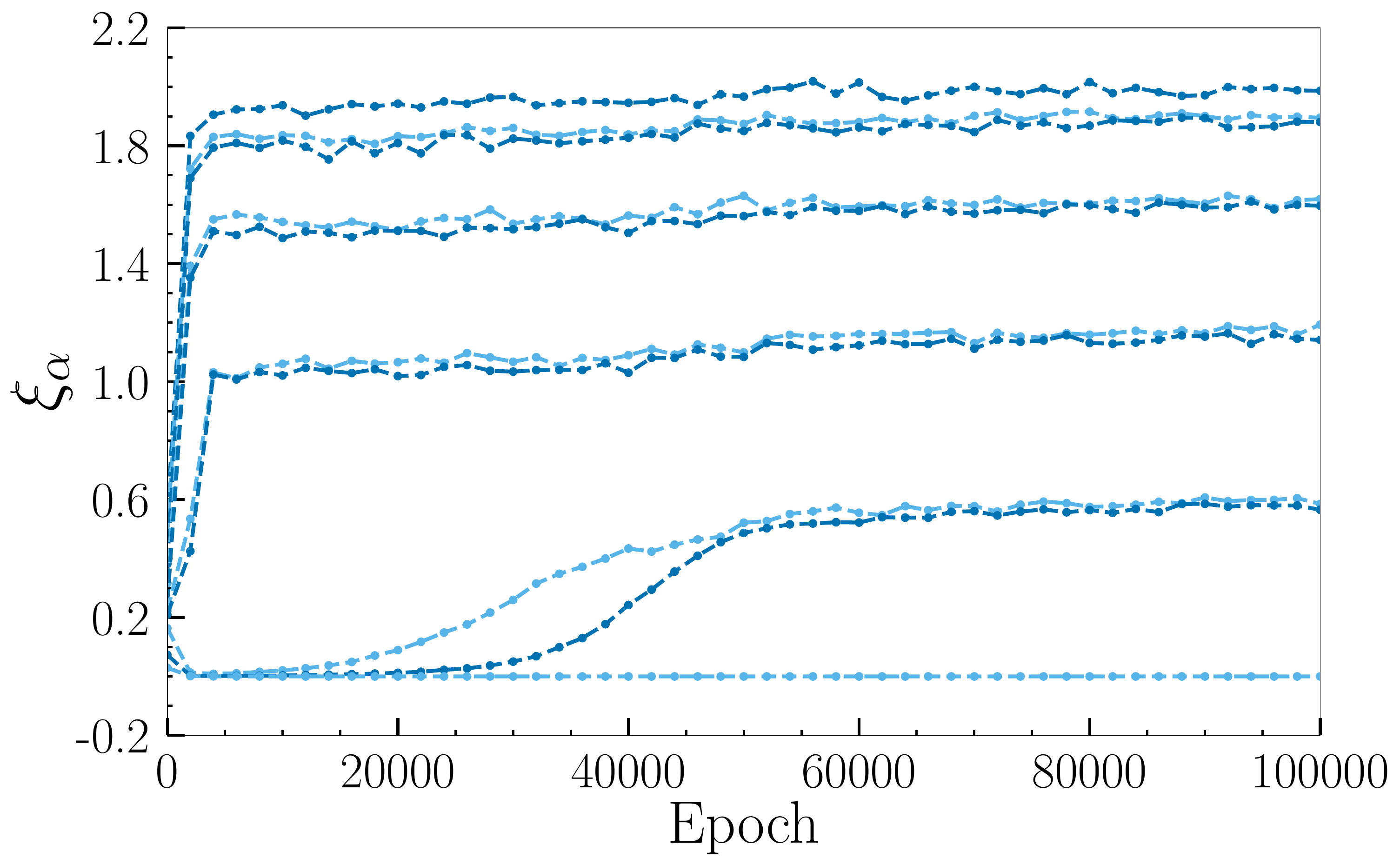}
    \caption{Left: evolution of the RBM eigenvalues and mass parameter $\mu^2$,
      with the latter initialised below $\kappa_{\rm max}$. Right: evolution of
      the singular values. Since the system is overparametrised, one of the
      singular values remains at the initial value.  }
    \label{fig:scalar_result_m1}
  \end{figure*}

  Since the elements of $W$ are initially relatively small, the corresponding
  singular values $\xi_\alpha$ are small as well and the RBM eigenvalues
  $\lambda_\alpha = \mu^2-\sigma_h^2\xi_\alpha^2$ are close to $\mu^2$
  initially. They quickly evolve to the target values $\kappa_\alpha$. The
  order in which the modes are learnt (or thermalised) can be understood
  easily. Referring back to Sec.\ \ref{sec:SVD}, we consider Eq.\
  (\ref{eq:XiXi}) for the singular values and Eq.\ (\ref{eq:Lambda}) for the
  driving term. Assuming we are on the correct eigenbasis, the latter reduces
  to 
  \be
    \Lambda = \Lambda_d =  D_\phi^{-1} -D_K^{-1} 
    = \mbox{diag}\left(1/\kappa_\alpha - 1/\lambda_\alpha\right), 
    \ee
    where 
    $\lambda_\alpha = \mu^2-\sigma_h^2\xi_\alpha^2$. 
  Equation (\ref{eq:XiXi}) then becomes \cite{Decelle_2021}
  \be \label{eq:training_eq}
    \frac{d}{dt}\xi_\alpha^2 = 2\sigma_h^2
    \left( \frac{1}{\kappa_{\alpha}} - \frac{1}{\mu^2 - \sigma_h^2 \xi_{\alpha}^2} \right) 
    \xi_{\alpha}^2.
  \ee
  Note that this equation was encountered before (in a general basis) for
  $N_v=N_h=2$, see Eqs.\ (\ref{eq:xi1}), (\ref{eq:xi2}). During the initial
  stages, the term within the brackets is negative and largest for the smallest
  eigenvalues. Hence the corresponding singular values evolve quickest. At late
  times, one may linearise around the fixed point. In Sec.\ \ref{sec:SVD} we
  demonstrated for $N_v=2$ nodes that the convergence in the linearised regime
  is exponentially fast and that the rate of convergence is set by
  $\gamma_\alpha=(\mu^2-\kappa_\alpha)/\kappa_\alpha^2$. Hence the most
  infrared modes equilibrate fastest and the ultraviolet modes slowest. These
  aspects are demonstrated in Fig.\ \ref{fig:scalar_convergence}, where we have
  shown the evolution of both the singular values (left) and the eigenvalues
  (right) during the initial stages of the training (the largest singular
  values correspond to the smallest eigenvalues). We note the similarity with
  the case of $N_v=2$ modes in Sec.\ \ref{sec:SVD}, see in particular Fig.\
  \ref{fig:evol} (top row). 
  So far we have kept the RBM mass parameter $\mu^2$ fixed. However, it can
  also be treated as a learnable parameter using Eq.\ (\ref{eq:mu2update}).
  This is particularly useful if details of the target spectrum are not known.
  It provides then an additional degree of freedom. In Fig.\
  \ref{fig:scalar_result_m1}, the initial RBM mass parameter is initialised
  below $\kappa_{\rm max}$. It subsequently increases to match the largest
  eigenvalue, see Fig.\ \ref{fig:scalar_result_m1} (left).  Since the system is
  over-parametrised, one of the singular values remains at the initial value,
  see  Fig.\ \ref{fig:scalar_result_m1} (right).  Note the different timescale
  for equilibration compared to the case with a constant $\mu^2$, as it takes
  time for $\mu^2$ to find the correct value.
\begin{figure*}[t]
  \centering
    \includegraphics[width=0.48\textwidth]{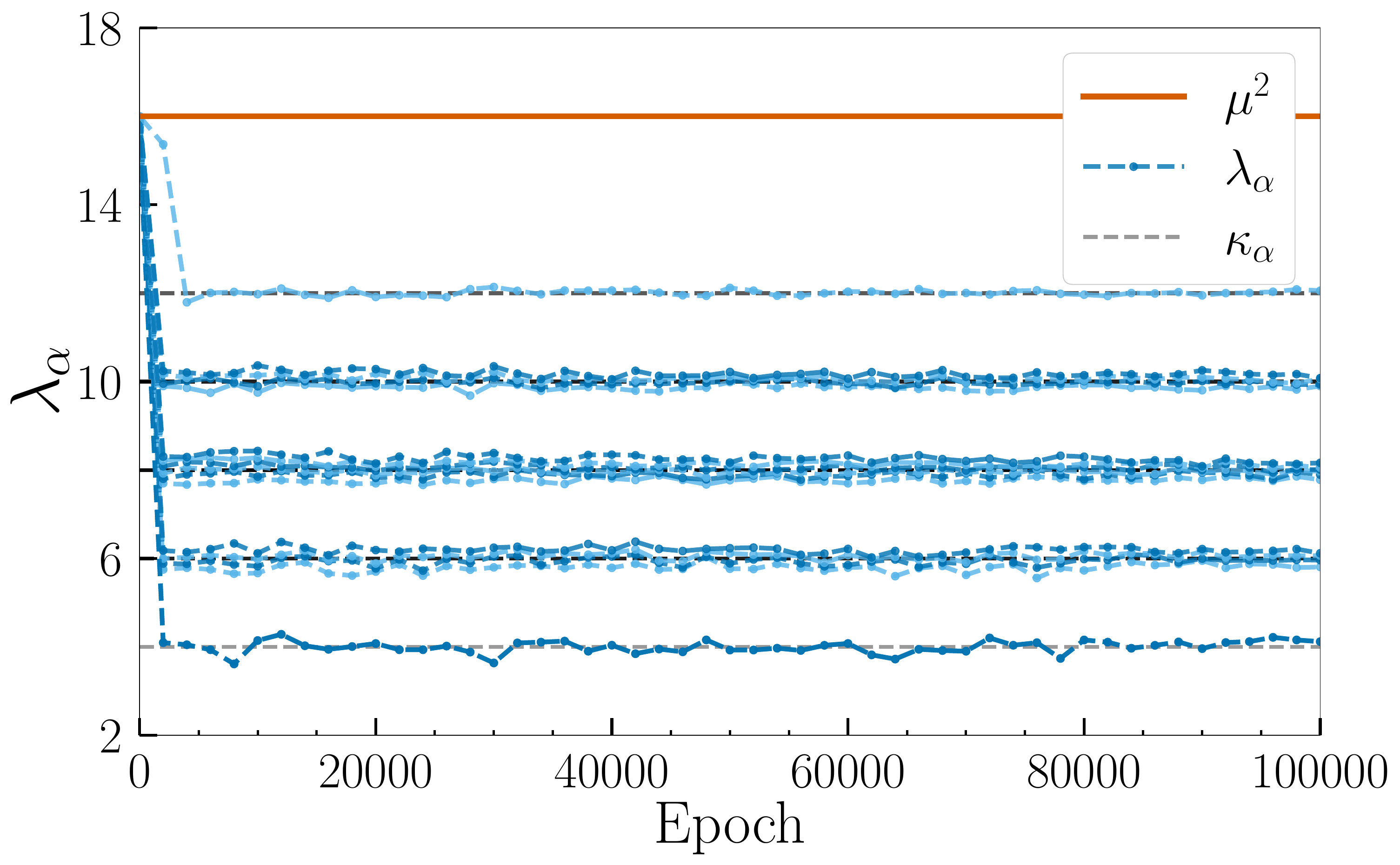}
    \includegraphics[width=0.48\textwidth]{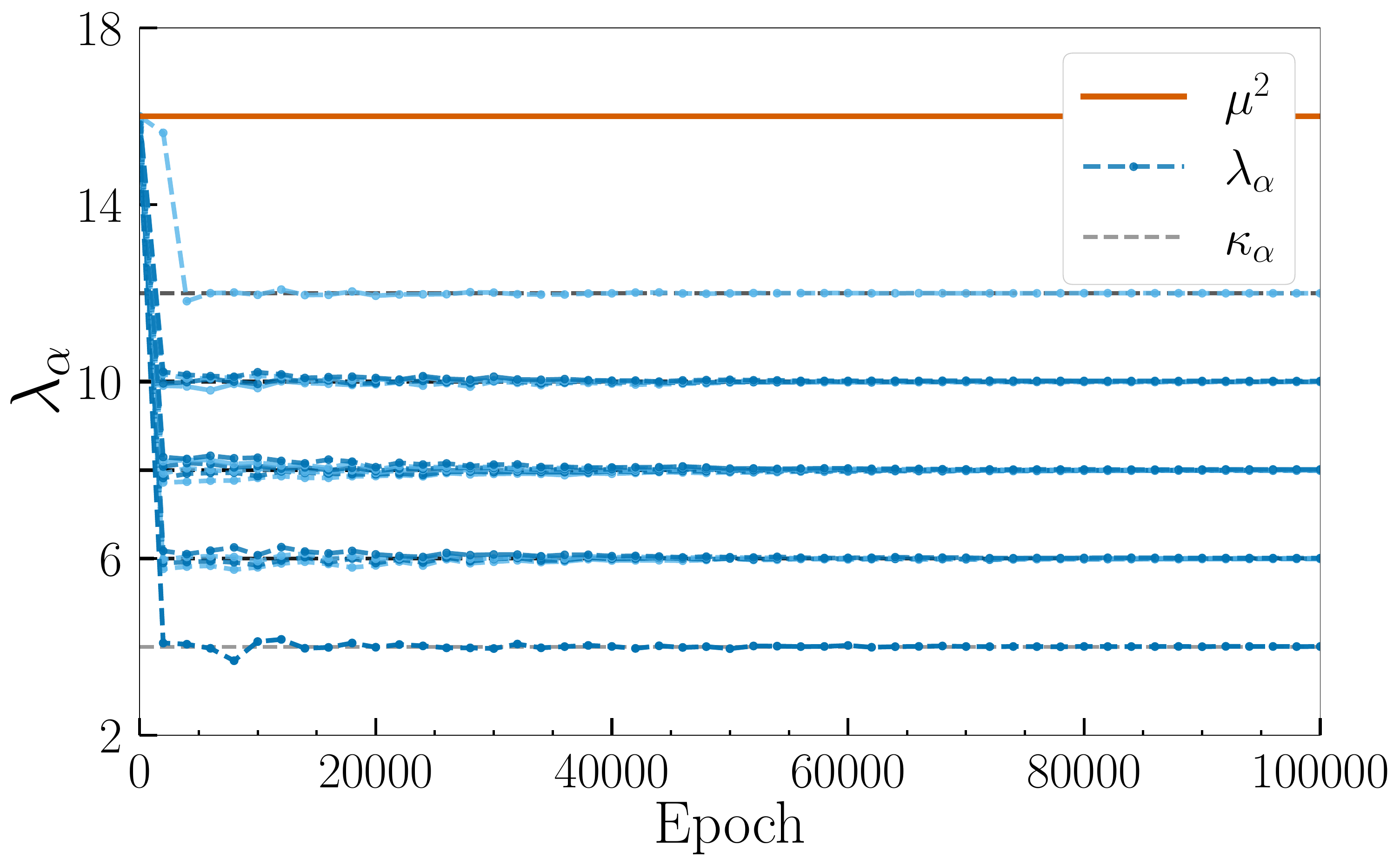}
    \caption{Evolution of the eigenvalues in the two-dimensional case during
      training, with constant $\mu^2=16$ and $N_v=N_h=16$, using a fixed (left)
      and a diminishing (right) learning rate. }
    \label{fig:learningrate}
  \end{figure*}
  \begin{figure}[t]
  \centering
    \includegraphics[width=0.8\columnwidth]{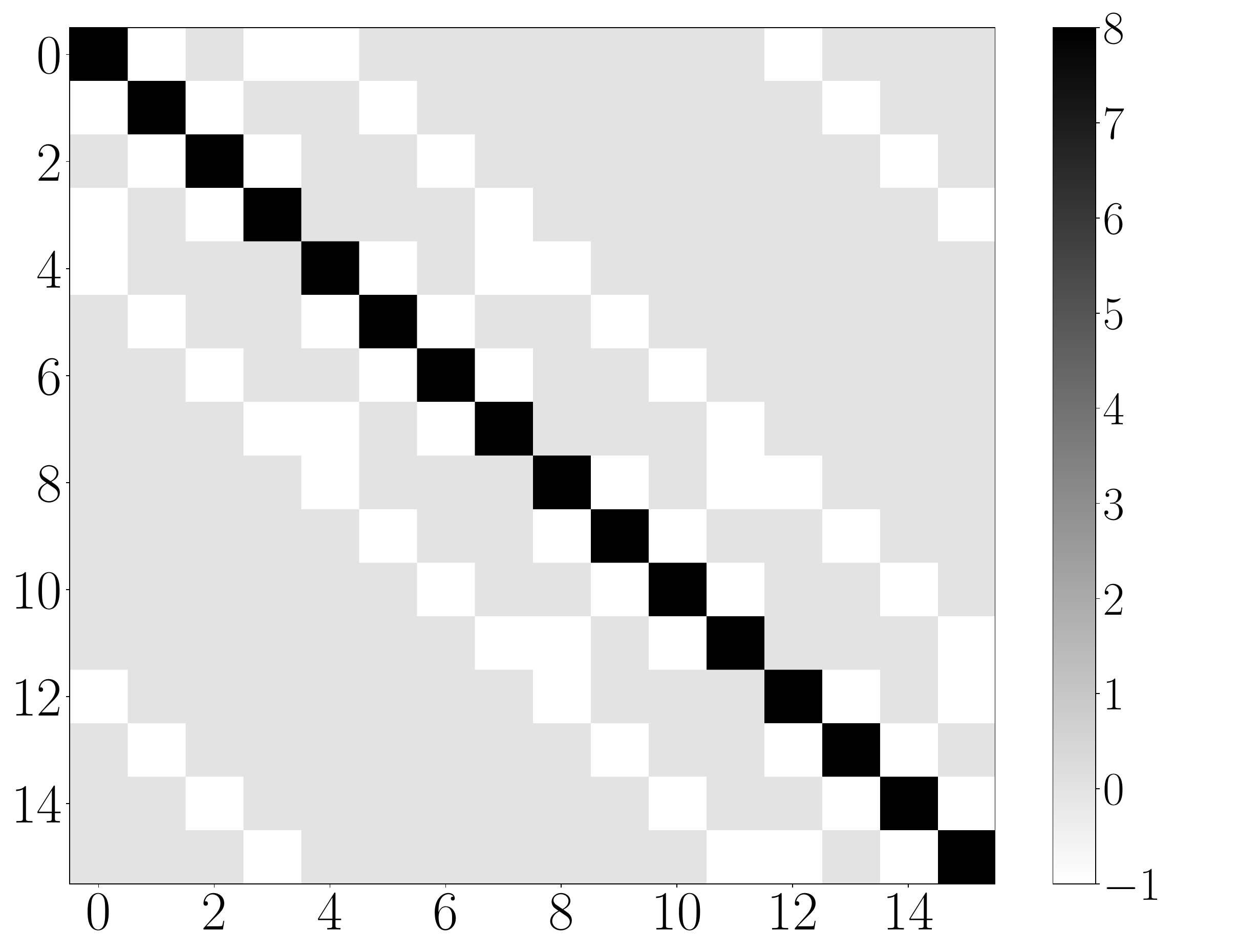}
    \caption{Flattened kernel for a two-dimensional scalar field theory on a
      lattice with $N_x\times N_y=4\times 4$ sites and $m^2=4$. Each site has four nearest
      neighbours. }
    \label{fig:2d}
 \end{figure}

  Up to now, we considered a scalar field in one dimension only. The
  generalisation to higher dimensions is interesting since the RBM does not
  know about the dimensionality {\it a priori}, with the $N_v$ visible nodes only
  connecting to the hidden nodes. We consider here two dimensions, using an
  $N_x\times N_y$ lattice. The eigenvalues of the target kernel are 
  \bea \label{eq:kappa2d}
    \kappa_\veck = &&\hm m^2 + p_{{\rm lat}, \veck}^2 \nn\\
    =&&\hm m^2 + 4-2\cos\left(\frac{2\pi k_x}{N_x}\right) 
    -2\cos\left(\frac{2\pi k_y}{N_y}\right),
\ \eea
  with $ -N_{x,y}/2 < k_{x,y} \leq N_{x,y}/2$. 
  In this case, there is a larger degeneracy of eigenvalues. The RBM has
  $N_v=N_x\times N_y$ visible nodes. The dimensionality has to be learnt and
  encoded in the coupling matrix $W$. The (target) kernel and two-point
  functions are $(N_x \times N_y) \times (N_x \times N_y)$-dimensional tensors.
  This two-dimensional structure can be flattened in a one-dimensional
  representation, where the kinetic term is decomposed into a tensor product of 
  two one-dimensional Laplacian operators,
  \begin{align}
    K^{\phi, 2d} = \;& m^2 \id + \Delta^{2d} \nonumber \\
    = \;& m^2 \id_{N_x\times N_x}\otimes \id_{N_y\times N_y} \nonumber \\
    & + \Delta^{1d} \otimes \id_{N_y\times N_y}  + \id_{N_x\times N_x}  \otimes \Delta^{1d},
\label{eq:tensor}
  \end{align}
  where in the last expression $\otimes$ is the Kronecker product and the sizes of the identity matrices are given explicitly. 
  
  Figure 9 shows an example of a flattened scalar field kernel for the two-dimensional case with $N_x=N_y=4$. Importantly, the spectrum of the flattened kernel and the original kernel are identical, since the boundary conditions are encoded correctly. The tensor product decomposition (\ref{eq:tensor}) allows one to see this explicitly.

  In Fig.\ \ref{fig:learningrate} (left), we show the evolution of the RBM
  eigenvalues. The RBM mass parameter  is $\mu^2=16> \kappa_{\rm max} = 12$.
  There should be four degenerate eigenvalues at 6 and 10, and six degenerate
  ones at 8. Yet it appears the eigenvalues only lie within a band close to the
  expected value. This is due to the fact that to obtain these results we have
  used a fixed learning rate (time step), which prevents the system from
  reaching high precision. This can be remedied by introducing an epoch
  dependent learning rate. This is explored in Appendix \ref{app:pcd}. We multiply
  the learning rate by a factor close to one, $r=0.99$, after a given number of
  epochs, $N_{\rm epoch}^{\rm rate}=128$. The virtue of having a diminished
  learning rate in the later stages is that it allows the model to be finely
  trained, with less statistical fluctuations. The result is shown in Fig.\
  \ref{fig:learningrate} (right), where we indeed observe precise agreement
  with the target spectrum. 

  \begin{figure*}[t]
  \centering
    \subfloat[$\mu^2=7.8$]{\includegraphics[width=0.48\textwidth]{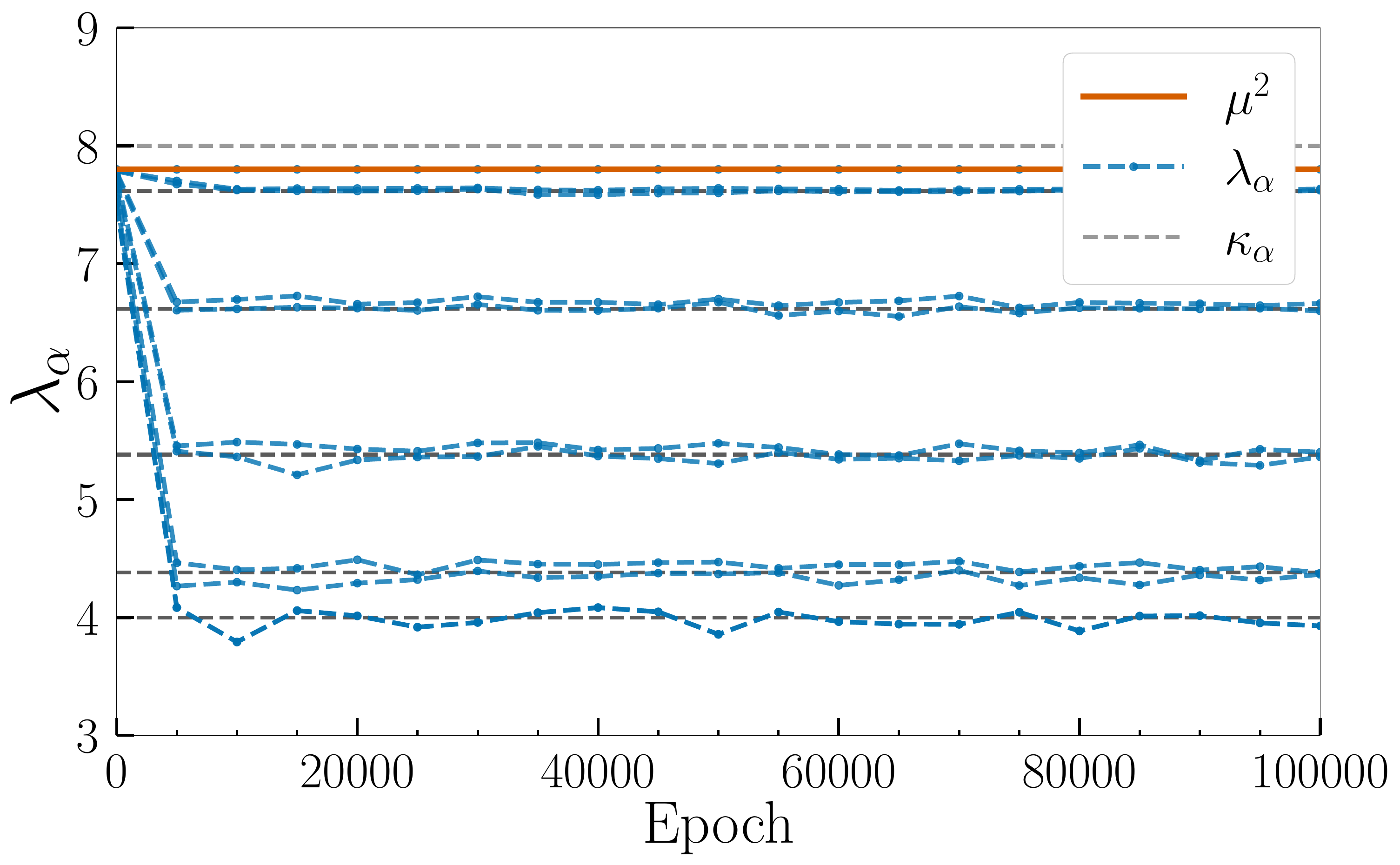}}
    \subfloat[$\mu^2=7.1$]{\includegraphics[width=0.48\textwidth]{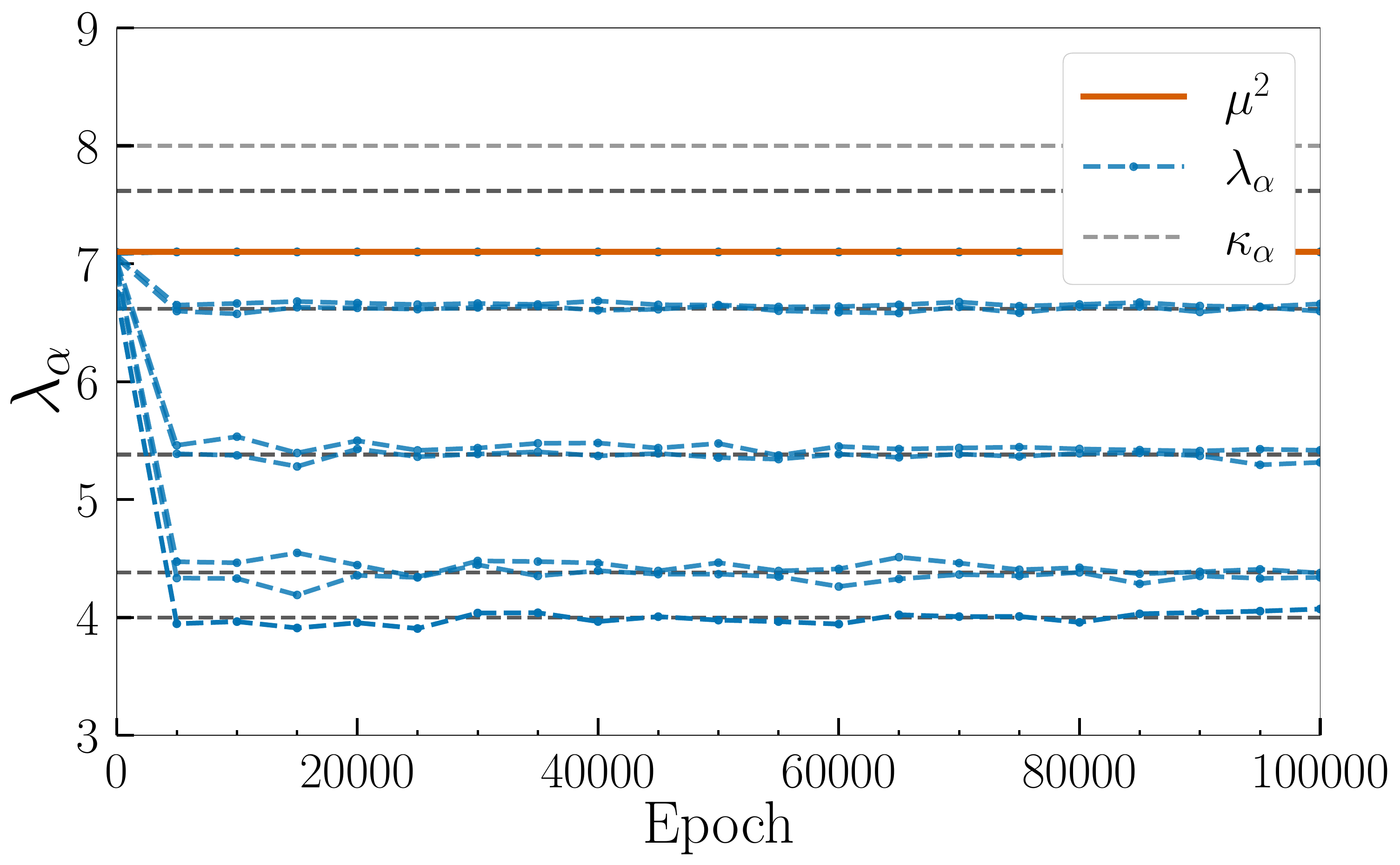}}
    \\
    \subfloat[$\mu^2=5.8$]{\includegraphics[width=0.48\textwidth]{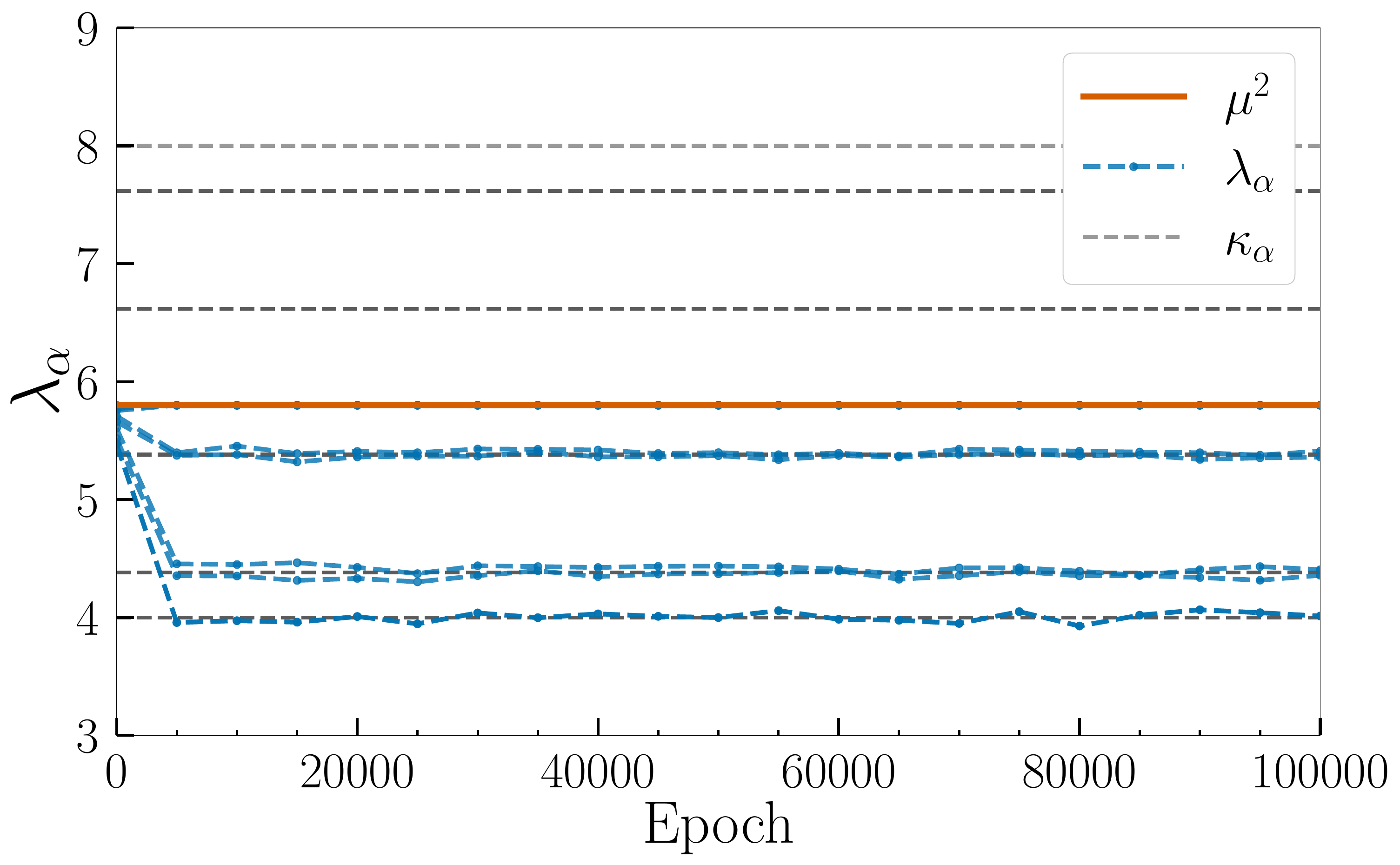}}
    \subfloat[$\mu^2=4.8$]{\includegraphics[width=0.48\textwidth]{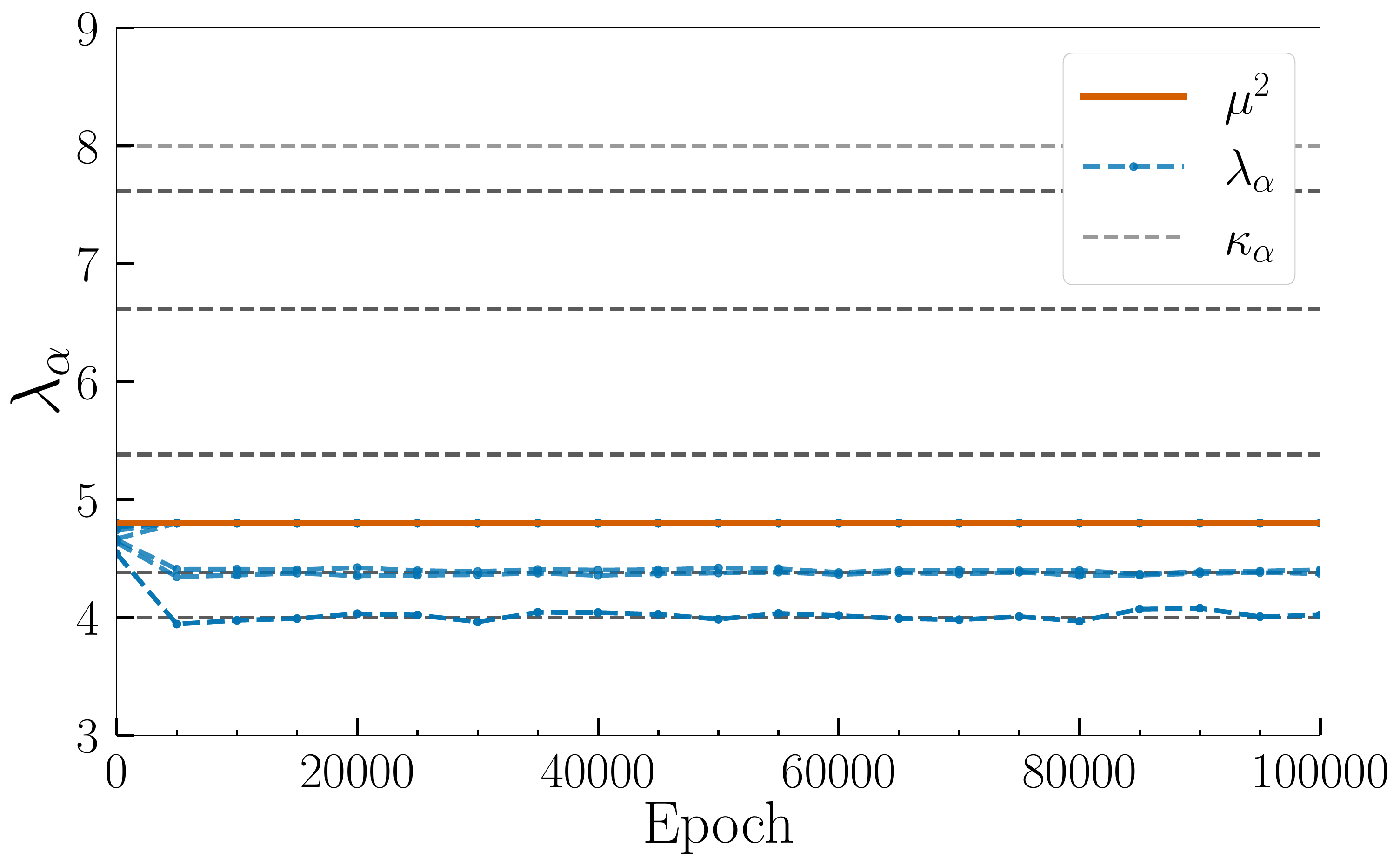}}
  \caption{Regularisation by RBM mass parameter $\mu^2$: evolution of the
    eigenvalues in the one-dimensional scalar field theory. Only the infrared
    part of the spectrum is reproduced.}
  \label{fig:m_cutoff}
  \end{figure*}

\subsection{Ultraviolet regularization by the RBM mass parameter}

  Up to now, we have considered the ideal ``architecture,'' namely $N_h=N_v$
  and $\mu^2>\kappa_{\rm max}$, for which Gaussian distributions can be learnt
  exactly, as we have demonstrated. In practice, one often chooses $N_h< N_v$
  and the maximum eigenvalue may not be known. Here we determine what this
  implies.

  We start with the case where $N_h=N_v$, but with $\mu^2$ fixed and less than
  $\kappa_{\rm max}$. We refer to Eq.\ (\ref{eq:training_eq}) for the evolution
  of the singular values in the eigenbasis. Take $\mu^2<\kappa_\alpha$. In this
  case, the term inside the brackets is always negative and the only solution
  is $\xi_\alpha=0$. The corresponding eigenvalue is then
  $\lambda_\alpha=\mu^2$. When $\mu^2>\kappa_\alpha$, the solution is given by
  the fixed point, $\sigma_h^2 \xi_\alpha^2 = \mu^2-\kappa_\alpha$, and
  $\lambda_\alpha = \mu^2-\sigma_h^2\xi_\alpha^2$. We hence conclude that the
  infrared part of the spectrum, with $\kappa_\alpha<\mu^2$, can be learnt,
  whereas the ultraviolet part, with $\kappa_\alpha>\mu^2$, cannot be learnt.
  Instead, the RBM eigenvalues take the value of the cutoff, $\mu^2$
  \cite{KARAKIDA201678}.

  This is demonstrated in Fig.\ \ref{fig:m_cutoff} for a one-dimensional scalar
  field theory with $N_v=N_h=10$ nodes. As the condition for exact training is
  violated, the RBM model can no longer faithfully reproduce the target data
  and distribution. The impact of this depends on the importance of the
  ultraviolet modes, as we will see below for the MNIST dataset.

  \begin{figure*}[t]
  \centering
    \subfloat[$N_h = 9$]{\includegraphics[width=0.48\textwidth]{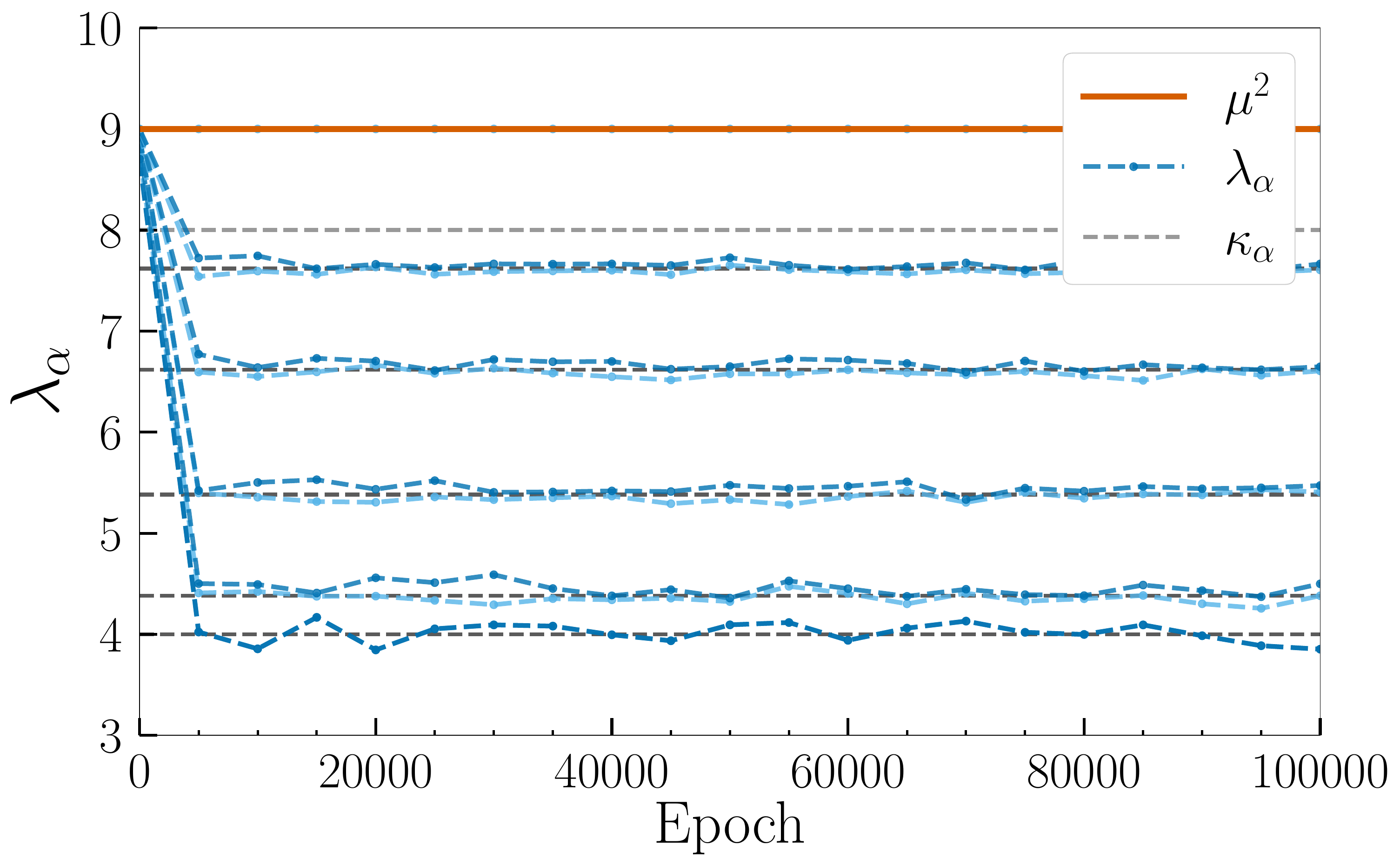}}
    \subfloat[$N_h = 8$]{\includegraphics[width=0.48\textwidth]{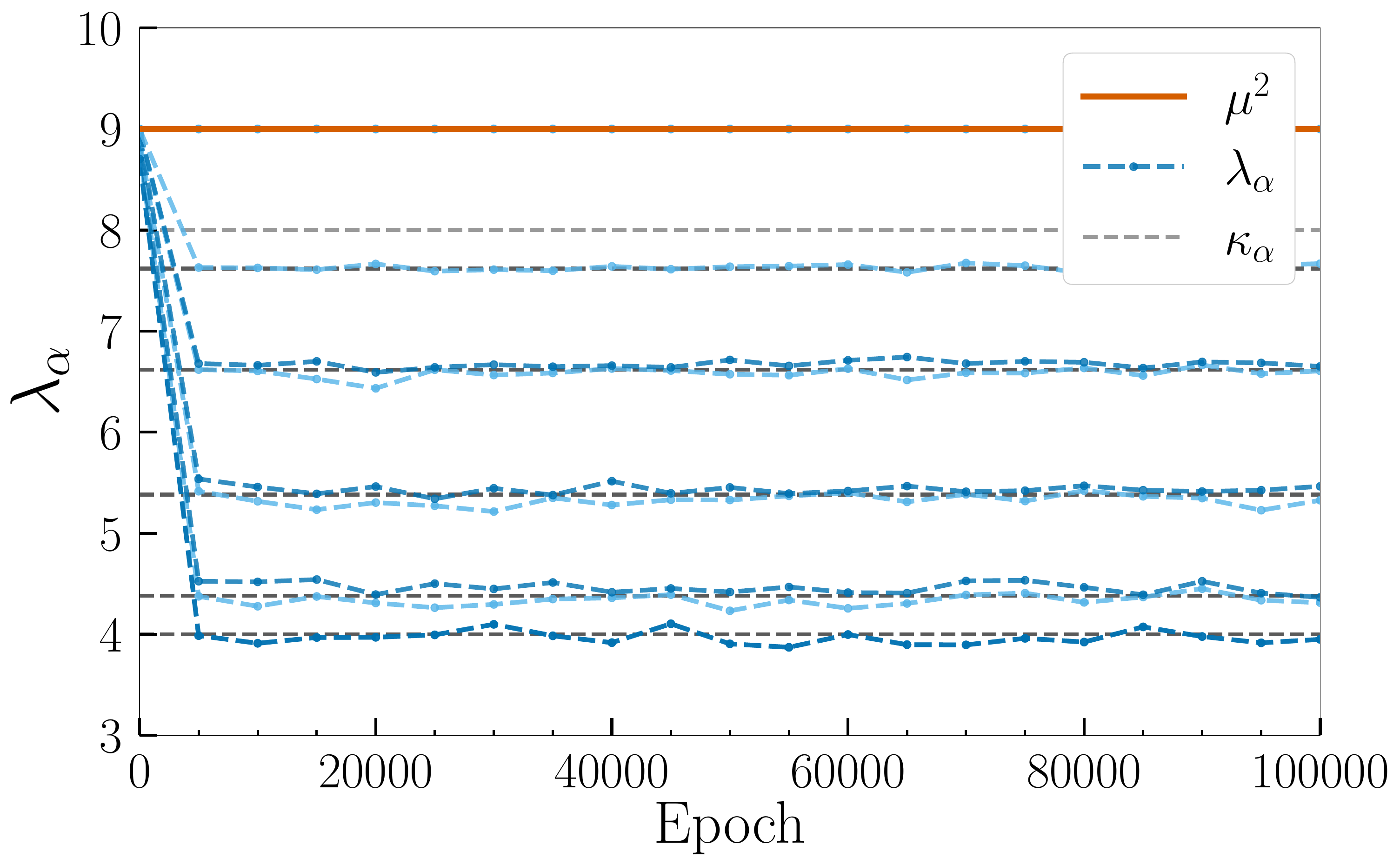}}
    \\
    \subfloat[$N_h = 7$]{\includegraphics[width=0.48\textwidth]{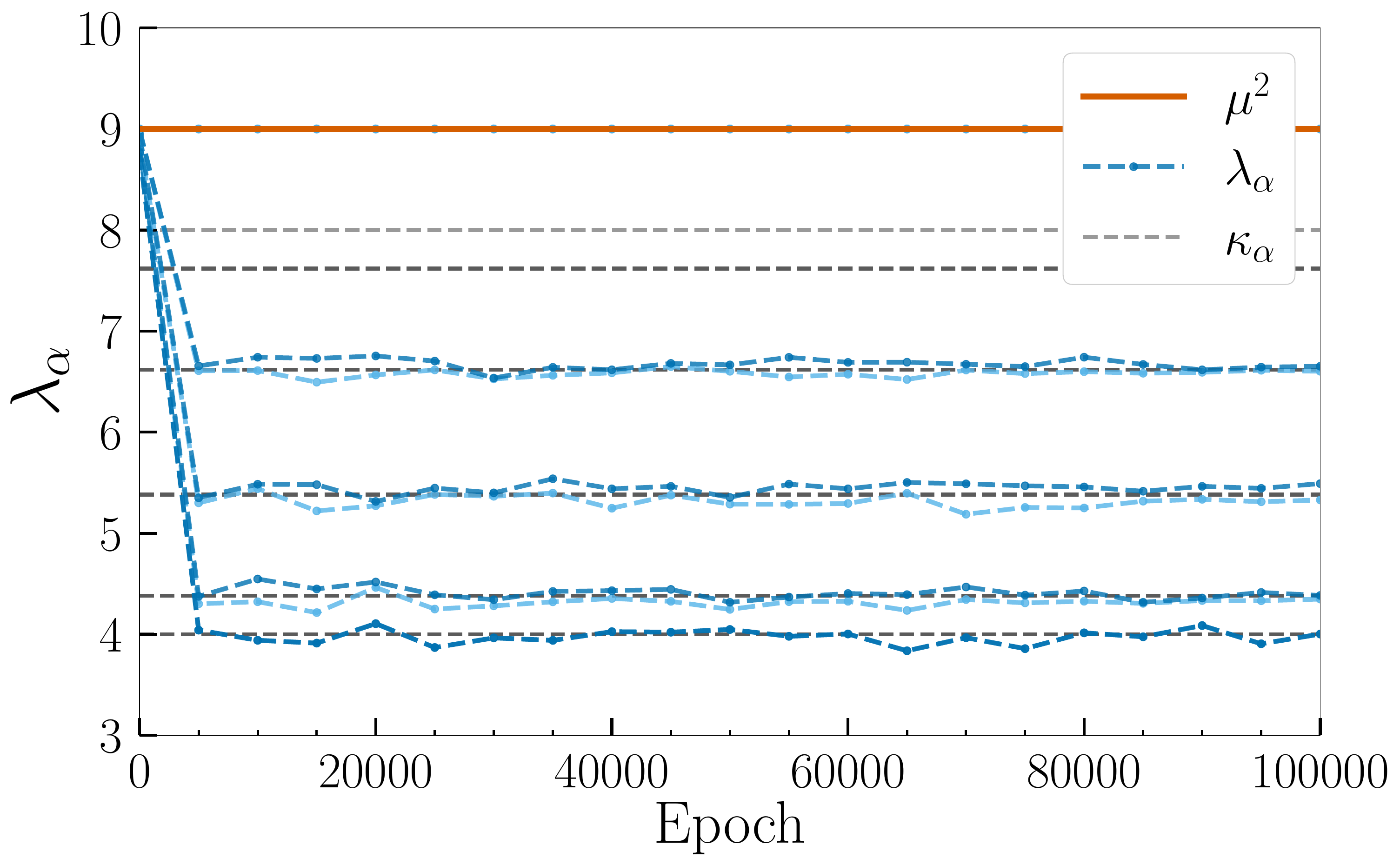}}
    \subfloat[$N_h = 6$]{\includegraphics[width=0.48\textwidth]{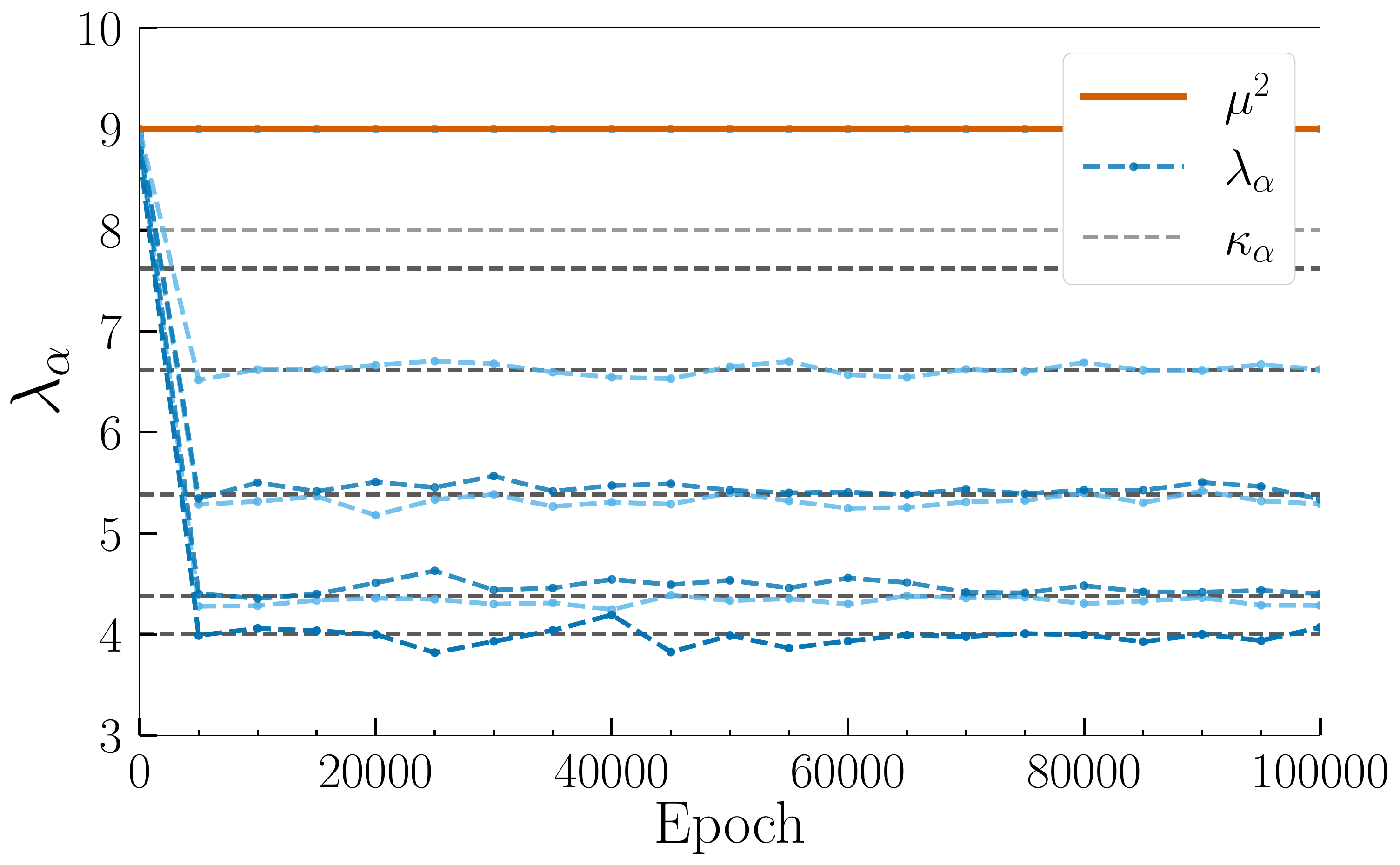}}
  \caption{As above, but with regularisation by the number of hidden nodes $N_h$. 
  }
  \label{fig:N_K}
  \end{figure*}
  
  
  \begin{figure*}[t]
  \centering
    \includegraphics[width=0.48\textwidth]{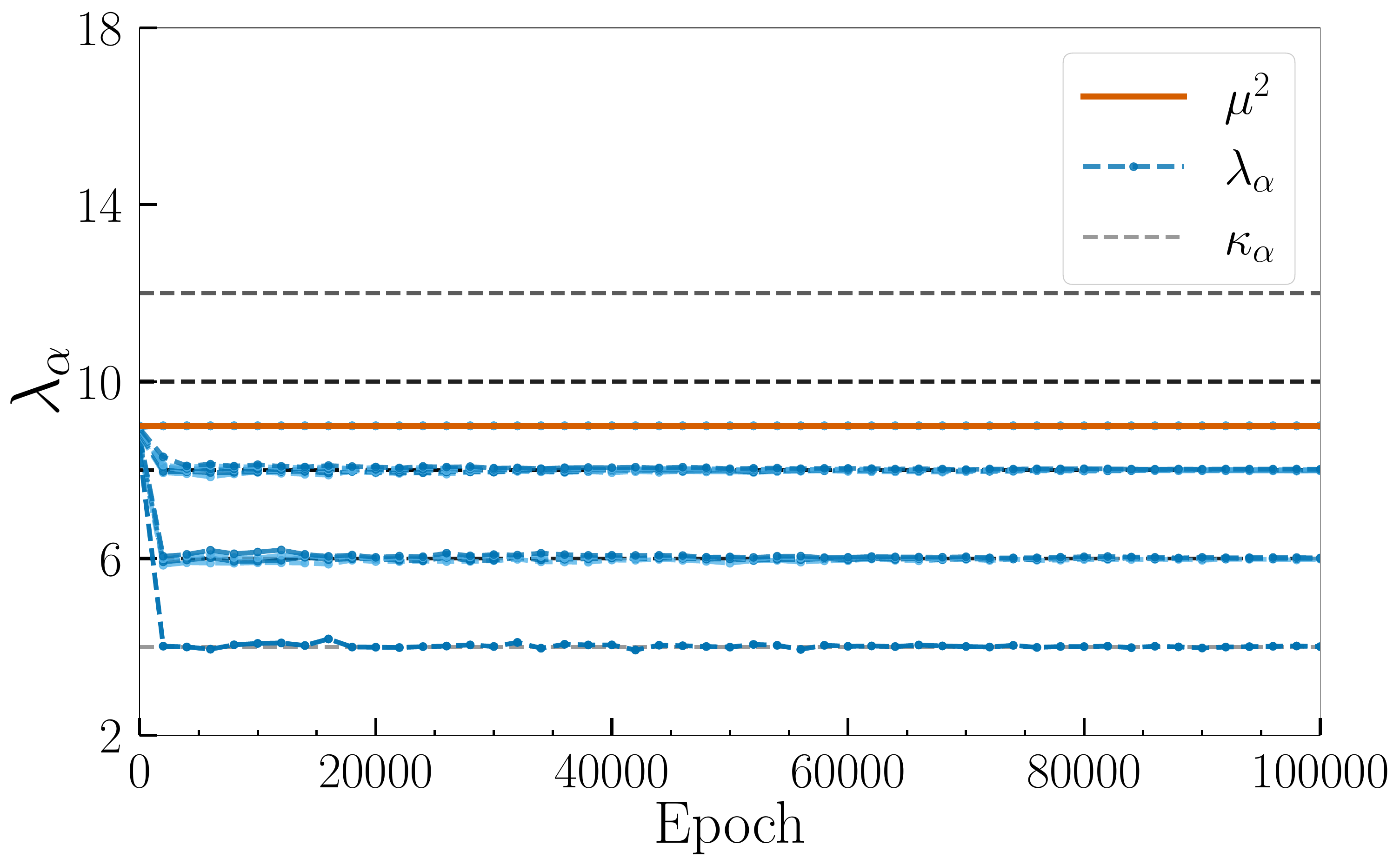}
    \includegraphics[width=0.48\textwidth]{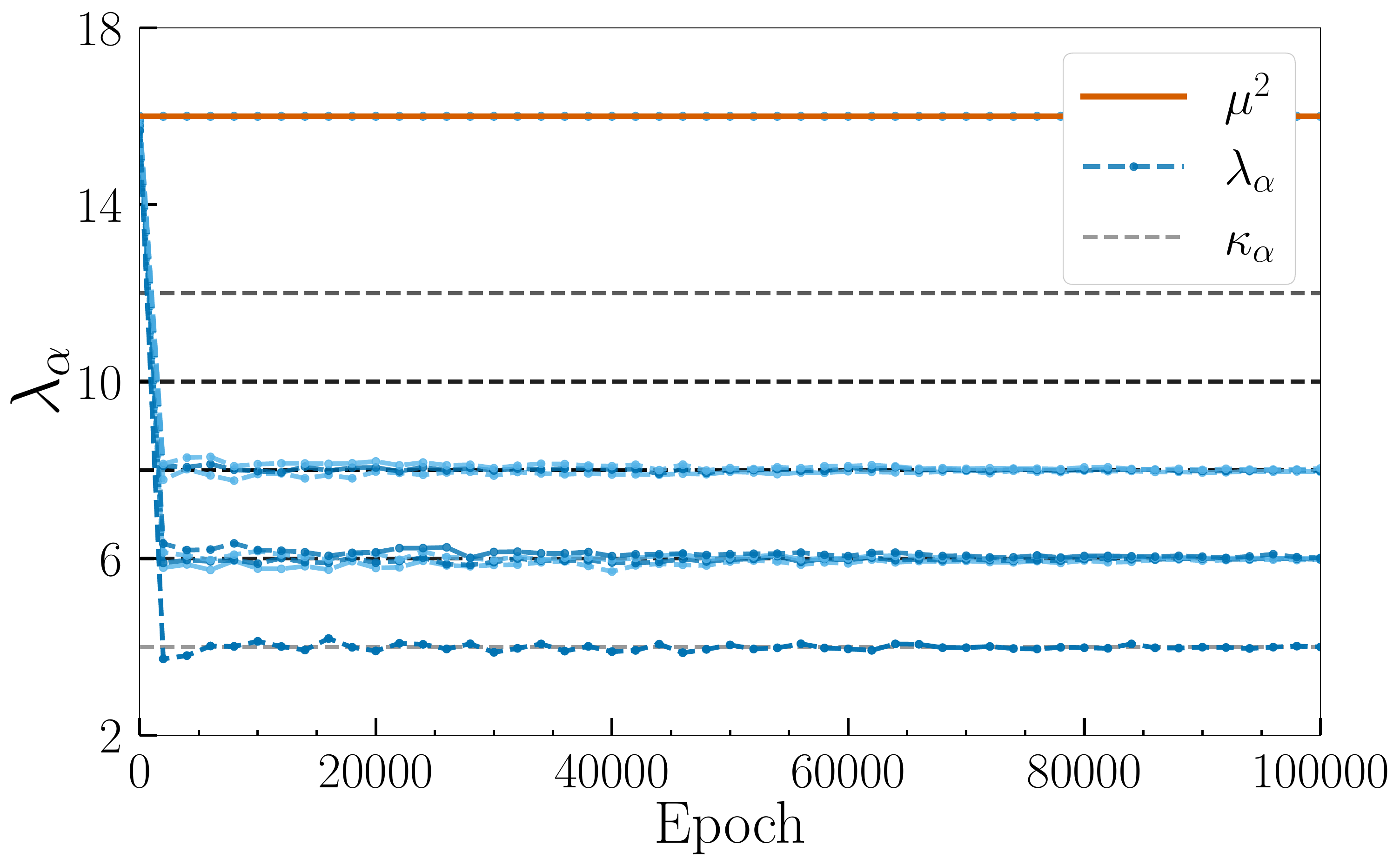}
    \caption{Regularisation by the RBM mass parameter $\mu^2=9$ (left) and by
      the number of hidden modes $N_h=8$ (right) in the two-dimensional scalar
      field theory with $N_x\times N_y=4\times 4$. }
    \label{fig:reg2d}
  \end{figure*}

\subsection{Ultraviolet regularisation by the number of hidden nodes}

  Next, we consider the case with $N_h < N_v$. This is straightforward, as
  there are only $N_h$ singular values, leading to the RBM eigenvalues 
  \be
    \lambda_\alpha = 
      \begin{cases}
        \mu^2 - \sigma_h^2 \xi_{\alpha}^2
        & \quad 1\leq \alpha \le N_h,
        \\
        \mu^2  & \quad N_h< \alpha \leq N_v,
      \end{cases}
  \ee
  see e.g.\ Eq.\ (\ref{eq:DK}). Again we note that the infrared part of the
  spectrum can be reproduced, whereas the ultraviolet part is fixed at $\mu^2$,
  irrespective of the actual value of the target eigenvalue.

  This is shown in Fig.\ \ref{fig:N_K} for the one-dimensional case with
  $N_v=10$ and $N_h=9, 8, 7, 6$. Note that all eigenvalues, except the minimal
  and maximal ones, are doubly degenerate. Hence in the case of $N_h=8$ and
  $6$, one of the degenerate eigenvalues remains and one is removed. 

  Finally, in Fig.\ \ref{fig:reg2d} we give two examples in the two-dimensional
  scalar theory, using $\mu^2=9<\kappa_{\rm max}$ on the left and
  $N_h=8<N_v=16$ on the right.

\section{MNIST dataset}
\label{sec:MNIST}

 \begin{figure*}[t]
  \centering
    \includegraphics[width=0.7\textwidth]{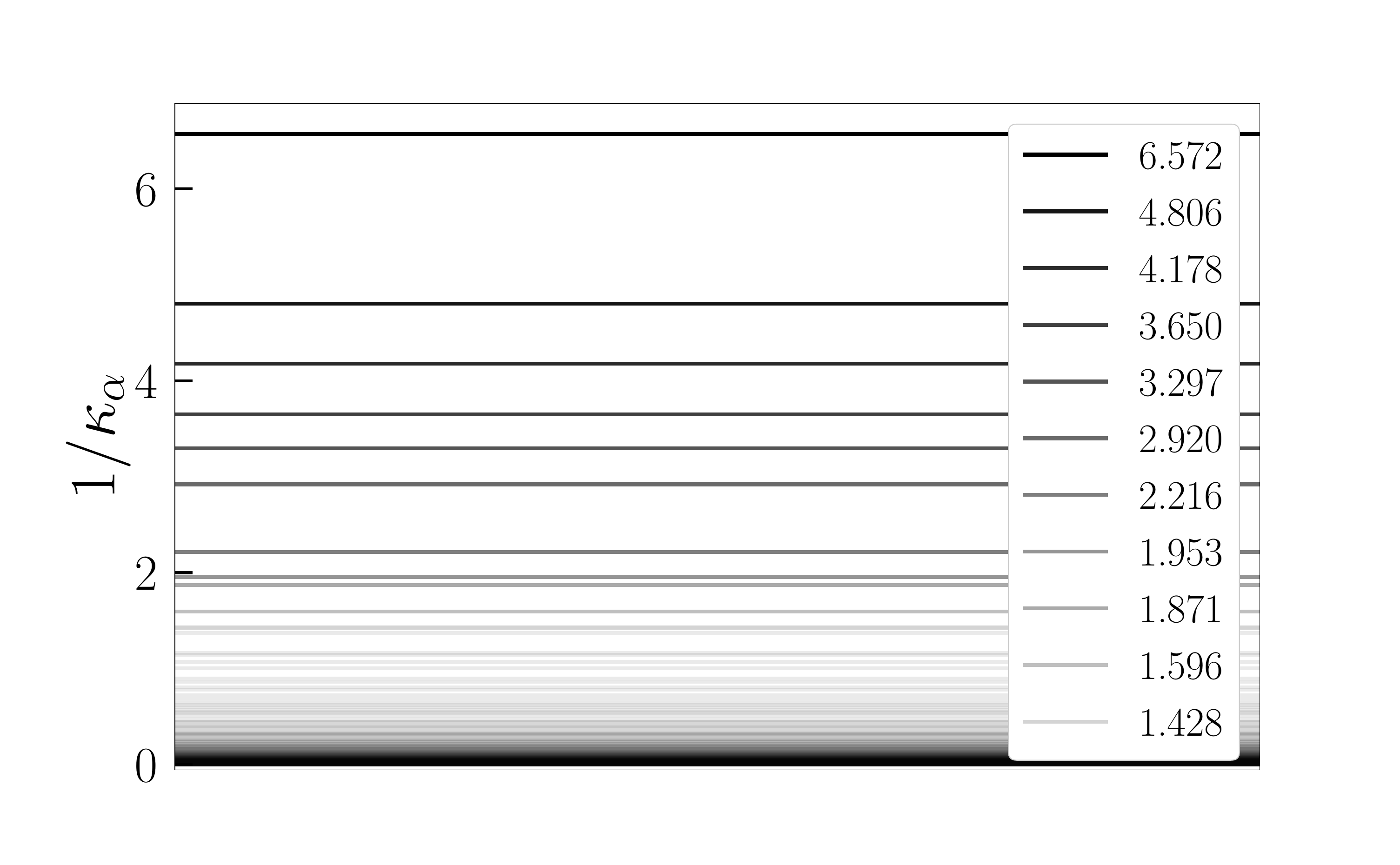}
    \caption{Eigenvalues of the correlation matrix $\expv{\phi_i \phi_j}$ of the
      MNIST dataset. Note that many eigenvalues are close to zero. The values of
      the ten largest eigenvalues are indicated.}
    \label{fig:mnist}
  \end{figure*}

  It is important to ask whether the considerations above are also relevant for
  realistic datasets commonly used in ML. We consider the MNIST dataset
  \cite{MNIST}, consisting of 60,000 $28\times 28$ images of digits. Hence
  $N_v=784$, substantially larger than what we have considered so far.

  Unlike in the case of scalar fields, the probability distribution function is
  not known. However, we may still obtain the correlation matrix
  $\bra\phi_i\phi_j\ket$ by summing over the 60,000 realisations. The MNIST
  (target) kernel is then given by its inverse, 
  \be
    K^{\rm MNIST} = \expv{\phi_i \phi_j}_{\rm MNIST}^{-1}.
  \ee
  The eigenvalues of the correlation matrix are the inverse of the eigenvalues
  of the kernels discussed so far and we hence denote them as
  $1/\kappa_\alpha$. The 784 eigenvalues are shown in Fig.~\ref{fig:mnist}.
  Many eigenvalues are close to zero. In the language of the previous sections,
  these correspond to the ultraviolet part of the spectrum of the quadratic
  kernel and hence the MNIST dataset can be said to be ultraviolet divergent.
  The values of the ten largest eigenvalues of the correlation matrix are
  listed explicitly on the right. These correspond to the infrared part of the
  spectrum of the quadratic kernel. Since these are finite, the MNIST dataset
  is infrared safe. This terminology reflects the ordering of the eigenvalues $\kappa_\alpha$ encoding the quadratic correlations in the MNIST data, from small (infrared) to large (ultraviolet).
As in the two-dimensional scalar case, the eigenvalues do not depend on the flattening of the indices.

\begin{figure*}[t]
  \centering  
    \includegraphics[width=0.56\textwidth]{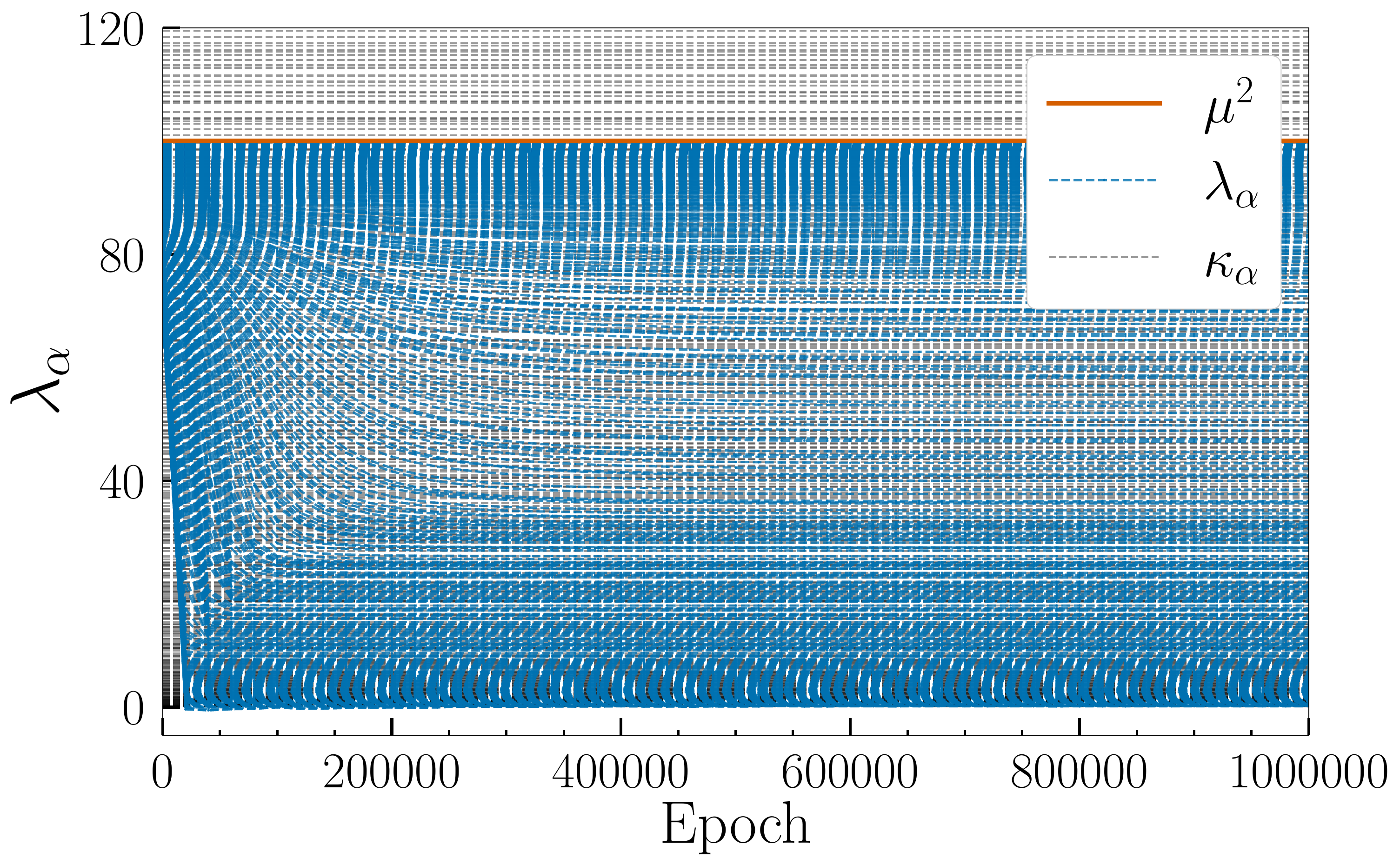}
    \includegraphics[width=0.39\textwidth]{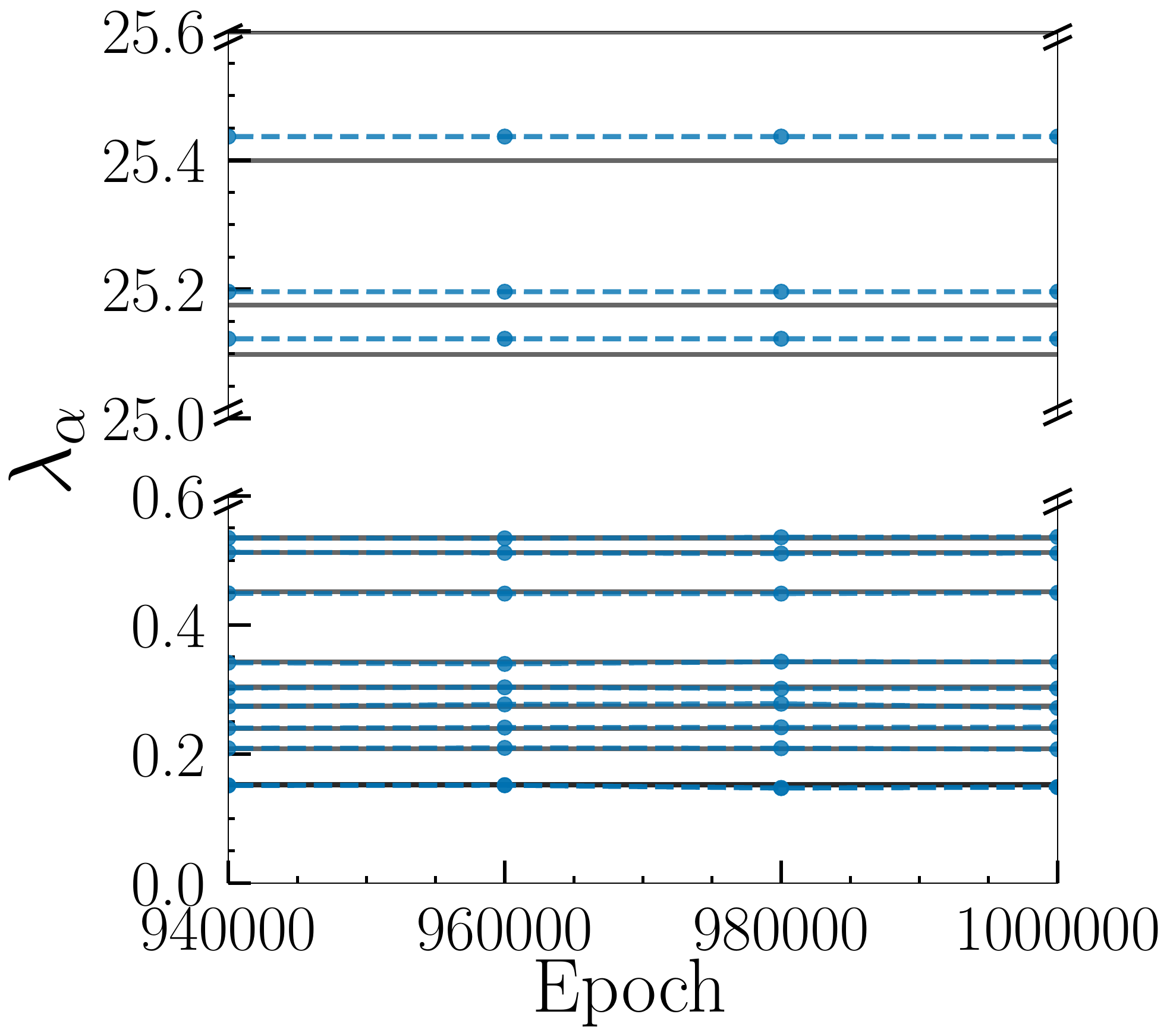}
    \caption{Left: evolution of the eigenvalues for the MNIST dataset with fixed
      RBM mass parameter $\mu^2=100$, and $N_v=N_h=784$. Right: evolution during
      the last few training epochs. The lowest eigenvalues have already matched
      their target values but the higher modes are still being trained, albeit at
      a very small learning rate.}
    \label{fig:MNIST_K_sm}
  \centering
    \vspace*{0.4cm}
    \includegraphics[width=0.56\textwidth]{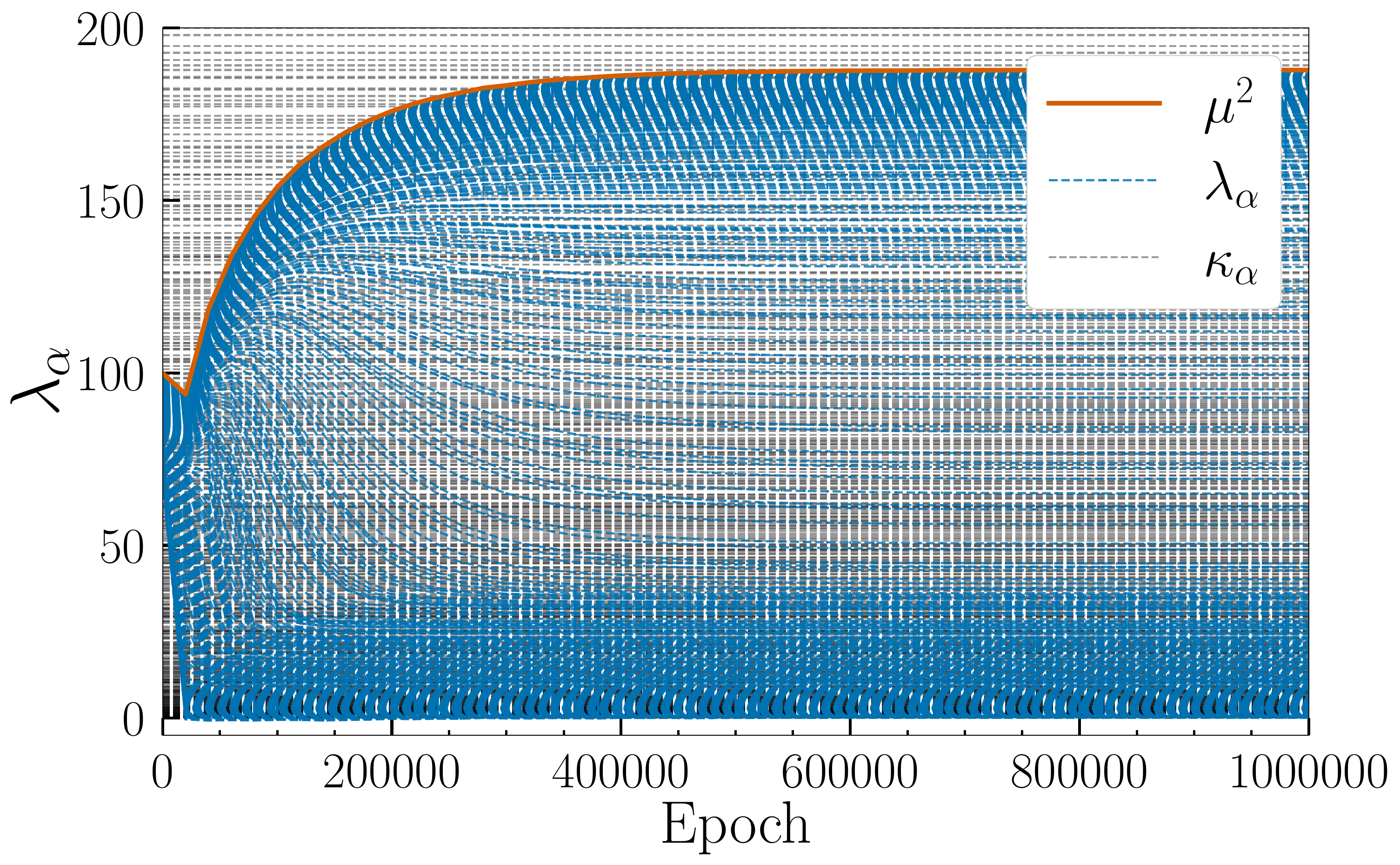}
    \includegraphics[width=0.39\textwidth]{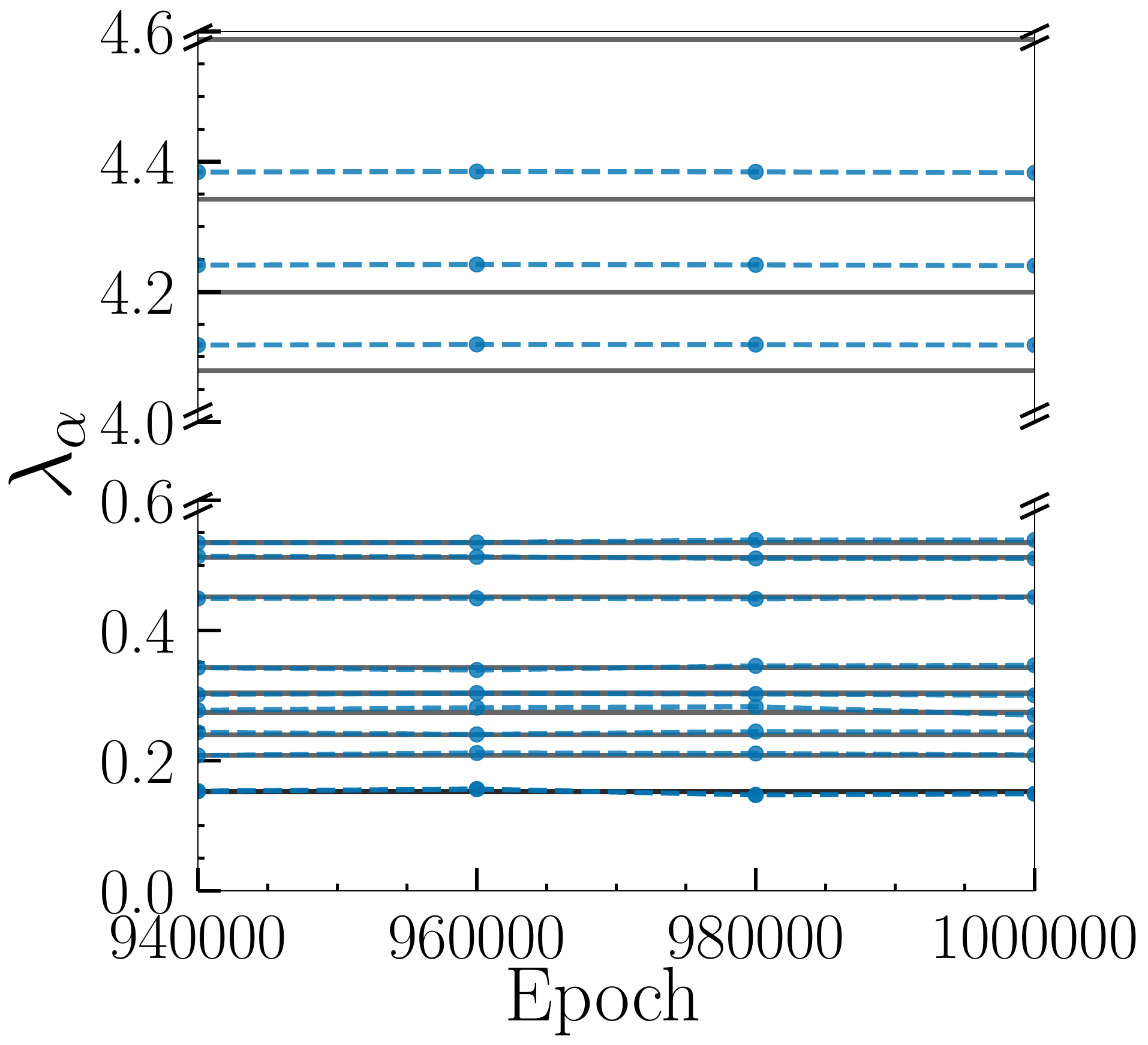}
    \caption{As above, but with a learnable RBM mass parameter $\mu^2$.}
    \label{fig:MNIST_K_dm}
 \end{figure*}

  We will now train the scalar field RBM on the MNIST dataset, starting with
  $N_h=N_v$ and a fixed RBM mass parameter $\mu^2=100$. The result is shown in
  Fig.\ \ref{fig:MNIST_K_sm} (left). As before, the horizontal dashed black
  lines are the target eigenvalues, obtained from the MNIST correlation matrix.
  The blue lines are the RBM eigenvalues. The initial values are close to
  $\mu^2$ and hence they become smaller during the learning stage. As above the
  infrared part of the spectrum is learnt quickest. This is further illustrated
  in Fig.\ \ref{fig:MNIST_K_sm} (right), where the evolution during the final
  60,000 epochs are shown (out of one million). The smallest eigenvalues agree
  with the target values but the larger ones have essentially stopped learning
  before reaching the correct value, due to the reduced learning rate. We note
  that the RBM mass parameter $\mu^2=100$ regulates the number of modes here.
  In fact, there are 289 modes below the cutoff set by $\mu^2$. Hence the
  number of hidden modes, $N_h=784$, can be reduced without a loss of quality.
  We come back to this below.

Without knowledge of the target spectrum, the (constant) value for $\mu^2$
  may be chosen to be on the small side; as is obvious in Fig.\
  \ref{fig:MNIST_K_sm} (left), there are many eigenvalues above $\mu^2=100$.
  This can be remedied by promoting $\mu^2$ to a learnable parameter. This is
  demonstrated in Fig.\ \ref{fig:MNIST_K_dm}, where $\mu^2$ increases such that
  the target spectrum can be better learnt. The learning dynamics employs a
  diminishing learning rate, see Appendix \ref{app:pcd}, which slows down the
  increase of $\mu^2$ but also hinders the learning of the spectrum beyond the
  infrared modes. As stated, larger eigenvalues give smaller contributions to
  the update equations, leading a slower learning.

 \begin{figure*}[t]
  \centering
    \subfloat[$N_h=784$]{\includegraphics[width=0.32\textwidth]{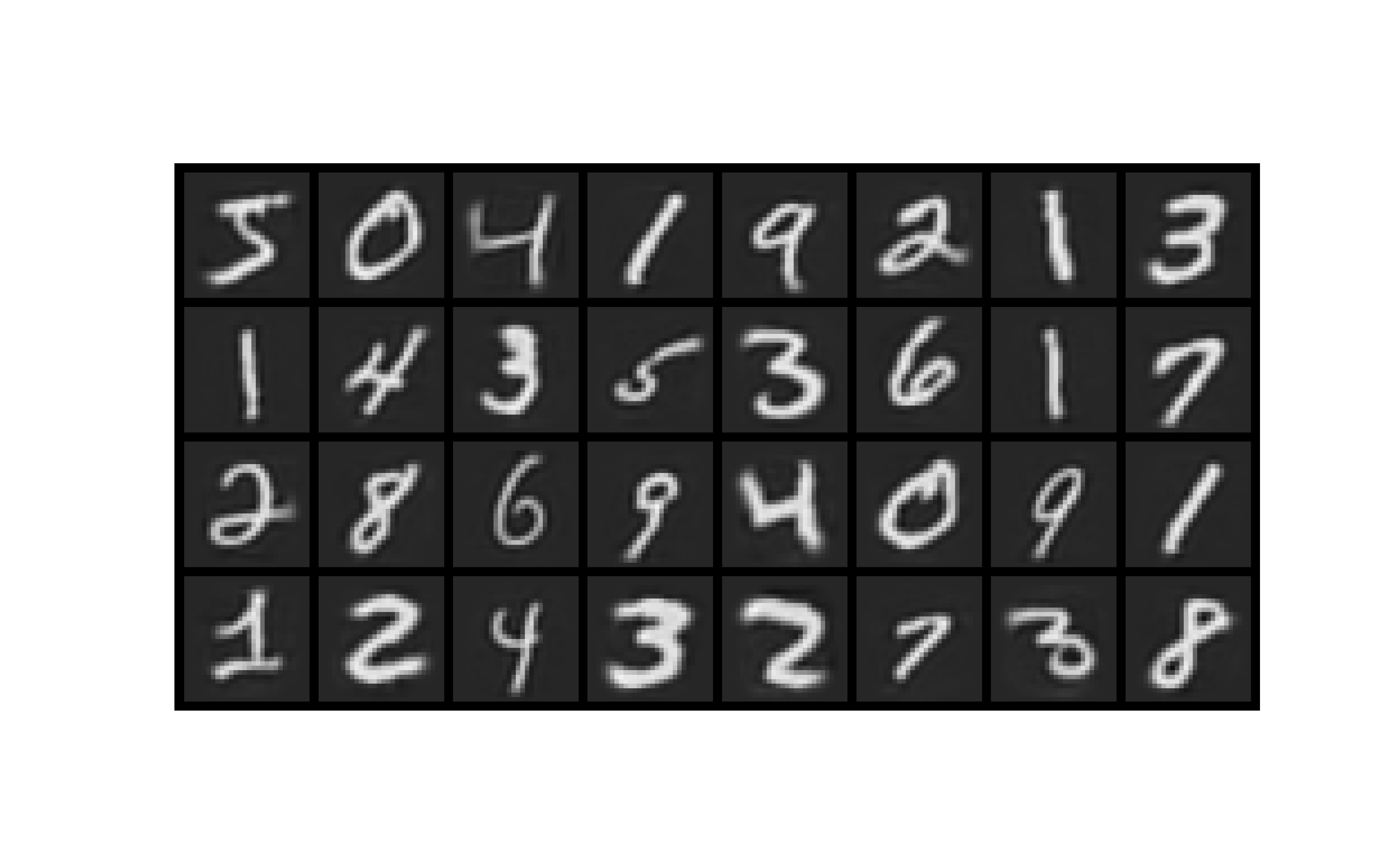}}
    \subfloat[$N_h=225$]{\includegraphics[width=0.32\textwidth]{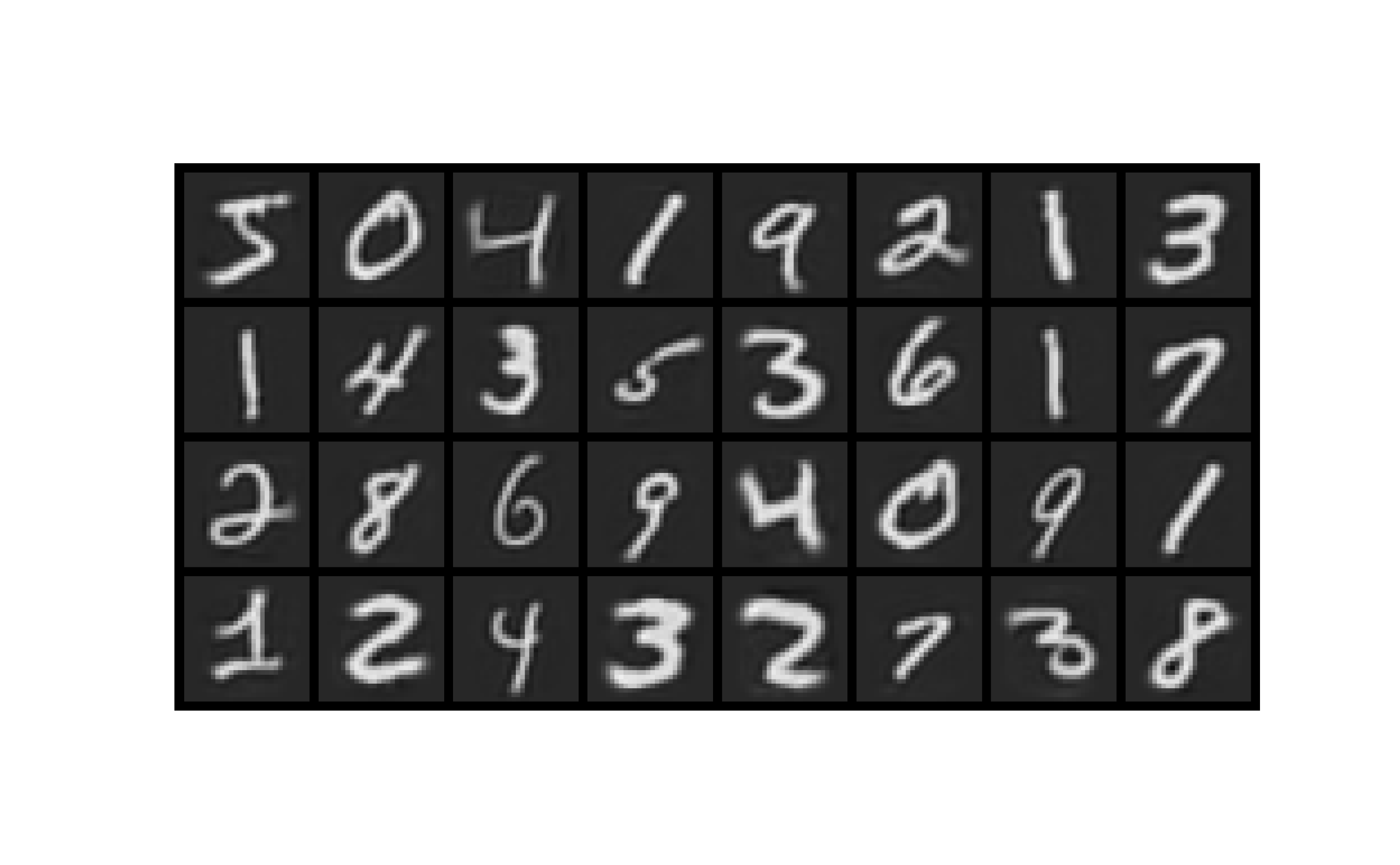}}
    \subfloat[$N_h=64$]{\includegraphics[width=0.32\textwidth]{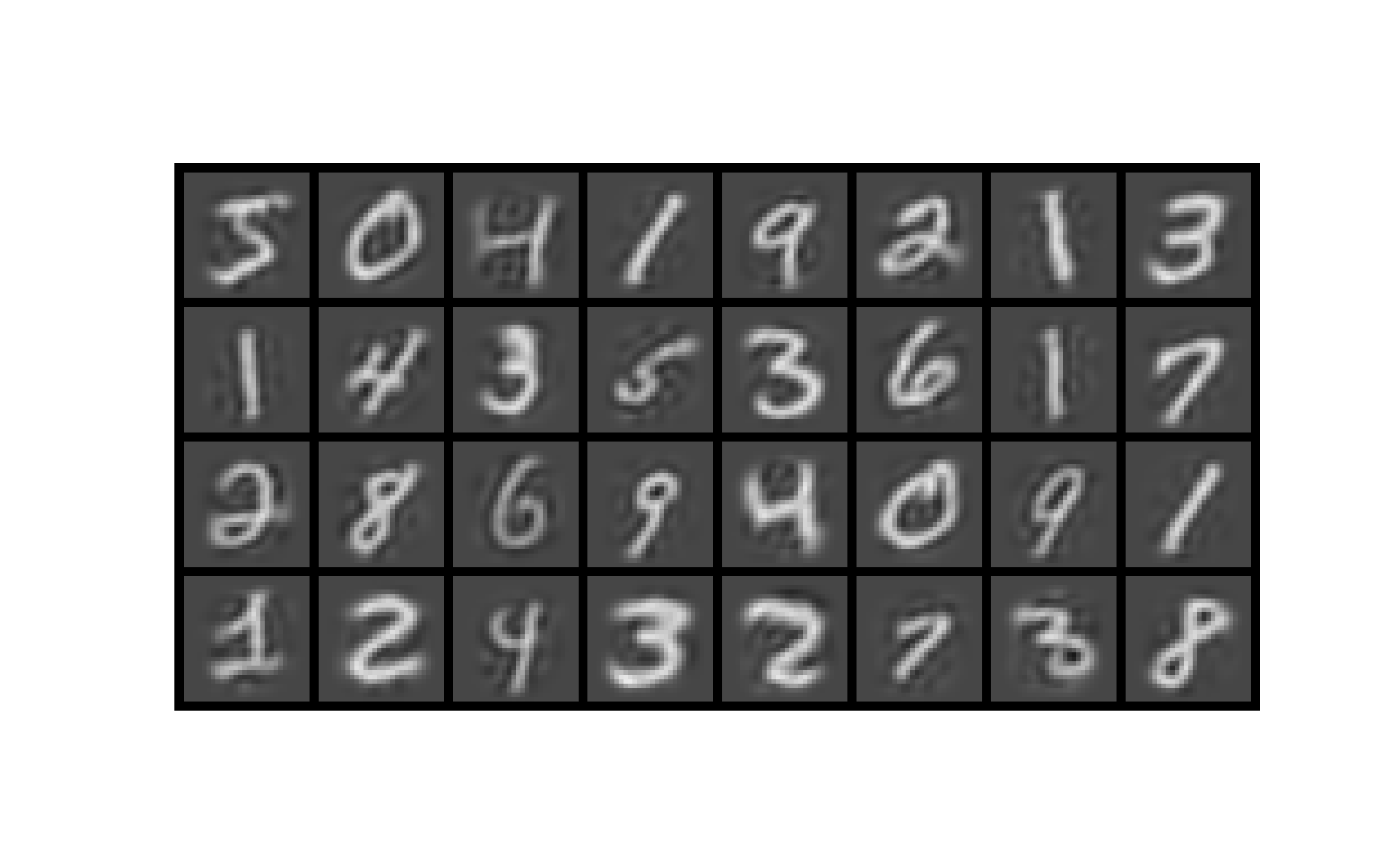}}
    \\
    \subfloat[$N_h=36$]{\includegraphics[width=0.32\textwidth]{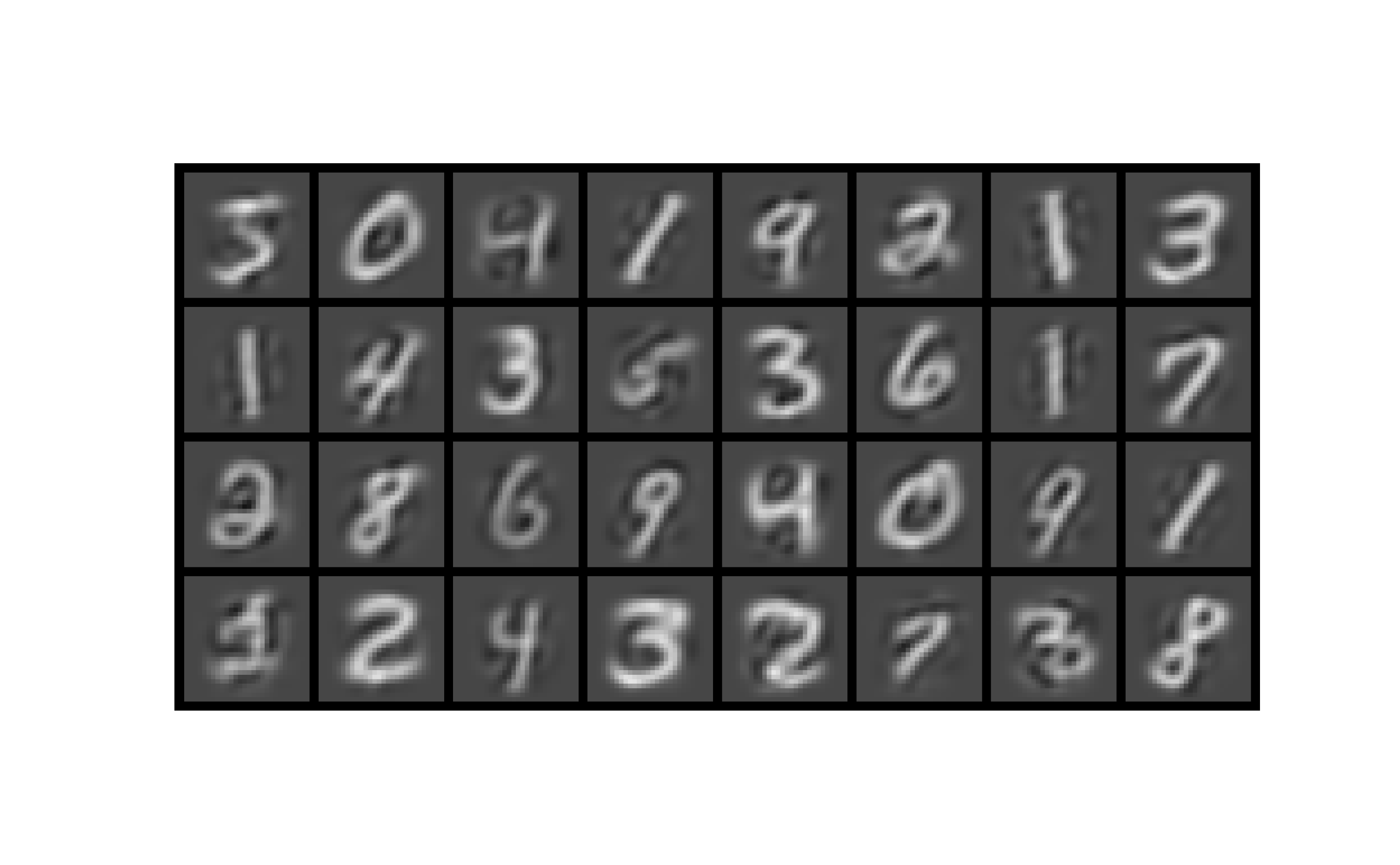}}
    \subfloat[$N_h=16$]{\includegraphics[width=0.32\textwidth]{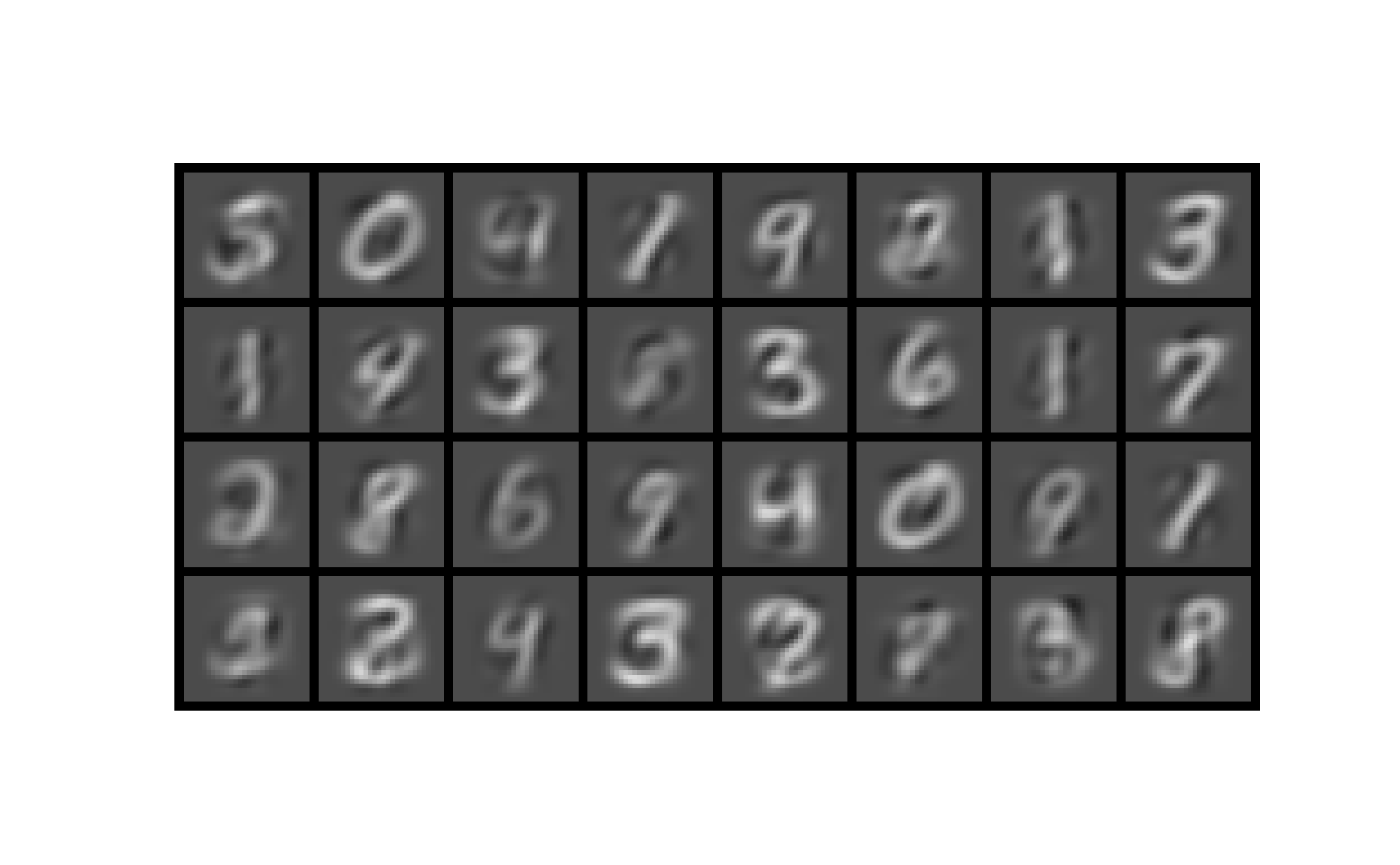}}
    \subfloat[$N_h=4$]{\includegraphics[width=0.32\textwidth]{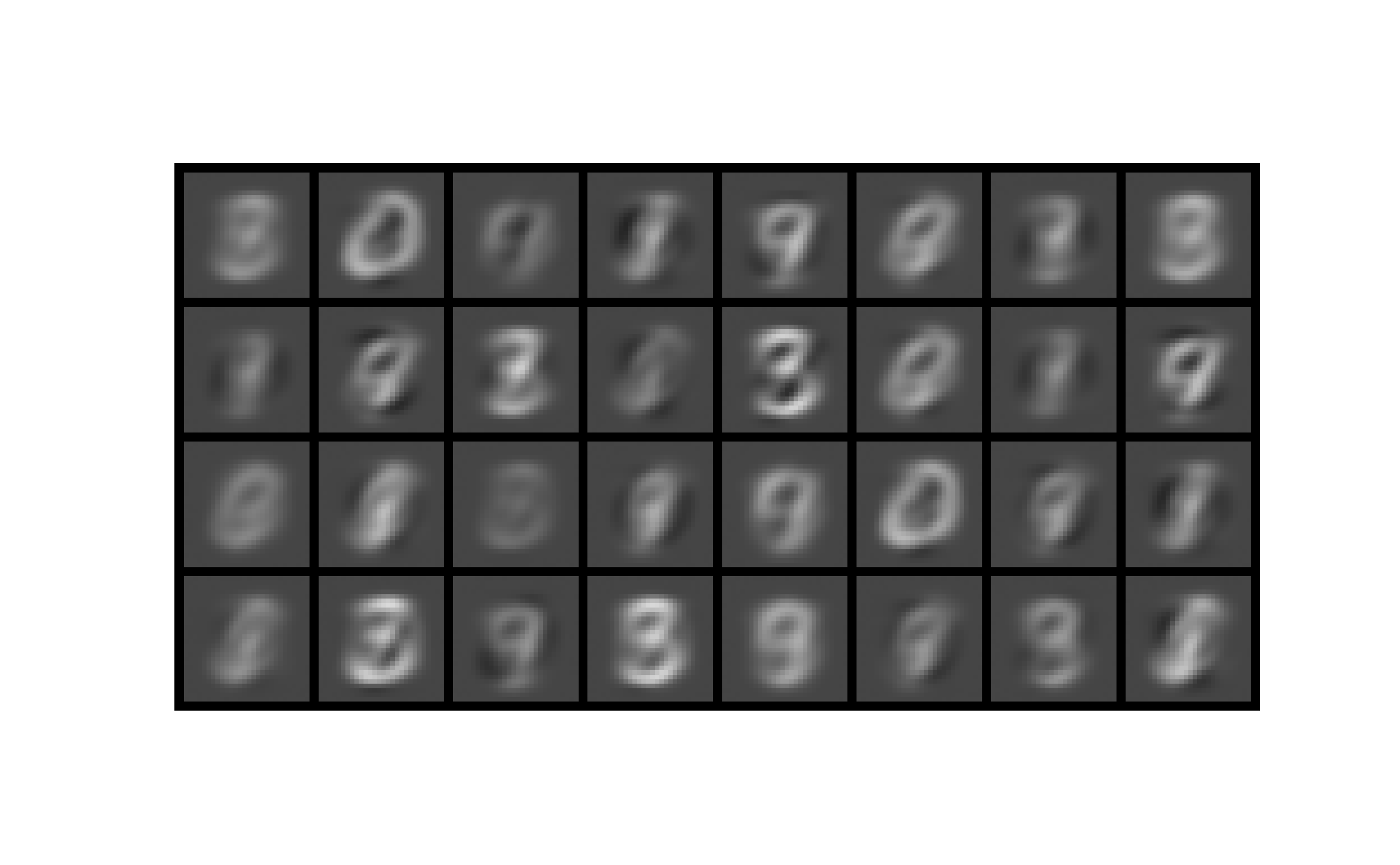}}
    \caption{Quality of regenerated images with different numbers of hidden
      nodes. As the number of hidden nodes decreases, the regeneration quality
      decreases.}
    \label{fig:MNIST_nh}
  \end{figure*}

  \begin{figure}[t]
    \centering
    \includegraphics[width=0.9\linewidth]{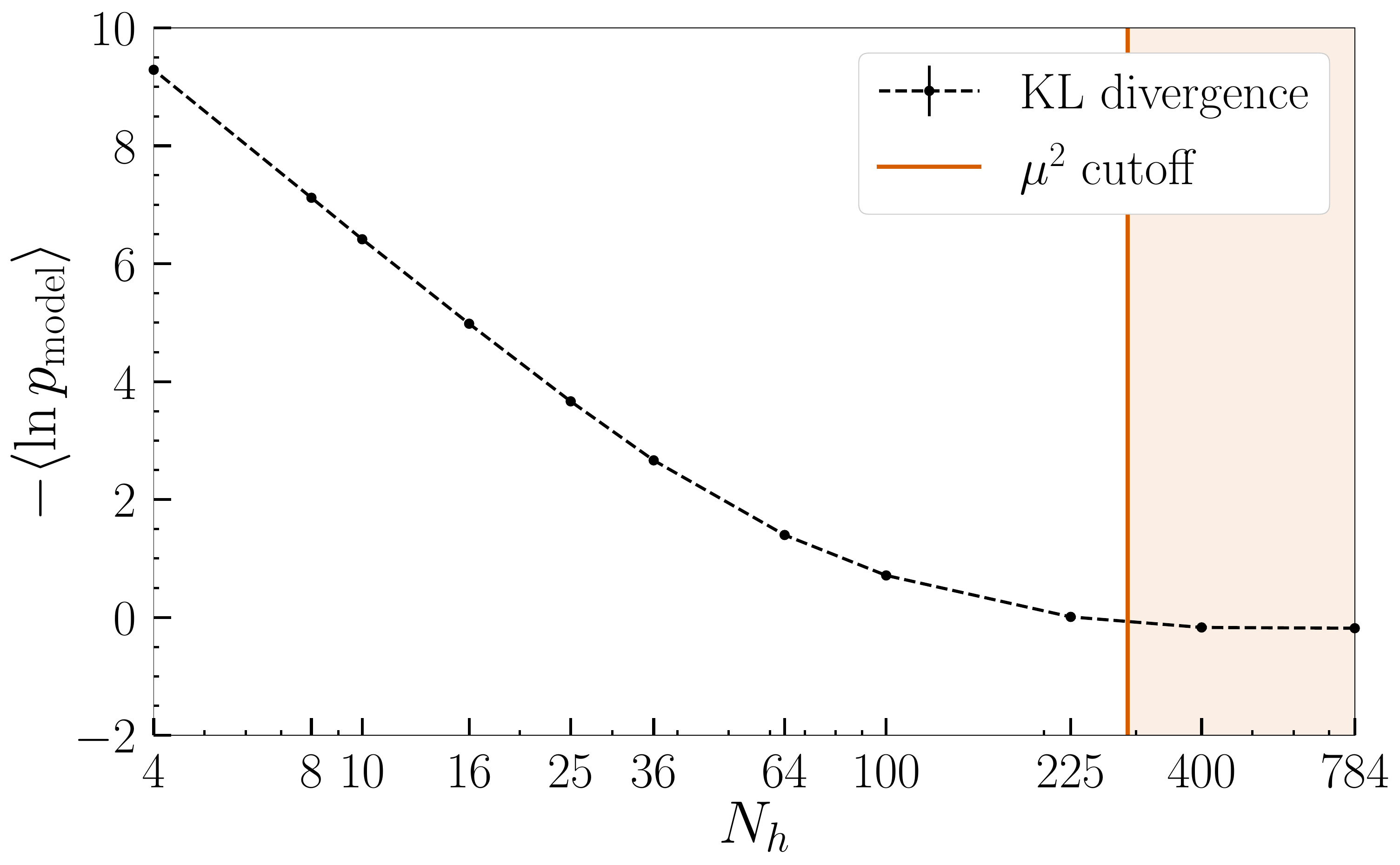}
    \caption{Data-averaged KL divergence for the trained RBM, as a function of
      the number of hidden nodes $N_h$ (on a logarithmic scale) at fixed
      $\mu^2=100$. For this value of $\mu^2$, the maximal number of modes
      included is $N_{\rm modes}^{\rm max} = 289$ and hence increasing $N_h$
      above this value does not lead to additional improvement. }
    \label{fig:KL}
  \end{figure}

  As in the scalar field case, we can regularise the MNIST data by
  choosing $N_h < N_v$. In this case, the number of modes that can be learnt
  depends on the number of modes in the spectrum with an eigenvalue less than
  $\mu^2$, and the number of hidden nodes.  Figure \ref{fig:MNIST_nh} shows
  the quality of regenerated images after one passes forward and backward
  through the trained RBM. Using the fixed RBM mass parameter $\mu^2=100$
  limits the maximal number of modes to be included to $N_{\rm modes}^{\rm
  max}=289$, the number of modes with an eigenvalue less than $\mu^2$. We
  observe that one needs at least around 64 hidden nodes to have
  an acceptable generation, which is considerably smaller than the maximal
  possible number. This illustrates that the ultraviolet part of the spectrum
  can be ignored.

  To give a more quantitative measure of the quality of regeneration, we have
  computed the data-averaged KL divergence for the trained model,
  \begin{widetext}
  \begin{align}
    KL(p_{\rm data} || p_{\rm model}) 
    =  -\frac{1}{\Nconf}\sum_{d=1}^\Nconf \log p_{\rm model}(\phi^{(d)}, \theta^*) 
    + \mbox{cst.} 
    =  \frac{1}{\Nconf}\sum_{d=1}^\Nconf \half {\phi^{(d)}}^T K \phi^{(d)} 
    +\log Z_{\rm model} + \mbox{cst.},
  \end{align}
  \end{widetext}
  where the constant ``cst'' term is independent of the model distribution. The
  result is shown in Fig.~\ref{fig:KL}. We indeed observe the KL divergence
  between the target distribution and the model distribution starts to increase
  as more modes are excluded. Adding modes beyond the cutoff imposed by the
  choice of $\mu^2$ does not increase the quality, as expected. As concluded
  ``by eye'' above, around 64-100 hidden nodes are required for a reasonable
  quality of regeneration.

\section{Interactions}
\label{sec:interact}
 
  Strictly speaking, the Gaussian-Gaussian RBM can only be exact if the target
  distribution is Gaussian as well. To go beyond this, one needs to introduce
  interactions. There are (at least) two ways of doing so. Motivated by the
  notion of QFT-ML \cite{Bachtis:2021xoh}, one may add local interactions via
  potentials defined on nodes, see Eqs.\  (\ref{eq:action}),
  (\ref{eq:potential}). A simple choice would be to add a $\lambda\phi^4$ term
  to each node on the visible layer since such systems are well understood and
  allow one to study e.g.\ spontaneous symmetry breaking in the context of the
  learning process. Of course, sampling the hidden layer requires a more costly
  sampling method than for the Gaussian case.

  One may also change the nature of the hidden nodes from continuous to
  discrete, taking e.g.\ $h_a=\pm 1$. This leads to the Gaussian-Bernoulli RBM
  (see e.g.\ Ref.\ \cite{Decelle_2021}), with the distribution and action
  \bea
    p(\phi, h) =&&\hm \frac{1}{Z} \exp(-S(\phi, h)),
    \\
    S(\phi,h) =&&\hm V_\phi(\phi) - \sum_{i,a} \phi_i W_{ia} h_a 
            + \sum_a \eta_a h_a,
  \eea
  where
  \be
    Z = \int D\phi\, \prod_{a=1}^{N_h}\sum_{h_a = \pm 1}\exp(-S(\phi, h)).
  \ee
  This gives the following induced distribution on the visible layer, 
  \be
   p(\phi) = \frac{1}{Z} \exp (-S(\phi)), 
   \ee
   with the effective action
   \be
   S(\phi) = V_{\phi}(\phi) - \sum_a \ln \left(2\cosh(\psi_a) \right),
  \ee
  where $\psi_a = \sum_i \phi_i W_{ia} - \eta_a$. A formal expansion in
  $\psi_a$ then yields, up to a constant, the effective action on the visible
  layer, 
  \bea 
    S(\phi) &&\hm= V_{\phi}(\phi) 
    - \sum_a \ln \left( 1 + \sum_{n=1}^{\infty} \frac{\psi_a^{2n}}{(2n)!} \right) \nn \\
    &&\hm= V_{\phi}(\phi) 
    +  \sum_a \sum_{n=1}^{\infty} (-1)^n \frac{c_n}{(2n)!}\psi_a^{2n},
  \eea
  with easily determined coefficients $c_n$. This is a highly nonlocal action.  

  To make the connection with the previous sections, it is straightforward to
  see that the quadratic ($n=1, c_1=1$) term yields the same structure as
  above, 
  \be
    -\sum_a \half\psi_a^2 = 
    -\frac{1}{2} \phi^T WW^T\phi - \frac{1}{2} \eta^T \eta  + \phi^T W\eta,
  \ee
  which, when combined with the RBM mass parameter $\mu^2$ on the visible
  layer, gives the same kernel, $K=\mu^2\id - WW^T$, and source, $J=W\eta$.

  Quartic interactions are generated at the next order. Taking for simplicity
  $\eta_a=0$, such that only even terms in $\phi_i$ are present, one finds the
  $n=2$ term ($c_2=2$),
  \be
    \sum_a \frac{1}{12}\psi_a^4 = \frac{1}{12}\sum_{ijkl} \lambda_{ijkl} \phi_i\phi_j\phi_k\phi_l, 
 \ee
 with the coupling
 \be 
 \lambda_{ijkl} = \sum_a W_{ia}W_{ja}W_{ka}W_{la}.
  \ee
  This is indeed a quartic term but with an a priori highly nonlocal
  coupling, generated by the all-to-all coupling to the hidden layer. From a
  QFT perspective, it would be of interest to study such a theory, which we
  postpone to the future.

\section{Conclusion}
\label{sec:conclude}

  In this paper, we have studied scalar field restricted Boltzmann machines
  from the perspective of lattice field theory. The Gaussian-Gaussian case can
  be understood completely. We have demonstrated, using analytical work and
  numerical experiments, that the scalar field RBM is an ultraviolet regulator,
  regulating the ultraviolet part of the spectrum of the quadratic operator of
  the target theory. This is also the case when the target probability
  distribution is not known, such as in the MNIST case, but where the spectrum
  can be extracted from the data-averaged correlation matrix. The cutoff is
  determined by the choice of the RBM mass parameter or the number of hidden
  nodes. This provides a clear answer to generally difficult questions on the
  choice of ML ``architecture", namely what are the consequences of choosing a
  particular setup. At least in this simple case the answer is straightforward
  and concerns the (in)sensitivity of the generative power of the RBM to the
  ultraviolet modes compared to the infrared modes.

  We have also shown that infrared modes are learnt the quickest. This is of
  interest for models which suffer from critical slowing down, for which
  infrared modes are usually hard to handle. Indeed, many ML (inspired)
  generative approaches have surprisingly short autocorrelation times, which
  is worth exploring further. 

  As an outlook, we note that in the final section, we have indicated two ways
  to go beyond the Gaussian-Gaussian case. The QFT-ML approach, in which local
  potentials are added to nodes on e.g.\ the visible layer, is convenient for
  LFT practitioners since the resulting models are well understood. Replacing
  the continuous hidden degrees of freedom with binary ones (Gaussian-Bernoulli
  RBM) yields models of a very different character, involving highly nonlocal
  interaction terms to all orders. It would be of interest to understand
  these constructions further using QFT methods.

  The data generated for this manuscript and simulation code are available from Ref. \cite{code}.

\vspace{0.2cm}

\section*{Acknowledgements}
G. A. and B. L. are supported by STFC Consolidated Grant No. ST/T000813/1. C. P. is supported by the UKRI AIMLAC CDT EP/S023992/1.
We thank ECT* and the ExtreMe Matter Institute EMMI at GSI, Darmstadt, for support and the participants of the ECT*/EMMI workshop {\em Machine learning for lattice field theory and beyond} in June 2023 for discussion during the preparation of this paper.

\appendix

\section{Details of the algorithm}
\label{app:pcd}

  The training equations for the RBM parameters $\theta$ read, schematically, 
  \bea
    \hspace*{-0.6cm} \theta_{n+1} &&\hm= \theta_n +\eta_n  \frac{\partial{\cal L}}{\partial\theta}, \nn \\
    \hspace*{-0.6cm} \frac{\partial{\cal L}}{\partial\theta} &&\hm= 
    -\half\left\bra \phi^T \frac{\partial K}{\partial\theta} \phi\right\ket_{\hspace*{-0.15cm} \rm target}
    + \half\left\bra \phi^T \frac{\partial K}{\partial\theta} \phi\right\ket_{\hspace*{-0.15cm} \rm model},
  \eea
  where $\eta_n$ is the learning rate. The first term on the rhs can be easily
  computed from the given data or target theory. The second term needs to be
  sampled from the model distribution, which is nontrivial. In most cases,
  this term is approximated by generating a Markov chain and truncating it
  after $k$ steps, where $k$ is empirically chosen. This is known as
  Contrastive Divergence (CD)\cite{pmlr-vR5-carreira-perpinan05a}. For standard
  CD updates, the Markov chain is initialised from the input data and then the
  successive chains are sampled by Gibbs sampling. A more efficient update
  algorithm is Persistent Contrastive Divergence
  (PCD)\cite{10.1145/1390156.1390290} and is used in this paper. PCD
  initialises the Markov chain from the last state of the most recent update.
  Since this last state of the previous chain is already closer to the
  representative of the model distribution, the new Markov chain is initialised
  with a nearly thermalised state and only requires a small number of updates.

\begin{figure*}[t]
  \centering
    \includegraphics[width=0.48\textwidth]{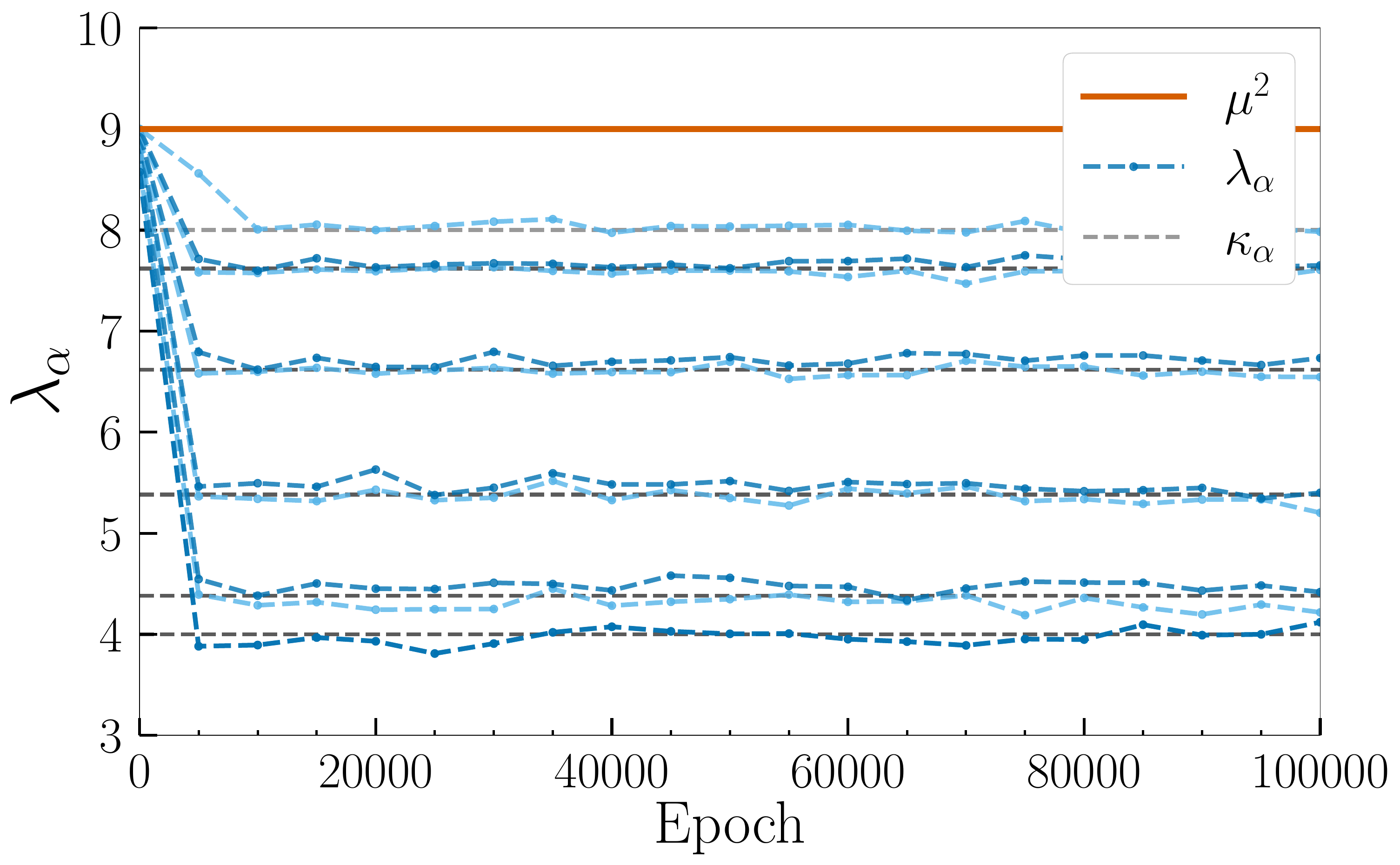}
    \includegraphics[width=0.48\textwidth]{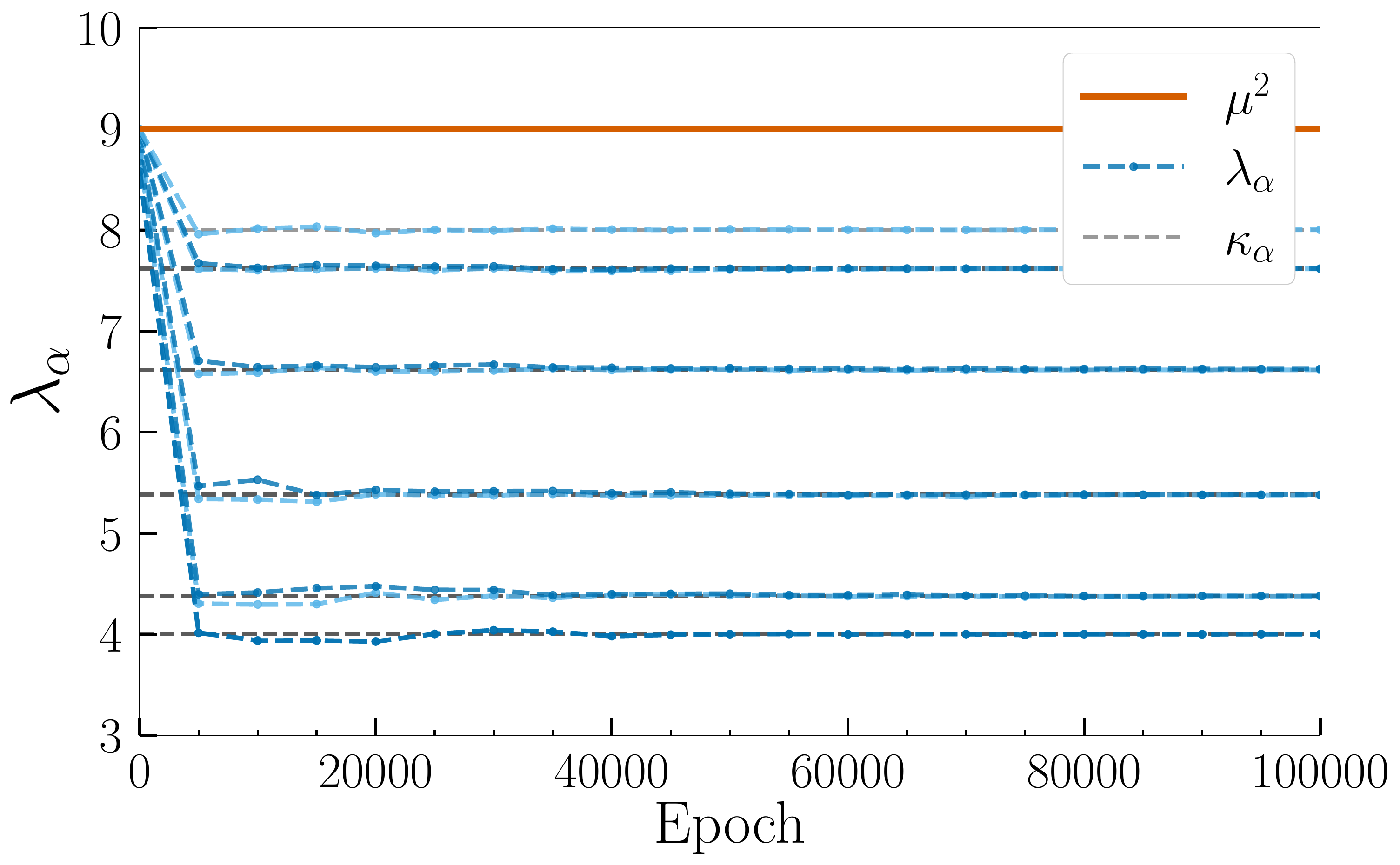}
  \caption{Scalar field RBM trained without (left) and with (right) learning
    rate decay, using $r=0.99, N_{\rm epoch}^{\rm rate}=128$, and a fixed RBM
    mass parameter $\mu^2=9$.}
  \label{fig:normal_m0_lrd}
  \end{figure*}

  \begin{figure}[t]
  \centering
    \includegraphics[width=0.9\columnwidth]{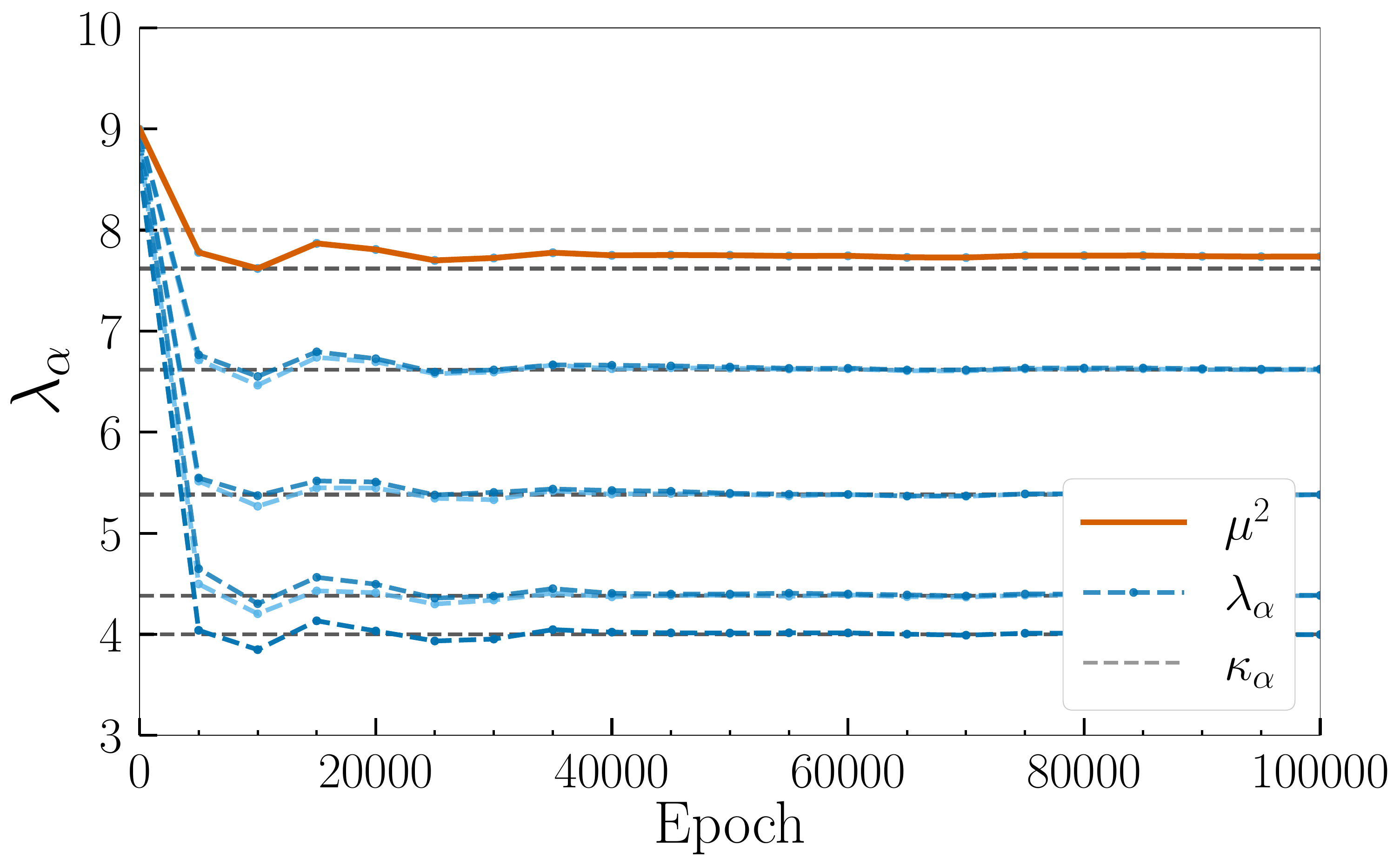}
  \caption{Scalar field RBM with trainable RBM mass parameter $\mu^2$ and
    learning rate decay as above. The model parameters are frozen before the
    RBM mass parameter reaches the (expected) largest eigenvalue of the target
    kernel.}
  \label{fig:lrd_etc}
  \end{figure}

  Alongside PCD, the gradient for each epoch is estimated by averaging over a
  minibatch. In the case of MNIST data, the training was done by using an
  effective correlation matrix obtained from the given dataset. Then 512
  parallel PCD Markov chains were generated to form a minibatch. For the scalar
  field theory case, the training was done by directly using the analytical
  form of the kernel matrix of the target distribution without predefined
  training data. Then for each training epoch, 128 parallel PCD Markov chains
  were generated to be averaged and used to estimate the gradient.

  The learning rate can be set to change during the training. For instance, one
  may multiply the learning rate by a factor of $r$ after every $N_{\rm
  epoch}^{\rm rate}$ epochs (e.g.\ $r=0.99$, $N_{\rm epoch}^{\rm rate}= 128$), 
  \be
    \eta_n = r\eta_{n-1} \quad\mbox{if} \mod(n, N_{\rm epoch}^{\rm rate}) = 0.
  \ee
  Hence the learning rate becomes smaller as more epochs have passed. The
  virtue of having a small learning rate during the later part of the training
  is that it allows the model to be finely trained and that it reduces
  statistical fluctuations.

  The effect of learning rate decay is shown in Fig.\ \ref{fig:normal_m0_lrd}.
  Two models are trained with the same hyperparameters and initialisation
  except for the learning rate decay parameters $r$ and $N_{\rm epoch}^{\rm
  rate}$. The first model shown in Fig.~\ref{fig:normal_m0_lrd} (left) is
  trained without learning rate decay. Fluctuation of the eigenvalues due to
  statistical noise remains. In contrast, the second model, shown in Fig.\
  \ref{fig:normal_m0_lrd} (right), uses learning rate decay with $r=0.99,
  N_{\rm epoch}^{\rm rate}=128$. Statistical fluctuations die off in the end,
  leading to a precise result.

  However, the values of $r$ and $N_{\rm epoch}^{\rm rate}$ should be chosen in
  a delicate manner. If the decay rate $r$ is too large, then the learning rate
  decreases too fast and the model freezes before it reaches the target
  destination. For example, in Fig.\ \ref{fig:lrd_etc}, the training flow of
  the scalar field RBM with the trainable mass parameter and $r=0.99, N_{\rm
  epoch}^{\rm rate}=128$ is shown (compare with Fig.\
  \ref{fig:scalar_result_m1}). The model does not suffer when it is learning
  infrared modes, which are learnt quickest, but it fails to learn the highest
  mode of the target kernel. The model parameter freezes out before it reaches
  the target. One can decide the learning rate decay parameters by observing
  the regenerated samples and measuring the goodness of those. Since the
  ultraviolet modes are less relevant compared to the infrared ones, one can
  accept a truncation of the higher modes provided a target goodness is
  achieved. One can also employ an adaptive learning rate decay.

  We have also looked at employing momentum optimisation and $L_2$
  regularisation of the coupling matrix but have found no need for these.

\section{Kullback-Leibler divergence}
\label{sec:KL}

  For completeness, we evaluate here the KL divergence in
  the case that both the target theory and the model are Gaussian, without
  linear terms. This allows us to compare it with the log-likelihood in the
  main text. We consider the KL divergence,
  \be
    KL(p || q) = \int D\phi\, p(\phi) \log\frac{p(\phi)}{q(\phi,\theta^*)},
  \ee
  with $p(\phi)$ the target distribution and $q(\phi,\theta^*)$ the trained
  distribution (hence the asterisk on $\theta$). We assume the learning process
  has found the correct eigenbasis, such that the distributions are
  \bea
    \hspace*{-0.6cm} p(\phi) = \frac{1}{Z_p} e^{-\half\sum_i a_i\phi_i^2}, 
    &&\ \hm Z_p = \prod_i\int d\phi_i\, e^{-\half a_i\phi_i^2}, \\
    \hspace*{-0.6cm} q(\phi, \theta^*) = \frac{1}{Z_q}e^{-\half\sum_i b_i\phi_i^2}, 
    &&\ \hm Z_q = \prod_i\int d\phi_i\, e^{-\half b_i\phi_i^2},
  \eea
  where all eigenvalues $a_i, b_i>0$. To make the connection with the scalar
  theory with $N_h<N_v$ in Sec.\ \ref{sec:gaussian}, we note that
  $i=1,\ldots,N_v$, and that after training,
  \be
    b_i = \begin{cases} \kappa_i &\ i\leq N_h, \\  \mu^2 &\ i>N_h.
    \end{cases}
  \ee
  It is then straightforward to evaluate the KL divergence. In particular,
  \be
    \log\frac{p(\phi)}{q(\phi,\theta^*)} 
    = -\half \sum_i \left(a_i-b_i\right)\phi_i^2 -\log\frac{Z_p}{Z_q},
  \ee
  with
  \be
    \log\frac{Z_p}{Z_q} = \half \sum_i \log\frac{b_i}{a_i}.
  \ee
  Putting everything together, one finds
  \be
    KL(p || q) = 
    \half\sum_i \left( -1+\frac{b_i}{a_i} - \log\frac{b_i}{a_i} \right) \geq 0.
  \ee
  Each term is non-negative, and $KL(p || q) \geq 0$, as it should be. The
  equality is achieved only when each eigenvalue is correctly determined. For
  the scalar theory in Sec.\ \ref{sec:gaussian}, this becomes
  \be
    KL(p || q) = \half\sum_{i=N_h+1}^{N_v} \left( -1+\frac{\mu^2}{\kappa_i} 
    - \log\frac{\mu^2}{\kappa_i} \right).
  \ee

\end{document}